# On a Coarse-Graining Concept in Colloidal Physics with Application to Fluid and Arrested Colloidal Suspensions in Shearing Fields

A dissertation submitted to
ETH ZURICH

for the degree of
Doctor of Sciences

presented by

Alessio Zaccone

Master of Science in Chemical Engineering,
Politecnico di Torino
born 7th of September, 1981
citizen of Italy

accepted on the recommendation of
Prof. Dr. M. Morbidelli (ETH Zurich), examiner
Prof. Dr. H. J. Herrmann (ETH Zurich), co-examiner
Dr. E. Del Gado (ETH Zurich), co-examiner

Zurich 2009

# Acknowledgements

I am very grateful to Prof. Morbidelli for having given me the opportunity to work at ETH Zurich, for his guidance, and for the freedom accorded to me in the choice of the research topics. I am deeply indebted to Dr. Hua Wu and Dr. Emanuela Del Gado for having guided me throughout my PhD research activities with their precious advice. Finally, a good part of this thesis would not be there without the valid assistance and hard work by Daniele Gentili. Special thanks go also to Dr. Marco Lattuada for many discussions.

The present work is dedicated to the memory of my grandfather, Cav. Uff. Guido Benzi.

> *"Hunc igitur terrorem animi tenebrasque necessest*
> *Non radii solis neque lucida tela diei*
> *Discutiant, sed naturae species ratioque.*
> *Principium cuius hinc nobis exordia sumet,*
> *Nullam rem e nilo gigni divinitus umquam."*
>
> T. Lucretius Caro, *De Rerum Natura* I, 146-150

> *"Contraddictio est regula veri"*
>
> G. F. W. Hegel, *Habilitationsschrift*

# Abstract

We poorly understand the macroscopic properties of complex fluids and of amorphous bodies in general. This is mainly due to the interplay between phenomena at different levels and length-scales. In particular, it is not necessarily true that the microscopic level (dominated by direct interactions) coincides with the level where the continuum description comes into play. This is typically the case in the presence of structural inhomogeneities which are inherent to all structurally disordered states of matter below close packing. As a consequence, the macroscopic response to external fields of either fluid or arrested disordered states is not well understood. In order to disentangle this complexity, in this work we build upon a simple yet seemingly powerful concept. This can be summarized as follows: the mesoscopic length-scale of structural inhomogeneities is assumed to be the characteristic length-scale of the effective building blocks, while the degrees of freedom of the primary particles are integrated out. Theoretical results are derived, in the present work, for the macroscopic response of fluid and dynamically arrested model colloidal states in fields of shear. The predictions of the coarse-grained theories and the applicability of the principle are tested in comparison with original simulation and experimental data.

# Sommario

Il corrente grado di comprensione delle proprieta' macroscopiche di fluidi complessi e solidi amorfi e' insoddisfacente. Cio' si deve soprattutto alla compresanza di fenomeni che hanno luogo a livelli e scale di lunghezza molto diversi. In particolare, nella maggior parte dei casi, non e' detto che il livello microscopico (dominato da interazioni dirette fra i costituenti primari) coincida con il livello di descrizione dei continui o, come si suol dire, idrodinamico. Tutto questo e' vero soprattutto a densita' intermedie fra il limite diluito (dove teorie a due corpi sono solitamente valide) e quello concentrato (laddove di solito domina la repulsione da volume escluso). Di conseguenza, finora non e' stato possibile comprendere appieno la risposta macroscopica a un campo esterno, ne' per sospensioni colloidali fluide ne' per sospensioni colloidali arrestate dinamicamente. Al fine di districare codesta complessita', in questo lavoro ci proponiamo di sviluppare un concetto semplice ma potenzialmente di grande impatto. Tale principio puo' riassumersi nel modo seguente: il livello microscopico e quello idrodinamico vengono considerati separatamente utilizzando il concetto di "clusters" come particelle effettive o "rinormalizzate". Percio', in buona approssimazione, il livello delle interazioni microscopiche influenza piu' che altro il processo di formazione di "clusters". Quindi, la descrizione a livello dei continui o idrodinamico puo' essere fatta applicandola direttamente a "clusters" trattati come fossero le particelle primarie costituenti il sistema e utilizzando come frazione volumica quella occupata dai "clusters". Combinando questo principio di "coarse-graining" con metodi standard della meccanica statistica e dei continui, oltre che con originali studi sperimentali e di simulazione numerica, possiamo finalmente spiegare e in alcuni casi anche descrivere quantitativamente la risposta macroscopica a campi di taglio di sospensioni colloidali interagenti sia nel regime fluido (nei termini della viscosita' del fluido) sia nel regime dinamicamente arrestato (nei termini del modulo elastico di taglio). Riportiamo anche osservazioni sperimentali, in sospensioni aggreganti semi-diluite, di fenomeni finora noti solo in sospensioni molto concentrate. Tali fenomeni ("shear-thickening", "shear-induced jamming", "yield-stress") possono essere spiegati e interpretati solo utilizzando il concetto base di "coarse-graining" proposto e sviluppato in questa sede.

# Table of Contents

# 1. Introduction

## *1.1 Fluid colloidal suspensions under shear flow*

### 1.1.1 The Smoluchowski equation, shear-induced aggregation, and the microscopic level

As opposed to polymer melts, the rheology of colloidal suspensions has remained a challenging issue for statistical mechanics till nowadays. This may be partly due to the higher sensitivity to shear of colloidal systems where even modest shear rates can result in significant perturbations. Another reason lies in the smaller length scales and geometrical environment which are relevant microscopically: since the 70's the dynamics of polymers has been effectively described by means of a "mean-field tube" with >$10^3$ primary constituents (chains) [1], whereas the number of neighbours around a colloid particle, both in the liquid and in the solid state, is always <20. This makes impractical any mean-field description of many-particle effects under non-dilute conditions. Another source of complexity is certainly the interplay between Brownian motion (the essential feature of colloidal suspensions), the imposed field of shear, and the interparticle interactions. The latter ones are often reduced to the mere hard-core repulsion (hard-sphere potential) which greatly simplifies analysis. The hard-sphere (HS) system is the simplest one where many-body thermodynamic and hydrodynamic interactions are evident and it therefore represents the essential test of theories which aim at capturing these effects. Nonetheless, the HS model presents on the other hand peculiarities which make it rather exotic in certain respects (just think of the fact that energy is totally absent as a parameter) [2]. Further, most colloidal systems do interact through more complex potentials which may be comprised of an attractive part (usually due to van der Waals forces), a longer range repulsive component (Coulomb or more frequently screened-Coulomb, as in aqueous systems), or a superposition of the two giving rise to a barrier in the interaction potential. Thus, such suspensions are often referred to as "charge-stabilized". This is by far the most important case both in Nature and in industrial practice. Its mathematical form was proposed around 1940 by Derjaguin and Landau in the USSR and by Verwey and Overbeek in the Netherlands, independently, and the corresponding interaction potential will be referred to from now on simply as "DLVO" [3]. It would be highly desirable to incorporate the microscopic complexity brought about by such direct interactions into the many-body dynamics which ultimately determine the macroscopic properties. A significant step in this direction would be to first identify the level and mechanisms where many-body dynamics are dominant over microscopic details and viceversa. An attempt along this line is presented in this work with the focus in the first part being on the viscosity of aggregating charge-stabilized suspensions in shear.

It is well known to anybody working with colloidal suspensions that the imposition of shear may or may not result in aggregation phenomena and that the assessment of colloidal stability under shear is elusive [3]. This is a crucial problem in colloid science and has a high practical relevance because colloids in most applications have to flow through tubes and channels. What is ubiquitously observed in industrial processes is that flowing suspensions may remain completely stable for an almost unlimited time or, most frequently, may completely coagulate after a certain induction time which varies enormously from system to system in a hitherto unpredictable fashion. The change in the suspension viscosity is often so dramatic that the system stops flowing. This phenomenon, which goes under the denomination of rheopexy or anti-thixotropy (=viscosity increasing with time under steady shear), is a lucid example of how intimately the macroscopic rheological properties of colloidal suspensions are connected to what is going on at the microscopic level (i.e. aggregation, in this case). Such abrupt increase of viscosity and eventual solidification is not only well known in industrial practice but is commonly observed also in natural and biological contexts. Rheopexy (anti-thixotropy), as a consequence of aggregation of the primary colloids under shear, occurs indeed in biological fluids such as protein solutions (e.g. the synovial fluid which lubricates mammalian freely moving joints) and blood plasmas [4,5]. Further, it has been shown that rheopexy, again as a result of clustering and aggregation under shear, plays a key role in the formation of spider silk within the spider's spinneret, where conditions of elongational and shear flow are present which enhance protein aggregation leading to large intermediate aggregates to be further extruded into a light material with formidable mechanical properties [5,6]. The induction period is observed in all systems with a long-range repulsive component of interaction (e.g. charge-stabilization), such as proteins like in the mentioned examples, and is highly suggestive of an activation mechanism for aggregation driven by shear, as was speculated in [7] on the basis of experiments on emulsions in simple shear where the induction period was found to decrease exponentially with the shear rate (see Figure 1-1). One reason for the great variability in the induction time preceding the viscosity explosion may reside in the fact that such systems are driven far from equilibrium so that their properties depend much on the initial conditions and on the history. Nevertheless, as suggested also by the work of Guery et al. [7], it should be possible to identify the probabilistic laws governing these phenomena. This is certainly possible by resorting to an appropriate, and possibly effective, statistical mechanics formulation of the problem. For example, a theoretical framework which is successful in describing the salient features of the macroscopic rheological properties of dense Brownian suspensions (as well as of soft glassy materials) is Mode-Coupling theory [8]. This is based on solving the equations of motion for the density correlators (within the Zwanzig-Mori formalism) with suitable approximations for the coupling constants containing the information about the microscopic phyics. The fact that hydrodynamic interactions (HI) are neglected in MCT (which is a liquid theory) seems to suggest that hydrodynamics may not be necessarily the crucial point [9].

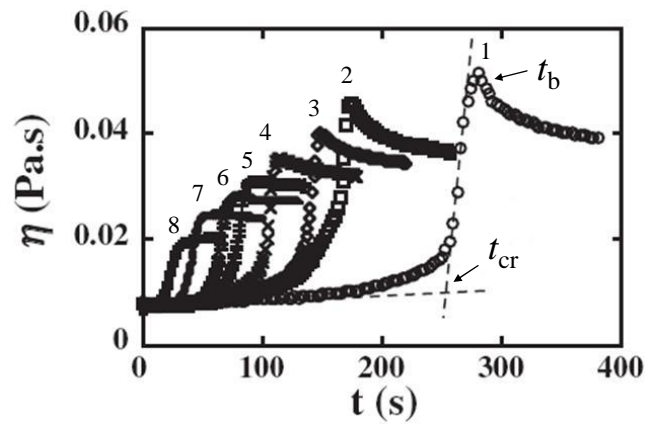

**Figure 1-1: Viscosity as a function of time for sheared charge-stabilized emulsions from Ref. [7]. $t_{cr}$ and $t_b$ refer to the onset times of self-accelerating kinetics and breakup, respectively. Increasing numbers on top of each curve correspond to increasing values of applied shear rate in the rheometer. As shown in Ref. [7], $t_{cr}$ decreases exponentially with the applied shear rate, which is suggestive of an activation process.**

Other approaches, based on numerics, such as Stokesian Dynamics (SD), suggest on the other hand that hydrodynamic interactions (which mediate interactions in colloidal systems and represent the only non-trivial difference with respect to molecular ones) may be important in the rheology of non-dilute colloidal suspensions. All these approaches, however, neglect another aspect which plays almost always a role in "real world" systems: the so-called direct or interparticle interactions between colloids. Indeed, the above mentioned methods exclusively deal, so far, with hard-sphere systems. Instead, the problem that we are tackling is clearly much affected by the direct interactions which are likely to play a major role since there the viscosity is a direct result of the aggregative stability and aggregation kinetics of the system (these being in turn a direct result of the interactions). Thus, for real systems, it is first necessary to provide a basic understanding of what is going on at the microscopic level and then try to include this information within an effective hydrodynamic framework to arrive at the macroscopic properties. The first step means, essentially, to unbundle the interplay at microscopic level between: 1) Brownian motion, 2) shear, 3) direct interactions. These elements need to be considered within the appropriate equation of motion for the phase space coordinates of Brownian particles. The most general description is given in terms of the set of positions and momenta of the Brownian particles [10]. This is equivalent to considering the solvent molecules as already at thermal equilibrium and that the relaxation times of the relevant process (i.e. Brownian motion) are thus long as compared with the time scale of the individual events causing its dynamics (i.e. the collisions with the solvent molecules) [10]. This is the assumption underlying the Fokker-Planck equation [10,11]. The latter is the equation of motion for the probability density function of the momentum and position coordinates of all the Brownian particles in the system. The corresponding time scale is known as the Fokker-Planck time scale and is such that changes in both momentum and position coordinates are important. A further simplification is to consider a time scale where the momenta of the Brownian particles are relaxed to equilibrium with the thermal bath of the solvent molecules. Thus the equation of motion is written in this case only for the probability density function in the set of position coordinates of the Brownian particles. It goes under the name of Smoluchowski equation and the pertinent time scale is the Smoluchowski or diffusive time scale. For an *N*-particle system it reads [10]

$$\frac{\partial}{\partial t} g(\mathbf{r}_1...\mathbf{r}_N,t) = \hat{\mathcal{L}}_S g(\mathbf{r}_1...\mathbf{r}_N,t) \tag{1.1}$$

where the Smoluchowski operator is given by

$$\hat{\mathcal{L}}_S(...) = \nabla \cdot \mathbf{D}(\mathbf{r}) \cdot [\beta[\nabla U](...) + \nabla(...)] - \nabla \cdot [\mathbf{\Gamma} \cdot \mathbf{r}(...) + \mathbf{C}'_s(\mathbf{r})(...)] \tag{1.2}$$

Under the approximations just mentioned above, differentiation is restricted over positional coordinates. $\mathbf{D}(\mathbf{r})$ is the microscopic diffusion matrix, $\mathbf{r}$ the interparticle distance vector, $U$ is the direct interaction potential, $\mathbf{\Gamma}$ is the external (linear) velocity gradient tensor, and $\beta = 1/k_B T$ is the Boltzmann factor. $\mathbf{C}_s'(\mathbf{r}) = \beta \sum_n \mathbf{D}_{jn} \cdot \mathbf{C}_n$ accounts for many-body hydrodynamic forces since the $\mathbf{C}_n$ are the hydrodynamic disturbance matrices as defined in the textbooks [10]. The basic type of velocity field is the linear one. In this case the velocity at a point $\mathbf{r}$ in space is given by $\mathbf{v}(\mathbf{r}) = \mathbf{\Gamma} \cdot \mathbf{r}$ and the corresponding velocity gradient tensor for simple shear, for a fluid moving in the x-direction with the velocity gradient in the y-direction, reads

$$\mathbf{\Gamma} = \dot{\gamma} \begin{pmatrix} 0 & 1 & 0 \\ 0 & 0 & 0 \\ 0 & 0 & 0 \end{pmatrix} \tag{1.3}$$

$\dot{\gamma}$ being the shear rate.

Solving Eqs. (1.1)-(1.3) for the set of N-particle positions as a function of time t clearly yields the complete information about the system and its temporal evolution. All the relevant physics is contained in the Smoluchowski operator where we can identify two main contributions: the term due to diffusion and direct interactions (the first term on the r.h.s.) and the term due to the flow effect (the second term on the r.h.s.).

Solving Eqs. (1.1)-(1.3) analytically is very challenging. The mathematical framework thus has to be simplified by means of inspired approximations in order to extract physical information. The first approximation is to reduce the N-particle problem to a two-particle problem. This is exact in the limit of very dilute systems, i.e. when the number concentration of particles $c \to 0$, and the probability density function reduces to the pair-correlation function $g(\mathbf{r}, t)$, where $\mathbf{r}$ is the interparticle distance vector. It has been recognized that such approximation still captures most of the physics of the problem, especially at the microscopic level. For example, recently it has been shown that with hard-spheres, the salient non-Newtonian rheological behaviours of concentrated HS suspensions seen in experiments and simulations seem to be indeed well captured by the two-body Smoluchowski equation [12]. Further, this approximation allows for incorporating the hydrodynamic effects directly into the velocity field for which analytical expressions are at hand. Since we are primarily concerned with the microscopic level at this stage, a further legitimate approximation is to regard the diffusion matrix as comprised of just the radial or isotropic term so that the mutual diffusion coefficient for two particles is given by $D = 2D_0$ where $D_0$ is the self-diffusion coefficient given by the familiar Stokes-Einstein formula. Under these approximations, the formalism simplifies to

$$\frac{\partial}{\partial t}g(\mathbf{r},t) = \hat{\mathcal{L}}_S g(\mathbf{r},t) \tag{1.4}$$

and the Smoluchowski operator is now given by

$$\hat{\mathcal{L}}_S(...) = \nabla \cdot D(r) \cdot [\beta[\nabla U](...) + \nabla(...)] - \nabla \cdot [\mathbf{v}(\mathbf{r})(...)] \tag{1.5}$$

where the effect of the flow disturbance induced by the second particle on the tagged one is included in velocity field $\mathbf{v}(\mathbf{r})$ by means of the corrections derived by Batchelor and Green [13]. However even this two-body simplified formulation poses significant analytical difficulties, even for noninteracting systems [12]. In chapter 2 we will present further physically-grounded approximations which allow for analytical solutions of the aggregation problem within the Smoluchowski equation framework and thus leading to a kinetic theory of colloidal aggregation under shear. It will also be shown how this microscopic-level result can be used to explain the sharp increase (rheopexy) in the time evolution of viscosity in charge-stabilized (DLVO-interacting) suspensions as is visible in Figure 1-1.

## 1.1.2 The mesoscopic level as the relevant input to the continuum treatment of macroscopic properties

This microscopic-level description is however not sufficient to quantitatively describe the viscosity of charge-stabilized (DLVO) suspensions. In the absence of a solid phenomenological background, this is a difficult problem for theory and even for computer simulation techniques. Hence the motivation to conduct extensive experimental investigations and use the experimental input to combine the Smoluchowski theory for aggregation with the mean-field hydrodynamic-level treatment of viscosity (accounting for many-body interactions in an approximate way) which applies effectively to the mesoscopic level of weakly interacting, hard-sphere like (being much larger than the primary particles) clusters produced by shear-induced aggregation. At such mesoscopic level, another important effect brought about by shear is the hydrodynamic rupture or breakup of the clusters which sees the interplay between hydrodynamics and the rigidity of the clusters. This phenomenon has also been studied in detail and the key result has been incorporated into the picture for the aggregating suspension viscosity. Thus we achieved in the end a semi-quantitative, phenomenological description of the viscosity of DLVO colloidal suspensions at high shear rate. Upon cessation of flow ($\dot{\gamma} \to 0$), after a shearing time at high $\dot{\gamma}$ sufficient to produce extensive aggregation, the system appears dynamically arrested and develops an elastic response dependent on the aggregation history. This solid-like state is due to the fact that at rest the mesoscopic aggregates formed under shear interconnect to form a spce-spanning structure which is rigid. This process can be understood at the mesoscopic level as the result of large clusters or aggregates growing due to shear-induced aggregation until they interconnect, a process similar to gelation under stagnant conditions [3]. By means of oscillatory rheology it has been possible to monitor the arising of rigidity upon cessation of flow as a function of the shearing time (or equivalently: of the extent of aggregation) and the process resembles in some respects an ordinary gelation process in the absence of flow. The fact that the system appears solid-like upon cessation of flow and fully liquid at high $\dot{\gamma}$ is investigated by experimentally studying the shear-rate dependent rheology of the suspensions after flow cessation. This highlights a pronounced shear-thinning behaviour followed, in a certain range of volume fractions occupied by the clusters (or equivalently: of shearing times), by shear-thickening. These observations and their interpretation lend further support to the coarse-graining and length-scale separation concepts lying at the heart of our analysis method.

## 1.2 Arrested colloidal suspensions as model amorphous solids

### 1.2.1 The short-range structure of amorphous solids and the structure-property relationships

The macroscopic properties of condensed matter systems are tightly connected with the underlying microscopic structure, in terms of spatial distribution of the building blocks (atoms, molecules, colloids, grains) constituting the material. This is very clear for example in the case of crystals, i.e. solids with both translational and rotational long-range order, where several macroscopic properties can be inferred on the basis of the symmetries of the crystal lattice. For disordered systems the structure-property relations are often not obvious. Liquids and amorphous solids possess only short-range order and the elementary topological measure of the latter is the mean coordination number or nearest neighbours number around a tagged particle. Here we encounter a first conceptual difficulty: how near should be the neighbour particles to be considered as nearest-neighbours? There is no generally valid answer to this question and different criteria are adopted depending on the particular system under consideration. For example, in the case of jammed granular packings, it is natural to define the coordination number as the number of particles in physical contact with the tagged one. In the case of liquids, although there are analogies with the short-range structure of random close packings of granular spheres (as first noted by Bernal [2]) the situation is very different because there are obviously no permanent contacts and the coordination number has to be defined in a different way (usually as the number of particles in a shell of width corresponding to the first minimum in the radial distribution function).

Whenever the relevant mean coordination can be meaningfully defined, in dense spatially homogeneous systems this simple measure may already provide the basis to assess several properties, such as thermal and electrical conductivity, and the mechanical response. If on the other hand the system is not structurally homogeneous, which is rather the rule at low and intermediate density (with the only exception of hard-sphere systems), the mean coordination of the primary particles may no longer be the key measure for determining the properties. What in this case assumes relevance is very often the mesoscopic structure of the system, which is usually in close relation to structural heterogeneity. The simplest example is the one where heterogeneity is manifest in the form of clusters, on which case we will mainly focus in this work. Then the relevant structural parameter, as is intuitive, will be the mean coordination of the clusters (treated once again as effective particles), rather than the mean coordination of the primary particles. This concept can be schematically exemplified as shown in Figure 1-2. In Figure 1-2 (a) the system is dense and spatially homogeneous: the relevant microscopic parameters (interaction, mean coordination etc.) are those relative to the primary particles. In Figure 1-2 (b), instead, the system is structurally heterogeneous and clusters are identifiable: in this case the relevant interaction, bonding, and mean coordination are those relative to the clusters.

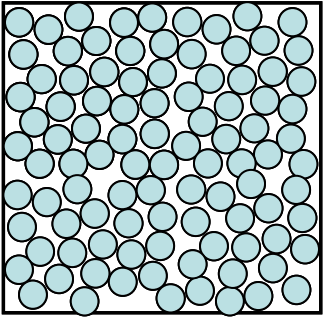 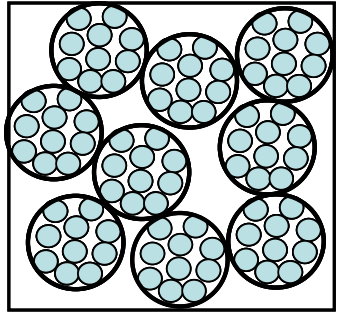

**Figure 1-2: (a) a dense spatially homogeneous system. (b) a structurally heterogeneous system with spherically shaped clusters.**

Thus, in order to study the macroscopic response of amorphous solids, consideration should be given also to the problem of quantifying and characterizing the short-range structure as well as the extent of structural heterogeneity. This has been done in the Appendix B.

### 1.2.2 Macroscopic response to shear deformation

The validity of our approach for describing the macroscopic properties of colloidal suspensions can be demonstrated by applying the same methodology also to dynamically arrested, solid-like colloidal systems. The relevant macroscopic response in this case is given in terms of the elastic modulus. Again, the basic type of deformation field is simple shear, on which we will focus our analysis. There are not many approaches available to describe the elasticity of dynamically arrested colloidal systems and amorphous solids in general. This is another longstanding problem in condensed matter physics if one thinks that glasses do not show appreciable structural differences with respect to liquids, yet liquids do flow whereas glasses do not (at least on measurable time scales) and have shear moduli up to $10^{10}$ Pa. In practice, only two very distinct limits have been given consideration so far. One is the limit of very strong attractive (irreversible) bonds and very low colloid concentration. Under these conditions (diffusion limited cluster) aggregation leads to clusters with fractal morphology which interconnect to form a rigid space-spanning network that is itself self-similar. Early approaches for random percolating systems allow for connecting the local elasticity of continuous chains of particles with bending resistance (constituting the network) to the macroscopic elasticity (the shear modulus) [14]. Using fractal scaling for the morphology of the chains leads to scaling expressions for the shear modulus as a function of the volume fraction which seem in good agreement with measurements. The opposite limit is that of zero-attraction hard-sphere systems at very high density, along the so-called metastable branch of the HS phase diagram up to random close packing [14]. Such hard-sphere glasses arise upon increasing the density (the only relevant parameter here since there is no energy scale with this potential) until a critical density is reached where the tagged particle cannot escape the cage constituted by its nearest-neighbours, and the structural relaxation time becomes infinite [14]. This transition is well captured by Mode Coupling Theory for what concerns the relaxation times [8]. However it remains unclear how rigidity arises in the absence of obvious structural differences between liquid and glass states [15].

In recent years another model colloidal system has been investigated extensively in order to draw more insights into the physics of amorphous states. It has been observed that high-density colloidal suspensions with a short-range attractive component of interaction (practically induced by polymer depletion) can form glassy states which are not driven by crowding, caging and geometric frustration (as it is the case for hard-spheres), rather by attraction itself [16]. Such glassy states have been named "attractive glasses", or sometimes, energy-driven glasses, and they make their appearance when the range of attraction is very short (less than 10%) compared to the colloid diameter and the attractive well of the potential sufficiently deep (or equivalently at sufficiently low temperatures) [16]. An interesting property of these systems is that the liquid-glass transition line is re-entrant (non-monotonic with the attraction strength) so that upon increasing the attraction strength one first melts the glass into a liquid (glass-liquid transition) and upon further increase in attraction the liquid solidifies again into the attractive glass. The latter state has been observed, for example, in technologically important systems such as protein suspensions. A major point of interest lies in the fact that attractive glasses can be used to better understand the effect of the attractive component of interaction on the macroscopic properties (the range and depth of attraction can be very easily tuned in experiments). This is a vital aspect since attraction is important for all amorphous solids in the real world. Furthermore, attractive glasses have the property that they can remain solid-like down to volume fraction much lower than the 0.58 value at which the hard-sphere glass transition is observed to take place. This fact raises an intriguing question regarding a possible continuation between glassy and gel states, the latter also being substantially dominated by attractive interactions. The linear and non-linear rheological response of these systems is currently garnering great attention in the colloidal physics literature [17].

In this work we use the dense attractive glass as the starting point to attempt a theoretical description of the elastic response of amorphous bodies. The main advantage is that, in the limit of strong attraction and high density, the established Cauchy-Born theory of solids in its formulation for disordered systems due to Alexander [18], can be adapted to yield an estimate of the elastic moduli, and in particular of the shear modulus. The physically motivated approximation, as will be discussed in detail in Chapter 8, is that the interparticle bonds due to the attractive interaction strongly hinder nonaffine relaxations which always accompany the elastic response of structurally disordered solids, at least in the first linear regime of response where, also based on experimental studies, the bonds are preserved unbroken. The affine approximation is shown to do a reasonably good job. The quantitative accuracy in this limit allows us to test the scaling principle put forward in this Thesis. Namely, the hypothesis that at lower density, where structural heterogeneity appears in the form of clusters, the latter ones can be treated as the effective building blocks which determine the material properties. To them the same (continuum) response formalism applies that successfully describes the macroscopic properties of the dense and homogeneous states. The main result out of this analysis is that the elastic response of arrested colloids with attractive interaction in a broad range of volume fraction (from 0.4 down to 0.2), which are usually classified as gel states, can be explained and sometimes even quantitatively described as "heterogeneous" glasses. This framework makes it possible to rationalize the kaleidoscopic variety of mechanical responses in the intermediate volume fraction range of arrested colloids.

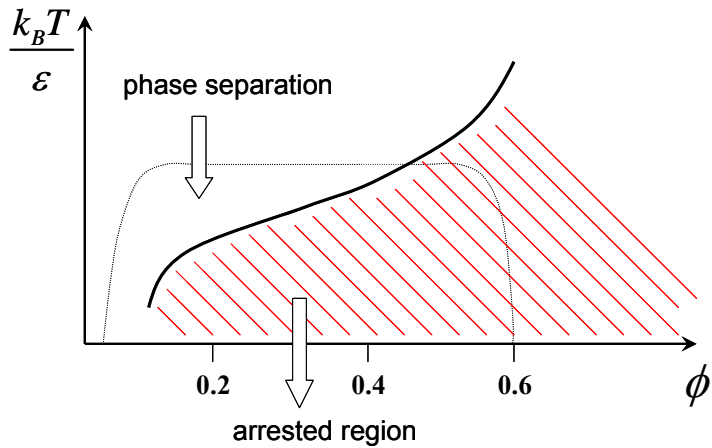

**Figure 1-3: schematic phase diagram of attractive colloidal suspensions. On the x axis is the inverse of the attraction strength, on the y axis the volume fraction. The dotted line represents the spinodal decomposition line. The solid line divides fluid states from rigid (arrested) ones. The red area represents the broad range of conditions where the elastic response is seen to display dramatic differences.**

The same approach has been applied to evaluating the rigidity, under the action of hydrodynamic stresses, of colloidal aggregates generated by shear-induced aggregation in dilute conditions. In particular it has been possible to explain the power-law correlation between the maximum mechanically-stable aggregate size and the applied shear rate (directly proportional to the hydrodynamic stress through the medium viscosity) observed in experiments [19]. It could therefore be understood that the maximum mechanically stable size of the aggregates depends uniquely upon the structure (fractal with generally high fractal dimension in sheared systems) of the aggregates and their bonding properties, and the applied shear rate [19]. Thus, in aggregating suspensions, the aggregation process will continue until this maximum size has been reached so that the suspension properties can be described from that point on solely in terms of clusters (effective particles) of that size. This key result is much used and discussed in chapters 5 and 6 in relation to the rheological properties of aggregating suspensions.

## 1.3 Overview

In the present work we have studied some of the problems outlined above in an attempt to understand the interplay between the microscopic and the mesoscopic level of structure and their relation with the macroscopic properties in colloidal suspensions with an attractive component of interaction. We have consistently applied a separation between the microscopic or single particle scale and the mesoscopic or cluster scale and focused on two states of colloidal suspensions under deformation: flowing (liquid) suspensions and arrested (solid) suspensions in external shear deformation fields. Our analysis shows that, in the presence of significant attraction, the macroscopic properties (viscosity, elastic modulus) can be *directly* described in a coarse-grained manner by treating the mesoscopic structures as the effective building blocks of the system, thus neglecting the degrees of freedom of the primary colloidal particles. On the other hand, we found that the direct interactions between primary colloidal particles play a vital role in determining the properties of the mesoscopic level and are therefore *indirectly* involved in the buildup of the macroscopic properties.

In the case of flowing (liquid) suspensions, we find that the formation of mesostructures is due to a microscopic interplay between the external field, the direct interaction and the Brownian motion, an interplay which obeys a Fokker-Planck description at the Smoluchowski level. In this framework, we provide an analytical theory which can describe and quantitatively predict the kinetic aggregation rate between Brownian particles in a flow field interacting via an arbitrary interaction potential. The theory is derived for dilute suspensions and subsequently extended to the case of arbitrary concentrations via an effective medium treatment accounting for collective hydrodynamics in a mean-field fashion.

In the case of solid, dynamically-arrested colloidal suspensions with strong interparticle attraction, we start with considering the elastic response of dense homogeneously disordered suspensions. At the highest densities (around 0.6 volume fraction), the primary colloidal particles represent the building blocks which determine the macroscopic properties, since structural heterogeneities are practically absent. For these systems we derive an analytical formula for the shear modulus starting from the free energy expansion. A strong attraction for a colloidal system is equivalent to the system being in a deep minimum of the potential energy landscape (which corresponds in practice to a deep quench of the supercooled liquid). Hence we introduce the approximation that, being the minimum very deep, the harmonic terms in the free energy expansion dominate over higher order (anharmonic) terms. Under this quasi-harmonic approximation, the particles in a deformed sample at small strain are displaced about the minimum of the interaction potential well and it is speculated that nonaffine displacements are also small. It turns out, from the comparison with simulations and with different sets of experimental data from the literature, that these approximations are indeed tenable in the case of strong interparticle attraction. At lower volume fractions, we demonstrate that the structural heterogeneity can be incorporated in the quasi-harmonic elastic theory by, again, separating the microscopic scale and the mesoscopic scale. It turns out also in this case that the mesoscopic structures (i.e. *clusters*) are those which effectively control the macroscopic response while their properties (in terms of cluster size, cluster rigidity etc.) are determined by the primary particles and their mutual interactions.

We hope that the resulting insights will enable the development of hierarchies of models for predicting how small-scale processes couple to the larger scales and how this coupling affects the macroscopic properties of complex disordered states of matter.

# 2. Kinetic theory of shear-induced aggregation of Brownian particles

## 2.1 Introduction

The macroscopic rheological behaviour of complex and biological fluids is governed, at the microscopic level, by the convective diffusion equation for interacting colloids, which is formally identical to the Smoluchowski equation in shear, Eq. (1.4)-(1.5). This partial differential equation is known to be analytically intractable for most problems of interest [22]. As a consequence, puzzling rheological behaviours of complex fluids (shear-thickening, thixotropy, rheopexy etc.) are poorly understood and their microscopic physics has remained unclear, in spite of their being ubiquitous (from paints and inks to blood plasmas) [20,21]. Here we have proposed an approximation scheme which allows for analytical solutions in linear flow fields (shear and extensional) and with arbitrary interaction potential between primary constituents. When an interaction barrier is present, an explicit (aggregation) rate-equation in Arrhenius form (which contains a term proportional to the Peclet number in the exponential factor) is derived. The predictive power of the approximation (with no fitting parameters) is verified against numerics for the case of DLVO-interacting spherical colloids in laminar extensional flow. The results explain the puzzling features observed in rheopectic complex fluids, including the majority of protein-based fluids, such as the presence of an induction period (which corresponds to an activation delay in our theory) followed by a sudden rise of the viscosity (explained here as due to a self-accelerating kinetics involving activated clusters) [5,7].

## 2.2 Derivation, results, and discussion

Let us consider a dispersion of diffusing particles interacting with a certain interaction potential. The (number-)concentration field will be called $c(\mathbf{r})$. The associated normalized probability density function $g(\mathbf{r}) \equiv \tilde{c}(\mathbf{r})$ for finding a second particle at distance $\mathbf{r}$ from a reference one is then normalized such that $c(\mathbf{r}) = c_0 \tilde{c}(\mathbf{r})$, where $c_0$ is the bulk concentration. The evolution equation thus reads

$$\partial c / \partial t = \text{div}(D\nabla c - \beta D\mathbf{K}c) \tag{2.1}$$

where $D$ is the mutual diffusion coefficient of the particles ($D = 2D_0 \mathcal{G}(r)$, $D_0$ being the diffusion coefficient of an isolated particle and $\mathcal{G}(r)$ the hydrodynamic function for viscous retardation), $\beta = 1/k_B T$ is the Boltzmann factor, and $\mathbf{K}$ is an arbitrary external force. According to this equation, the associated stationary current is given by

$$\mathbf{J} = (\beta D\mathbf{K} - D\nabla)c \tag{2.2}$$

Hence, when steady-state is reached, $\partial c(\mathbf{r})/\partial t = 0$ and

$$\text{div}(\beta D\mathbf{K} - D\nabla)c = 0 \tag{2.3}$$

If the dispersing medium is subjected to an external flow, the arbitrary external force field for our problem may be decomposed into two terms, a non-conservative one accounting for the drift caused by the flow velocity $\mathbf{v}(\mathbf{r}) = [\mathrm{v}_r, \mathrm{v}_\theta, \mathrm{v}_\varphi]$, and the term accounting for the conservative force field due to the two-body direct interaction between particles $-\nabla U(r)$. Introducing the hydrodynamic drag $b = 3\pi\eta_0 a$, one obtains

$$\mathbf{K}(\mathbf{r}) = -\nabla U(r) + b\mathbf{v}(\mathbf{r}). \tag{2.4}$$

where $\eta_0$ is the viscosity of the solvent and $a$ is the particle radius. Hence, the two-particle Smoluchowski equation with shear can be written as

$$\text{div}\left[\beta D(-\nabla U + b\mathbf{v}) - D\nabla\right]c = 0 \tag{2.5}$$

with the associated current given by

$$\mathbf{J} = \left[\beta D(-\nabla U + b\mathbf{v}) - D\nabla\right]c \tag{2.6}$$

We can now define the collision rate across a spherical surface of radius $r$, concentric with the stationary particle, as

$$\begin{aligned} G = \oint \mathbf{J} \cdot \hat{\mathbf{n}}\, dS &= \oint \left[D\nabla c + \beta D(\nabla U - b\mathbf{v})c\right] \cdot \hat{\mathbf{n}}\, dS \\ &= 4\pi r^2 D\left[\left(\frac{\partial \langle c \rangle}{\partial r} + \beta \frac{\partial U}{\partial r}\langle c \rangle\right)\right] + 4\pi r^2 b \langle \mathrm{v}_r^+ c \rangle \end{aligned} \tag{2.7}$$

where $dS = r^2 \sin\theta\, d\theta\, d\phi$ denotes the element of (spherical) collision surface, while $\langle .. \rangle$ denotes the angular average $(4\pi)^{-1} \int_0^{2\pi} d\varphi \int_0^\pi \sin\theta\, d\theta$. Recall that $G$, by definition, is the *inward* flux of particles through the spherical surface. Therefore, integration runs exclusively over those orientations (or, equivalently, those pairs of angles $\theta$, $\varphi$) such that $\mathbf{v} \cdot \mathbf{n} = -v_r$, which define the upstream region. Hence we will use $v_r^+(\mathbf{r})$ to denote the positive part of the radial component of the fluid velocity, $v_r(\mathbf{r})$. Thus

$$v_r^+(\mathbf{r}) = \max(v_r(\mathbf{r}), 0) = \begin{cases} v_r(\mathbf{r}) & \text{if } v_r(\mathbf{r}) > 0 \\ 0 & \text{else} \end{cases} \tag{2.8}$$

Under the approximation that the flow velocity and the concentration profile around the stationary particle are weakly correlated

$$\langle (v_r^+(\mathbf{r}) - \langle v_r^+(\mathbf{r}) \rangle)(c(\mathbf{r}) - \langle c(\mathbf{r}) \rangle) \rangle \approx 0, \tag{2.9}$$

i.e. $\langle v_r^+(\mathbf{r}) c(\mathbf{r}) \rangle \approx \langle v_r^+(\mathbf{r}) \rangle \langle c(\mathbf{r}) \rangle$. This approximation is expected to do a reasonably good job as long as the Brownian motion is effective in randomizing the particle concentration, i.e. when Brownian motion is not overwhelmed by convection.

Eq. (2.7), under this approximation, simplifies to

$$G = 4\pi r^2 D \left( \frac{\partial}{\partial r} + \beta \frac{\partial U}{\partial r} + b \langle v_r^+ \rangle \right) \langle c \rangle \tag{2.10}$$

Clearly, the collision rate given by Eq. (2.10) corresponds to the collision rate that we would have if the actual flow field $\mathbf{v}(\mathbf{r})$ were replaced by an *effective* flow field for aggregation where only the radial component, given by $\langle v_r^+(\mathbf{r}) \rangle$, is non-zero. The effective flow field will be denoted as $\mathbf{v}_{\text{eff}}(r) = [v_{r,\text{eff}}(r), 0, 0]$ with $v_{r,\text{eff}}(r) \equiv \langle v_r^+(\mathbf{r}) \rangle$. Therefore, a system where the collision rate (hence the colloidal aggregative stability) is identical to the one of the real system may be described, under the approximation Eq. (2.9), by the following *effective* two-particle Smoluchowski equation for the orientation-averaged concentration field $\langle c \rangle$

$$\text{div}\left[ \beta D (\nabla U + b \mathbf{v}_{r,\text{eff}}) + D \nabla \right] \langle c \rangle = 0 \tag{2.11}$$

which, for isotropic potential, can be written as an ordinary differential equation

$$\frac{1}{r^2} \frac{d}{dr} \left[ r^2 D \left( \beta \frac{dU}{dr} + b v_{r,\text{eff}} \right) \langle c \rangle + D r^2 \frac{d \langle c \rangle}{dr} \right] = 0 \tag{2.12}$$

**2.2.1 The far-field boundary condition with linear flow fields: boundary-layer analysis**

The boundary conditions for the irreversible aggregation problem are as follows. First, the reaction kinetics by which the particle irreversibly sticks to the reference one is taken to be infinitely fast at $r = 2a$, which corresponds to the familiar absorbing boundary condition $\langle c \rangle = 0$ at $r = 2a$. Second, the bulk concentration ($c/c_0 = 1$) must be recovered at a certain distance from the reference particle. This condition is often implemented at large distances, namely at $r \to \infty$, which is always possible for velocity fields which decay to zero at infinity. Classic examples are the problems of convective diffusion of a solute to a free-falling particle or convective diffusion to a rotating disk (cfr. Levich [22], Ch. 2). However, it is well known that in the case of *linear* velocity fields, application of the second (far field) boundary condition, $c/c_0 = 1$ at $r \to \infty$, is more complicated, due to a singularity at the domain boundary induced by $\mathbf{v} = \mathbf{\Gamma} \cdot \mathbf{r}$, where $\mathbf{\Gamma}$ is the velocity gradient tensor. In this case, the Smoluchowski equation becomes

$$\mathrm{div}\left[\left(-\beta D \frac{\partial U}{\partial r} + \mathbf{\Gamma} \cdot \mathbf{r}\right) - D\nabla\right]c = 0 \qquad (2.13)$$

As first diagnosed by Dhont [23] who used Eq. (2.13) to study the structural distortion of sheared non-aggregating suspensions [24], the $\mathbf{\Gamma} \cdot \mathbf{r}$ term, being linear in $\mathbf{r}$, overwhelms the other terms at sufficiently large separations, even for very small shear rates. It was shown that in the case of hard spheres the extent of separation $\delta$ where this occurs decreases with the Peclet number and is given by $\delta/a \sim Pe^{-1/2}$ [23]. In other terms, $\delta$ defines a boundary-layer width beyond which convection represents by far the dominant process [22,23]. To overcome the difficulty of having a singular term at the domain boundary which makes it impossible to enforce the far-field boundary-condition, specific considerations are required. For example, when the final goal is to determine the structure factor of a suspension, it is convenient to Fourier-transform the radial domain or equivalently to move to a reciprocal domain $q = 2/r$ (see e.g. [12]). Alternatively, within numerical studies in real space, the far-field boundary condition is usually applied at finite separations, after self-consistently identifying the location beyond which the concentration profile flattens out as a consequence of convection becoming predominant [25,26].

At this point, it is useful to consider the boundary-layer structure of the Smoluchowski equation and of the formally identical convective diffusion equation. In fact, as is well known within the theory of convective diffusion [22], due to the boundary-layer behaviour of Eq. (2.13), the convective flux dominates at sufficiently large interparticle separations (which has been verified numerically in [25]) since the other terms in the bracket in Eq.

(2.13) become negligible in comparison with $\mathbf{\Gamma} \cdot \mathbf{r}$. That region of space is thus described by

$$\operatorname{div}(\mathbf{v}c) = 0 \tag{2.14}$$

which admits the solution $c = const$ [22]. For colloidal particles in a linear flow field, Eq. (2.14) has been solved under account of hydrodynamic interactions by Batchelor and Green [13] who derived the following solution in terms of the pair-correlation function

$$g(r) = \frac{1}{1-A} \exp \int_r^\infty dr \frac{3(B-A)}{r(1-A)} \tag{2.15}$$

where the hydrodynamic functions $A(r)$ and $B(r)$ describe the nonlinear disturbance to the relative velocity due to the particle motion. Beyond a few particle radii distance both $A(r)$ and $B(r)$ decay to zero, and we have $g(r) = 1$ thus implying $c = const = c_0$. Hence, the mere effect of *convection* in a linear flow-field, in the absence of any perturbation (as could arise from diffusive motion or hydrodynamic interactions) is to *flatten out* the concentration profile. Therefore, if it is true that *convection* prevails at separations larger than the boundary-layer width, it must follow that assuming homogeneity, $c = const = c_0$, at separations larger than the boundary-layer is justified.

In this work we thus propose using $r = \delta + 2a$ (instead of $r \to \infty$) in the far-field boundary condition. In the language of matched asymptotic expansions, this corresponds to exactly determining the solution within the boundary-layer and matching it to the leading order in $Pe^{-1}$ expansion in the outer layer. (Note however that the problem of Eq. (2.13) in real space is substantially more complicated than standard singularly perturbed equations due to the aforementioned singular behaviour of $\mathbf{\Gamma} \cdot \mathbf{r}$ at the domain boundary, in addition to the standard singular behaviour for large values of the Peclet number, as first noted by Dhont [23].) Hence, the collision kinetics being uniquely determined by the inner solution, the only approximation involved is on the location of the far-field boundary-condition, which we estimate in the following as a function of Peclet number and interaction parameters. The good accuracy of the estimate is subsequently verified by quantitative comparison with numerics.

The width of the boundary-layer $\delta$, where the major change in the concentration profile occurs, and within which diffusion, convection and direct interactions are all important, can be uniquely determined from dimensional considerations. The governing parameters upon which $\delta$ depends are: $\lambda$, $a$, $\dot{\gamma}$ and $D_0$, where $\lambda$ denotes the range of (repulsive) interaction. Thus the dimensionless boundary-layer width $\delta/a$ is expressible as a dimensionless combination of the governing parameters (Rayleigh)

$$\delta/a \sim \lambda^l a^m \dot{\gamma}^n D_0^p \tag{2.16}$$

From the boundary-layer behaviour of the Smoluchowski equation for hard spheres, it is known that $\delta/a \sim Pe^{-1/2}$ [23], which identifies $n = -1/2$ and $p = 1/2$. Hence considering that $[\lambda] = [a] = L$ and $[\dot{\gamma}]^{-1/2}[D_0]^{1/2} = L$, it obviously follows

$$l + m + 1 = 0 \tag{2.17}$$

Application of the $\Pi$-theorem of dimensional analysis tells that $\delta/a$ has to be expressed in terms of two independent dimensionless groups, $\lambda/a$ and $\dot{\gamma}a^2/D_0$, as $\delta = a\Phi(\lambda/a, \dot{\gamma}a^2/D_0)$, because $a$ and $\dot{\gamma}$ are parameters with independent dimension [27]. This fixes the *m* value:

$$m = -3/2 \tag{2.18}$$

It is therefore concluded that the dimensionless boundary-layer width $\delta/a$ is approximately given by

$$\delta/a \sim \sqrt{(\lambda/a)/Pe} . \tag{2.19}$$

When Eq. (2.19) is compared to the case of hard spheres, $\delta/a \sim Pe^{-1/2}$, the effect of the colloidal interactions on the boundary-layer width enters through the interaction range $\lambda$, which for the screened-Coulomb repulsion is given by $\lambda = \kappa^{-1}$, where $\kappa^{-1}$ is the Debye length.

### 2.2.2 Approximate solution for the aggregation rate

Based on the above boundary value problem analysis, let us rewrite the effective two-particle Smoluchowski equation in shear, Eq. (2.12), in dimensionless form [28]

$$\frac{1}{Pe}\frac{1}{(x+2)^2}\frac{d}{dx}(x+2)^2\frac{dC}{dx} + \frac{1}{Pe}\frac{1}{(x+2)^2}\frac{d}{dx}\left((x+2)^2\frac{d\tilde{U}(x)}{dx}C\right)$$
$$+ \frac{1}{(x+2)^2}\frac{d}{dx}\left[(x+2)^2\tilde{v}_{r,\text{eff}}C\right] = 0 \tag{2.20}$$

with the boundary conditions

$$\begin{aligned} C &= 0 \quad \text{at} \quad x = 0 \\ C &= 1 \quad \text{at} \quad x = \delta/a \end{aligned} \tag{2.21}$$

where, for simplicity of notation, we have set $\langle \tilde{c}(\mathbf{r}) \rangle \equiv C(x)$, since the latter is a function only of the dimensionless separation $x = (r/a) - 2$. The tilde indicates dimensionless quantities. The Peclet number is given by $Pe = \dot{\gamma} a^2 / D = 3\pi\eta_0 \dot{\gamma} a^3 / k_B T$. Eq. (2.20) is formally identical to a stationary one-dimensional Fokker-Planck equation in spherical geometry with time independent drift and diffusion coefficients [29]. The concentration profile in dimensional form after application of the second boundary condition to Eq. (2.20) reads

$$\langle c \rangle = \left\{ \exp \int_{\delta/a}^{x} dx(-\beta dU/dx - Pe\tilde{v}_{r,\text{eff}}) \right\}$$
$$\times \left\{ c_0 + \frac{G}{8\pi a D_0} \int_{\delta/a}^{x} \frac{dx}{\mathcal{G}(x)(x+2)^2} \exp \int_{\delta/a}^{x} dx(\beta dU/dx + Pe\tilde{v}_{r,\text{eff}}) \right\} \quad (2.22)$$

The flux *G* is determined from the absorbing boundary condition at contact as

$$G = \frac{8\pi D_0 a c_0}{2 \int_0^{\delta/a} \frac{dx}{\mathcal{G}(x)(x+2)^2} \exp \int_{\delta/a}^{x} dx(\beta dU/dx + Pe\tilde{v}_{r,\text{eff}})} \quad (2.23)$$

Using $\delta/a = \sqrt{(\lambda/a)/Pe}$, Eq. (2.23) can be integrated with standard numerical techniques and can find direct application to colloidal systems under shear and straining flows, provided that an appropriate form of the velocity field is given. Note that the effect of hydrodynamic viscous dissipation is accounted for through the function $\mathcal{G}(x)$. For an axisymmetric extensional flow, the radial component of the velocity field is given by [25]

$$v_r = \frac{1}{2}\dot{\gamma} a(x+2)\left[1 - A_E(x)\right]\left(3\cos^2\theta - 1\right) \quad (2.24)$$

where $A_E(x)$ is the corresponding function accounting for the effect of hydrodynamic retardation. Based on how the rescaled effective velocity has been defined, we thus obtain

$$\tilde{v}_{r,\text{eff}}(x) \equiv \langle \tilde{v}_r^+(\mathbf{r}) \rangle = -\left(1/3\sqrt{3}\right)(x+2)\left[1 - A_E(x)\right] \quad (2.25)$$

Similarly, in the case of pure laminar shear we have

$$\tilde{v}_{r,\text{eff}}(x) \equiv \langle \tilde{v}_r^+(\mathbf{r}) \rangle = -\left(1/3\pi\right)(x+2)\left[1 - A_S(x)\right] \quad (2.26)$$

where $A_S(x)$ is the hydrodynamic retardation function for simple shear.

Hence, within this formulation it has been possible to account for hydrodynamic interactions (in the two-body limit) resulting from viscous dissipation, through $\mathcal{G}(x)$, and from hydrodynamic disturbance (retardation) induced by the second particle, through $A(x)$.

### 2.2.3 Irreversible aggregation kinetics and colloidal stability in shear

Since we have retained all terms in the governing equation (and built the solution exactly in the inner layer), Eq. (2.23) is valid for arbitrary thickness of the boundary layer $\delta$. In particular we observe that in the limit $Pe \to 0$, Eq. (2.23) reduces to the well-known Fuchs' formula for the aggregation rate constant (collision rate) in the presence of direct interaction forces but in the absence of flow, which reads [30]

$$G = \frac{8\pi D_0 a c_0}{2\int_0^\infty dx \frac{e^{\beta U(x)}}{\mathcal{G}(x)(x+2)^2}} \tag{2.27}$$

Comparing Eqs. (2.23) and (2.27) leads to the definition of a generalized stability coefficient which is valid for arbitrary $Pe$ numbers and interaction potentials

$$W_G = 2 \int_0^{\sqrt{(\lambda/a)/Pe}} \frac{dx}{\mathcal{G}(x)(x+2)^2} \exp \int_{\sqrt{(\lambda/a)/Pe}}^x dx(\beta dU/dx + Pe\tilde{v}_{r,\text{eff}}) \tag{2.28}$$

This, at $Pe=0$, reduces to the classic stability coefficient first derived by N. A. Fuchs in 1934 (cfr. [30])

$$W = 2\int_0^\infty dx \frac{e^{\beta U(x)}}{\mathcal{G}(x)(x+2)^2} \tag{2.29}$$

Thus, the combined effect of fluid motion (convection) and direct interactions can either reduce or augment the coagulation rate with respect to the case of non-interacting Brownian particles in a stagnant fluid by a factor equal to $W_G$.

### 2.2.4 Initial irreversible aggregation kinetics

In deriving Eq. (2.23) for the collision rate against a stationary particle we did not account for the diffusive motion of the latter. Accounting for that, leads to a factor two times the macroscopic number concentration of particles in the system, $c_0$. Hence, the kinetic equation for the rate of change of the concentration of particles reads

$$\frac{dc}{dt} = -\frac{16\pi D_0 a}{W_G} c^2 \qquad (2.30)$$

where $W_G$ is the generalized stability coefficient given by Eq. (2.28). Integrating Eq. (2.30) yields the law of variation with time of the particle concentration

$$c(t) = \frac{c_0}{1 + t/t_c} \qquad (2.31)$$

where

$$t_c = (16\pi D_0 a c_0 / W_G)^{-1} \qquad (2.32)$$

is the characteristic time of aggregation. Its reciprocal value defines the kinetic constant for the aggregation of primary particles

$$k_{1,1} = 16\pi D_0 a c_0 / W_G \qquad (2.33)$$

Since $W_G$, which is given by Eq. (2.28), brings about a complex dependence upon the particle size when $Pe > 0$, it can be anticipated that the evolution of the aggregation process will substantially differ from the purely Brownian case in a stagnant medium, with important consequences for the dynamics of new phase formation in sheared colloids.

### 2.2.5 Comparison with numerical results from the literature

Let us compare the theoretical predictions from Eq. (2.23) with numerical results where the full convective diffusion equation, Eq. (2.3), was solved numerically, by means of a finite difference method [25]. The colloidal system is composed of colloidal particles with $a = 100$ nm, with interactions via standard DLVO potential, and convection is induced by laminar axysimmetric extensional flow [25]. The numerically obtained values of $c(\mathbf{r})$ were then used to determine the aggregation rate constant from numerical evaluation of

$$G = \oint \mathbf{J} \cdot \hat{\mathbf{n}} \, dS = \oint \left[ D\nabla c + \beta D (\nabla U - b\mathbf{v}) c \right] \cdot \hat{\mathbf{n}} \, dS \qquad (2.34)$$

where the collision surface is taken as the spherical surface of radius $2a$.

The comparison is shown in Figure 2-1 for three different values of ionic strength and a fixed surface potential equal to -14.7 mV. The direct potential *U* for the same conditions of the numerical simulations, as well as the hydrodynamic functions *A(x)* and $\mathcal{G}(x)$, have been consistently calculated according to [25]. As shown in the figure, the theory is able to quantitatively reproduce the numerical data for all conditions. In particular, it is seen that the inflection point which marks the transition from a purely-Brownian like regime at $Pe<1$ to a shear-dominated regime at $Pe>10$ is very well captured by the theory. Some underestimation arises in the regime of high Peclet numbers $Pe>50$, which tends to become more important upon further increasing $Pe$. Such underestimation is related to the approximation $\langle (v_r^+(\mathbf{r}) - \langle v_r^+(\mathbf{r})\rangle)(c(\mathbf{r}) - \langle c(\mathbf{r})\rangle)\rangle \approx 0$ made in the derivation of Eq. (2.23). Clearly, the spatial correlation between the flow velocity and the concentration field around the stationary particle would become non-negligible at high Peclet numbers. In this regime, in fact, the randomizing effect of Brownian motion is progressively lost. The angular regions (relative orientations between particles), where the flow velocity is higher and inwardly-directed, tend instead to coincide with the regions where the probability of finding incoming particles is higher.

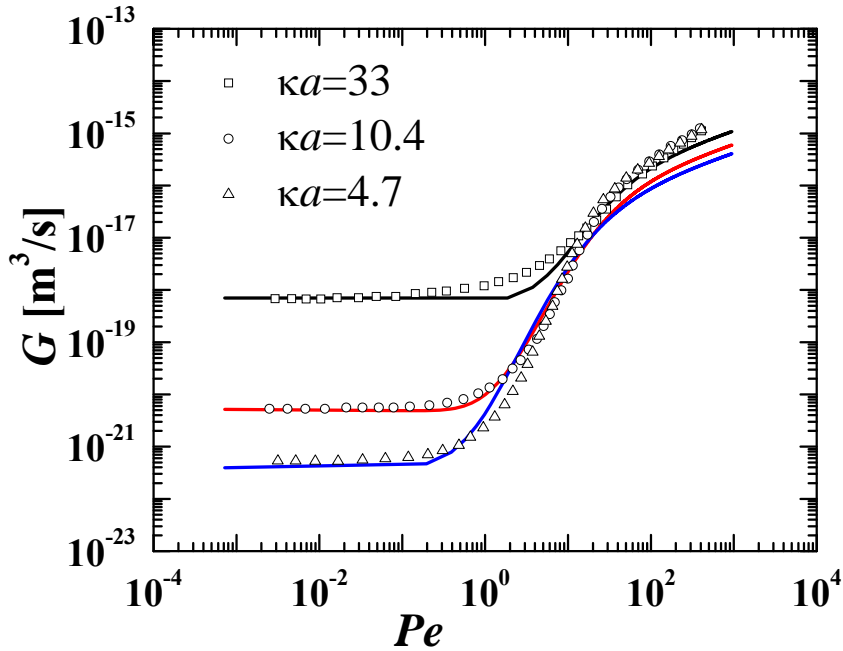

**Figure 2-1: Comparison between calculations based on the proposed theory (Eq. (2.23)) and the numerical simulations of the full convective diffusion equation (Eq. (2.3)) reported in [23] for three different ionic strength conditions. The colloid surface potential is fixed and equals -14.7 mV.**

### 2.2.6 Potential barrier crossing as a shear-driven activation process

Let us consider the case of a high potential barrier in the interaction between particles (as in the case of charge-stabilized colloids at low ionic strength, as predicted by DLVO theory). Further, we will neglect the effect of the hydrodynamic retardation on the velocity field, $A(x) = 0$ in Eqs. (2.25)-(2.26), so that the effective velocity reads $\tilde{v}_{r,\text{eff}} = -\alpha(x+2)$, where $\alpha$ is a numerical coefficient which depends upon the type of flow (e.g. $\alpha = 1/3\pi$ for simple shear). Then, the integrand in the second integral in the denominator of Eq. (2.23) reduces to

$$\int_{\sqrt{(\lambda/a)/Pe}}^{x} dx(\beta dU/dx + Pe\tilde{v}_{r,\text{eff}}) \approx \beta U - \beta U|_{x=\sqrt{(\lambda/a)/Pe}} - \frac{\alpha}{2}Pe(x+2)^2 + \frac{\alpha}{2}\lambda \quad (2.35)$$

When $Pe$ is not too high, $U|_{x=\sqrt{(\lambda/a)/Pe}} \approx 0$. Thus the denominator on the r.h.s. of Eq. (2.23) becomes

$$2\int_{0}^{\sqrt{(\lambda/a)/Pe}} \frac{dx}{\mathcal{G}(x+2)^2} \exp \int_{\sqrt{(\lambda/a)/Pe}}^{x} dx(\beta dU/dx + Pe\tilde{v}_{r,\text{eff}})$$

$$\approx 2e^{\alpha\lambda/2-\beta U|_{x=\sqrt{\lambda/Pe}}} \int_{0}^{\sqrt{(\lambda/a)/Pe}} \frac{dx}{\mathcal{G}(x+2)^2} \exp\left[\beta U - \alpha Pe(x+2)^2/2\right] \quad (2.36)$$

Since $U$ goes through a potential maximum (barrier) in $x \in [0, \sqrt{(\lambda/a)/Pe}]$, so does the function $\beta U - \alpha Pe(x+2)^2/2$. The argument of the exponential can thus be expanded near the maximum up to second order

$$\beta U - \alpha Pe(x+2)^2/2 \approx \beta U_m - \alpha Pe(x_m+2)^2/2 + (U_m'' - \alpha Pe)(x-x_m)^2 \quad (2.37)$$

where the subscript $m$ indicates quantities evaluated at the potential maximum. We can thus evaluate the remaining integral

$$\int_{0}^{\sqrt{(\lambda/a)/Pe}} \frac{dx}{\mathcal{G}(x)(x+2)^2} \exp\left[(\beta U_m'' - \alpha Pe)(x-x_m)^2\right] \quad (2.38)$$

by the method of steepest descent (or saddle-point method) to finally obtain

$$W_G \approx \sqrt{\frac{2\pi}{\alpha Pe - \beta U_m''}} \frac{2e^{\alpha\lambda/2-\beta U|_{x=\sqrt{(\lambda/a)/Pe}}}}{(x_m+2)^2 \mathcal{G}(x_m)} e^{\beta U_m - \alpha Pe(x_m+2)^2/2} \quad (2.39)$$

More precise approximations can be obtained by considering higher order terms in the expansion Eq. (2.37). From Eq. (2.39), in view of being $x_m + 2 \approx 2$, the effect of the

Peclet number and potential barrier on the two-particle aggregation rate constant, $k_{1,1}$, is given by

$$k_{1,1} \propto \sqrt{\alpha Pe - \beta U_m''} e^{-\beta U_m + 2\alpha Pe} = \sqrt{\frac{3\pi\alpha\eta_0\dot{\gamma}a^3 - U_m''}{k_B T}} e^{-(U_m - 6\pi\alpha\eta_0\dot{\gamma}a^3)/k_B T} \quad (2.40)$$

It is interesting to observe that Eq. (2.40) appears to be in Arrhenius form, with the pre-exponential or frequency factor $\sqrt{(3\pi\alpha\eta_0\dot{\gamma}a^3 - U_m'')/k_B T}$ and the activation energy $U_m - 6\pi\alpha\eta_0\dot{\gamma}a^3$. Note that due to $U_m$ being the potential maximum, $U_m''$ is negative. It follows that the quantity under the square-root in the pre-factor is always positive. In both parameters the shear rate $\dot{\gamma}$ plays a prominent role. Increasing $\dot{\gamma}$ leads to increase the collision rate, through the prefactor, and at the same time to decrease in the activation energy (thus increasing the fraction of successful collisions). Of course, for substantially large $\dot{\gamma}$ values, $\dot{\gamma}$ present in the activation energy has the dominant effect on the aggregation rate, due to the exponential form. Further, a critical value of the shear rate can be defined, which corresponds to vanishing a activation energy barrier

$$\dot{\gamma}^* = U_m / 6\pi\alpha\eta_0 a^3 \quad (2.41)$$

when $\dot{\gamma} \ll \dot{\gamma}^*$, the interaction barrier plays the dominant role, and $k_{1,1}$ increases as $U_m$ decreases. When $\dot{\gamma} \gg \dot{\gamma}^*$, instead, the shear-induced aggregation takes over the dominant role, and $k_{1,1}$ increases as $\dot{\gamma}$ increases. Thus, this critical shear rate marks the transition from a slow aggregation regime with an activation delay due to non-zero potential barrier to a fast aggregation regime, with no activation barrier. In fact, if $\dot{\gamma}^*$ is also such that the pre-exponential factor is of order unity, then the resulting kinetics will be of the same order of purely Brownian diffusion-limited aggregation (DLA) in a stagnant fluid at $\dot{\gamma} = \dot{\gamma}^*$. Any further increase of the shear rate above $\dot{\gamma}^*$ will then produce higher coagulation rates.

The coagulation constant $k_{1,1}$ defines a characteristic aggregation time

$$\tau = \frac{1}{k_{1,1}} \propto \frac{1}{\sqrt{(3\pi\alpha\eta_0\dot{\gamma}a^3 - U_m'')/k_B T}} e^{(U_m - 6\pi\alpha\eta_0\dot{\gamma}a^3)/k_B T} \quad (2.42)$$

An exponential dependence on $\dot{\gamma}$ for the aggregation time has been recently observed for the shear-induced aggregation of charged non-Brownian suspensions in simple shear [7].

## 2.2.7 Shear-driven self-accelerating kinetics, rheopexy, and shear-induced gelation

The aggregation time and the rate constant in the presence of shear display a very strong dependence upon the particle size, as it is evident from Eqs. (2.41)-(2.42). For instance, in the case where the potential is fixed, the dependence on the colloid radius reads

$$\tau \propto \frac{1}{\sqrt{3\pi\alpha\eta_0\dot{\gamma}a^3/k_BT}} e^{-6\pi\alpha\eta_0\dot{\gamma}a^3/k_BT} \tag{2.43}$$

For example, with $\mu = 0.001$ Pa·s, $\alpha = 1/3\pi$ and $\dot{\gamma} = 500$ s$^{-1}$, the expression in Eq. (2.43) amounts to 2.26 if $a = 100$ nm and to 0.14 if $a = 200$ nm. Doubling the colloid radius leads to a reduction of the characteristic time for binary aggregation by an order of magnitude.

This effect becomes of paramount importance when one considers the long time evolution of the coagulation process. Let us consider, e.g., a system of Brownian drops. The evolution equation for size classes characterized by their size $i$ (where $i = 1, 2, 3, ..., \infty$), is given by [31]

$$\frac{dc_k}{dt} = \frac{1}{2}\sum_{i+j=k} k_{i,j} c_i c_j - c_k \sum_{j=1}^{\infty} k_{j,k} c_j \tag{2.44}$$

Based on the above considerations, the rate constants will be of the order

$$k_{i,j} \propto \sqrt{\frac{3\pi\alpha\eta_0\dot{\gamma}(a_i+a_j)a_i a_j - U''_m}{k_BT}} e^{[-U_m + 6\pi\eta_0\dot{\gamma}(a_i+a_j)a_i a_j / k_BT]} \tag{2.45}$$

where the mutual diffusivity of two drops of size $i$ and $j$ respectively is defined as $D_{i,j} = k_BT(1/a_i + 1/a_j)/6\pi\eta_0$, leading to $Pe_{i,j} = 3\pi\eta_0\dot{\gamma}(a_i+a_j)a_i a_j / k_BT$. In comparison with the diffusion-limited case in stagnant fluids, where the kinetics slows down as the particles size grows by coalescence (the size dependence of the kinetics is dictated by diffusion), we can see from Eq. (2.45) that under shear, instead, larger drops coalesce much faster, thus leading to a *self-accelerated* kinetics. In particular, at fixed $\dot{\gamma}$, an *activated* size, $a^* = (U_m/6\pi\alpha\eta_0\dot{\gamma})^{1/3}$, can be defined, and it corresponds to a vanishing activation barrier. This may help explain the explosive rise of viscosity in non-Brownian suspensions, reported in [7], as well as in Brownian suspensions, as will be shown in the next Section, in terms of a self-accelerating kinetics setting in as the linear size of the growing structure (*clusters*, with non-coalescing colloids) reaches the activated value $a^*$. The extremely rapid growth of viscosity is indeed tightly connected to the rapid growth in size which causes an increase in the volume occupied by the clusters. Associated with the increase in the effective volume fraction of clusters is the increase in (many-body) hydrodynamic interactions which are in turn responsible for the increase in the high-shear suspension viscosity.

## *2.3 Overview*

Starting from the two-body Smoluchowski equation for interacting particles in shear, we have derived an approximate solution in analytical form which enables one to quantify the aggregation rate between colloidal particles interacting with an arbitrary colloidal potential, in linear (laminar) flow fields, and at arbitrary Peclet. The predictions from the approximate solution, with no fitting parameter, are in good agreement with numerical simulations from the literature, up to very high Peclet numbers (over $\sim 10^2$). When an interaction barrier is present (as for DLVO-interacting systems) an explicit rate theory for the kinetic constant of aggregation has been derived which exhibits the typical Arrhenius form and consists of a frequency factor (proportional to the square root of the shear rate $\dot{\gamma}$) multiplying an exponential factor which gives the fraction of successful collisions. The exponential factor reads $e^{-(U_m - 6\pi\alpha\eta_0\dot{\gamma}a^3)/k_BT}$, with the evident role played by the shear rate in diminishing the activation barrier $U_m$, thus enhancing the aggregation kinetics which shows a rise in the aggregation kinetic constant as large as 5 to 6 order of magnitudes, in correspondence of a critical Peclet which erases the barrier. This result generalizes Kramers' rate theory [32] to activated barrier-crossing processes driven by shear. Further, it offers the key to explain the induction period followed by self-accelerating kinetics observed in the rheopectic behaviour of many complex fluids (e.g. protein-based biological fluids [4,5]) where the constituents interact via an interaction barrier (which stabilizes them against aggregation in the absence of shear).

In the next chapter we are going to extend the microscopic theory valid for dilute suspensions presented in this chapter in order to include the effects of colloid concentration. We will also compare the predictions with measurements on nondilute colloidal suspensions.

# 3. Aggregation kinetics in nondilute colloidal suspensions under shear flow: theory and experiments

## 3.1 Introduction

The two-body theory presented in the previous chapter is able to qualitatively account for the observed induction delay and exponential dependence of the characteristic aggregation time on the shear rate observed, e.g., in [7]. However, two-body theory applies rigorously only in the limit of infinite dilution ($\phi \to 0$), and neglects important density-dependent effects. This is a strong limitation to the applicability of the theory.

In this chapter we extend the theory to arbitrary colloid volume fractions (relevant for both biological and industrial applications) by generalizing the Smoluchowski problem with shear (dealt with in the previous chapter) through an effective medium approach which allows us to account for the effects of colloid concentration such as collective hydrodynamics. The modified theory is thus able to make quantitative predictions for nondilute systems which can be verified experimentally. Hence in this chapter we also report experimental data of the characteristic aggregation time in nondilute charge-stabilized colloidal suspensions that are compared with the theoretical predictions.

The colloidal suspensions used in our experiments are very stable with respect to aggregation for a very long time (i.e. months) since the ionic strength of the solution is kept at values very much below the critical coagulation concentration (which marks the transition from reaction limited and diffusion limited aggregation) of the electrolyte used to make up the ionic background. The colloidal stability is thus imparted by the weakly screened double layer repulsion (resulting in an interaction barrier of ~60$kT$). The suspensions are then subjected to simple shear flow in the Couette geometry of a rheometer (see Figure 3-1) at a fixed shear-rate kept constant in time. The viscosity of the suspension is measured as a function of the time of shearing (step-rate measurement). What is observed under all investigated conditions of ionic strength, shear rate, particle size and surface charge, is that the viscosity remains constant and equal to the zero-time value during an *induction* time after which a very sharp increase of the viscosity is observed. The sharp increase is used to estimate the characteristic aggregation time to be compared with the theory.

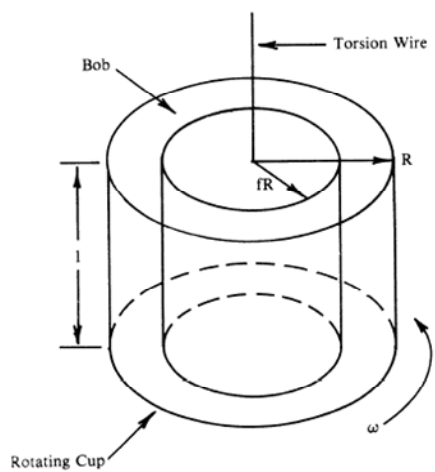

**Figure 3-1: schematic illustration of the Couette geometry used in the experiments.**

Our findings indicate that collective hydrodynamics has a strong effect in reducing the characteristic aggregation time due to the increased effective Peclet number which results from many-body hydrodynamic interactions.

## *3.2 Experimental*

### 3.2.1 Materials and methods

The colloidal system used to perform these experiments is a surfactant-free colloidal dispersion in water, constituted by styrene-acrylate copolymer particles supplied by BASF (Ludwingshafen, Germany) and produced by emulsion polymerization. The nearly monodisperse particles have mean radius of $a = 60 \pm 1$ nm, and were characterized by both dynamic and static light scattering (using a BI-200 SM goniometer system, Brookhaven Instruments, NY). In order to avoid contamination, a thorough cleaning of the suspensions by mixing with an ion-exchange resin (Dowex MR-3, Sigma-Aldrich) was performed. To check that the suspensions were free of impurities after the cleaning procedure, we measured the surface tension by means of the Wilhelmy plate method with a DCAT-21 tensiometer (Dataphysics, Germany) and only suspensions with surface tension $\geq 71.7$ mN/m were used for the investigations. For the shearing experiments, a small amount of electrolyte (NaCl) was added to make up the ionic background. In fact, with de-ionized suspensions, the shearing time at which viscosity rises due to aggregation would be very long (on the order of days). This can seriously affect the system and the reproducibility of the experiments due to solvent evaporation. However, the final NaCl concentration in the sample (17mM) is well below the critical coagulation concentration (50mM with NaCl). The long-time stability of each suspension after adding the NaCl solution was checked by light scattering. To induce the shear flow under shear-rate control and to simultaneously measure the viscosity of the flowing suspension, a strain-controlled ARES rheometer (Advanced Rheometric Expansion System, TA Instruments, Germany) with Couette geometry has been employed. The gap between the outer cylinder and the inner one is 1 mm and the diameter of the latter is 34 mm. The outer cylinder is temperature controlled at $T = 298 \pm 0.1$ K and, in order to prevent evaporation, a solvent trap has been fixed on the outer rotating cylinder. In all the experiments we used deionized water (milli-Q, Millipore) and the mixing of the latex suspensions with NaCl solutions was done in such a way to avoid heterogeneities in the concentration field which could cause the aggregation kinetics to speed up in locally more concentrated regions. It is worth noting that the sampling of all the NaCl-solution/latex mixtures was done carefully with a top-cut pipette in order to avoid any local shearing during the sampling that could induce aggregation. In order to ensure reproducibility, each time the shearing was switched on 7 minutes from the time of mixing between latex and background NaCl solution. For each point in the $\phi$-$\dot{\gamma}$ plane investigated, at least three repetitions were done. The experimental error on the aggregation time $\tau$ has never been found to exceed 15% of the mean value. $\zeta$-potential measurements were carried out using a Zetasizer Nano instrument (Malvern, UK), on dilute suspensions ($\phi = 5 \times 10^{-4}$) at the same ionic background concentration (17mM NaCl) used for the shearing experiments.

### 3.2.2 Statistical data treatment of viscosity versus time curves

Due to the experimental error, the $\eta(t)$ curves are distributed around an average critical time for aggregation, $\langle t_c \rangle$, where $\langle ... \rangle$ now indicates the arithmetic average. For each condition, we performed three repetitions (terminating with the overload of the rheometer) in order to average the data and to build a master or average $\eta(t)$ curve.

The idea is simply to shift each point of the three curves along the abscissa (t-axis), by applying the following transformation

$$t_{new,i} = t_i \pm (t_{c,i} - \langle t_c \rangle) \tag{3.1}$$

where subscript $i$ identifies the curve and $t_{new,i}$ is the new abscissa (for each experimental point of curve $i$). Thus the procedure to create the master curve can be summarized as follows

1) determine $t_{c,i}$ for each curve $i$;
2) average the $t_{c,i}$ values to obtain $\langle t_c \rangle$;
3) collapse the curves using Eq. (3.1).

As an illustration of this procedure, we show the three repetitions, in Figure 3-2 (a), and the resulting averaged master curve, in Figure 3-2 (b).

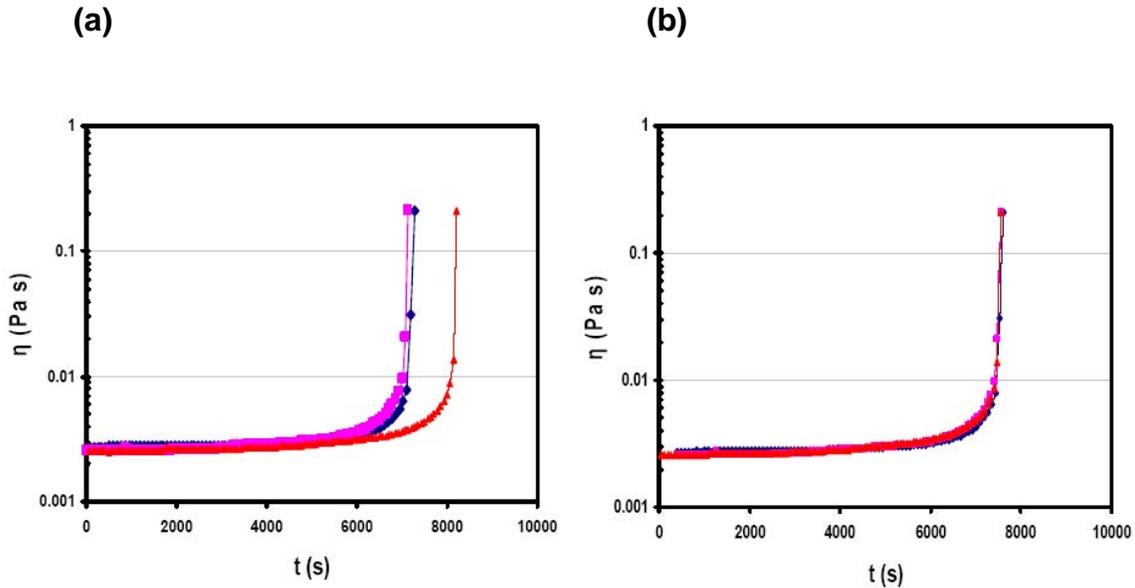

**Figure 3-2: illustration of the data treatment to average different repetitions. (a) three different repetitions at $\phi = 0.21$, 17mM of NaCl, and $\dot{\gamma} = 1500\,\text{s}^{-1}$. (b) master (average) curve obtained from the collapse of the three curves in (a) by using Eq. (3-1).**

## 3.3 Effective Medium Theory

Let us recall that the governing equation for the steady-state in the pair-correlation function $c(\mathbf{r}) = c_0 g(\mathbf{r})$ (where $c_0$ is the bulk concentration) is the following two-body Smoluchowski equation with convection (Eq. (2.5))

$$\text{div}\left\{\beta D[-\nabla U(r) + b\mathbf{v}(\mathbf{r})] - D\nabla\right\} c(\mathbf{r}) = 0 \tag{3.1}$$

where $D = 2D_0 \mathcal{G}(r)$ ($D_0$ being the self-diffusion coefficient and $\mathcal{G}(r)$ the hydrodynamic correction for viscous retardation), $\mathbf{v}(\mathbf{r})$ is the imposed fluid velocity field, $b = 3\pi\eta_0 a$ is the friction coefficient on a tagged particle (with $\eta_0$ the solvent viscosity and $a$ the colloid radius), and $U(r)$ is the isotropic pair-interaction potential. For concentrated systems, Eq. (3.1) should be rewritten by taking into account that the friction that an individual particle experiences under *nondilute* conditions has an additional contribution from the hydrodynamic interactions transmitted by the other Brownian particles. To this aim, we adopt here an *effective medium* approach where the solvent is replaced with an *effective* fluid having the macroscopic properties of the suspension [10]. We start by introducing an effective friction coefficient $b_{\text{eff}}$ (see Dhont [10] pp. 356-357) defined via the following Einstein relation

$$D_{\text{eff}} = (\beta b_{\text{eff}})^{-1} \mathcal{G}(r) \tag{3.2}$$

where $D_{\text{eff}}$ is the relative (long-time) diffusion coefficient for two particles embedded in the effective fluid. Within this approach, the effective friction coefficient is equal to [10]

$$b_{\text{eff}} = 3\pi\eta a \tag{3.3}$$

where $\eta$ is the concentration-dependent *effective* (macroscopic) viscosity of the suspension. We can thus rewrite the Smoluchowski equation with shear for two interacting Brownian particles moving in an effective medium representing the suspension as

$$\text{div}\left\{\beta D_{\text{eff}}(r,\phi)[-\nabla U(r) + b_{\text{eff}}(\phi)\mathbf{v}(\mathbf{r})] - D_{\text{eff}}(r,\phi)\nabla\right\} c(\mathbf{r}) = 0 \tag{3.4}$$

where $D_{\text{eff}}$ is given by Eqs. (3.2)-(3.3) and is now also a function of the colloid volume fraction $\phi$ due to the $\phi$-dependence of $\eta$. In the previous chapter we have proposed a novel scheme which allows one to obtain an analytical solution to Eq. (3.1) under the absorbing boundary condition ($c = 0$) at contact ($r = 2a$) and the far-field boundary condition ($c = c_0$) implemented at the hydrodynamic boundary-layer, for an arbitrary direct interaction potential $U(r)$. Now, we can generalize the approach to arbitrary volume fractions, within the effective medium approximation, by solving Eq. (3.4) with the same scheme introduced in the previous chapter. The concentration-dependent coagulation constant [22] for binary encounter rate that we obtain reads

$$k_{1,1} = \frac{8\pi[\beta b_{eff}(\phi)]^{-1} a c_0}{\int_0^{\xi(\phi)} \frac{dx}{\mathcal{G}(x)(x+2)^2} \exp \int_{\xi(\phi)}^{x} ds(\beta dU/ds + Pe_{eff}(\phi)\langle \tilde{v}_r^+(\mathbf{r})\rangle)} \quad (3.5)$$

with $x = (r/a) - 2$, and $\xi(\phi) = \sqrt{(\lambda_D/a)[Pe_{eff}(\phi)]^{-1}}$, where $\lambda_D$ is the range of the direct interaction (the Debye length in our case). According to the effective medium approach, the effective Peclet number is given by

$$Pe_{eff}(\phi) = \dot{\gamma} a^2 / D_{eff}(\phi) \quad (3.6)$$

Furthermore, the orientation-averaged inward (relative) velocity $\langle \tilde{v}_r^+(\mathbf{r})\rangle$ depends uniquely upon the type of flow (see the previous chapter). The integrals in Eq. (3.5) have to be evaluated numerically. However, as shown in the previous chapter, the generalized expression for the binary encounter rate can be significantly simplified in the case of an interaction potential which goes through a maximum (potential barrier). Applying the steepest-descent method, we arrive at the following activated-rate formula for the reaction-limited aggregation coagulation constant between two Brownian particles in a flowing system at colloid volume fraction $\phi$

$$k_{1,1} \approx 8\pi D_{eff}(\phi) a c_0 (\alpha Pe_{eff}(\phi) - \beta U_m'')^{1/2} e^{-\beta U_m + 2\alpha Pe_{eff}(\phi)} \quad (3.7)$$

$\alpha$ is a coefficient related to the type of flow ($\alpha = 1/3\pi$ in simple shear). $U_m$ is the interaction potential value at the local maximum (i.e. the potential barrier), while $U_m''$ denotes the second derivative evaluated at the maximum. Eq. (3.7) generalizes Kramers' rate theory to the presence of shear and to concentrated conditions.
The associated characteristic time for a binary encounter is given by

$$\tau_{1,1} = \frac{1}{k_{1,1}} \approx \frac{(8\pi D_{eff}(\phi) a c_0)^{-1}}{\sqrt{\alpha Pe_{eff}(\phi) - \beta U_m''}} e^{\beta U_m - 2\alpha Pe_{eff}(\phi)} \quad (3.8)$$

It is worth noting that hydrodynamic interactions due to the disturbance of the shear field induced by the relative motion of the particles are accounted for in the analytical treatment as long as the derivation of the rate expression Eq. (3.5) is concerned. In the subsequent steepest-descent approximation leading to Eqs. (3.7)-(3.8) these two-body hydrodynamic interactions are neglected. However, we have checked numerically that the importance of these hydrodynamic corrections is very minor. Indeed the difference between rates calculated using the full expression accounting for them, Eq. (3.5), and the approximate one, Eq. (3.7), is practically negligible. This is somewhat expected with DLVO interactions since the interparticle distance range where this hydrodynamic effect is important overlaps with the range where van der Waals attraction dominates. This leads to the hydrodynamic effect being masked, as it was originally observed by Smoluchowski [34]. Even if the total potential energy of the system can be expressed as a sum over pair-interactions, the force between two particles in a concentrated system is not equal to $-\nabla U(r)$. It is instead given by $-\nabla U_{\text{eff}}(r)$, where $U_{\text{eff}}$ is the potential of mean force between two particles which contains contributions from the remaining particles [10]. In the following we neglect this effect in the comparison with experimental data since in our system $\kappa a = \lambda_D^{-1} a = 24.74$ and $0.19 \leq \phi \leq 0.23$. Hence, the average separation between a particle and its nearest-neighbours is about 1.7-1.8 times the diameter [35], whereas the range of the screened-Coulomb repulsion is only about 1.04 times the diameter. Further, for a system at equilibrium, the potential of mean force equals $U_{\text{eff}}(r) = -k_B T \ln g_{\text{eq}}(r)$, where $g_{\text{eq}}(r)$ is the pair-correlation function at equilibrium. Under driven, non-equilibrium conditions, the actual pair-correlation function can differ substantially from $g_{\text{eq}}(r)$. These additional effects represent a formidable unresolved issue and here are simply omitted.

In order to close the model, we need expressions for the volume fraction-dependent effective viscosity of the medium, $\eta(\phi)$. For hard-spheres, an improved differential viscosity model has been recently proposed by Mendoza and Santamaria-Holek [36], which yields accurate expressions throughout the entire volume fraction spectrum, from the dilute limit up to close packing, and for both the low and high-shear viscosity. In the case of charge-stabilized particles, the effective viscosity can be further enhanced due to the effective enlargement of the colloid size induced by electrostatics. This effect is especially important at low shear rates but it becomes less and less significant as the shear rate goes up [37]. Since in the following we are going to compare our model predictions with experiments at substantially high shear rates ($\dot{\gamma} \sim 10^3 \text{ s}^{-1}$) we will neglect this effect and use the viscosity expressions for hard-spheres of Ref. [36].

## 3.4 Comparison with experimental data

Experimental curves of effective suspension viscosity ($\eta$) as a function of the elapsed shearing time ($t$) are plotted in Figure 3-1 for three different volume fractions: $\phi = 0.19$ (a), $\phi = 0.21$ (b), and $\phi = 0.23$ (c).

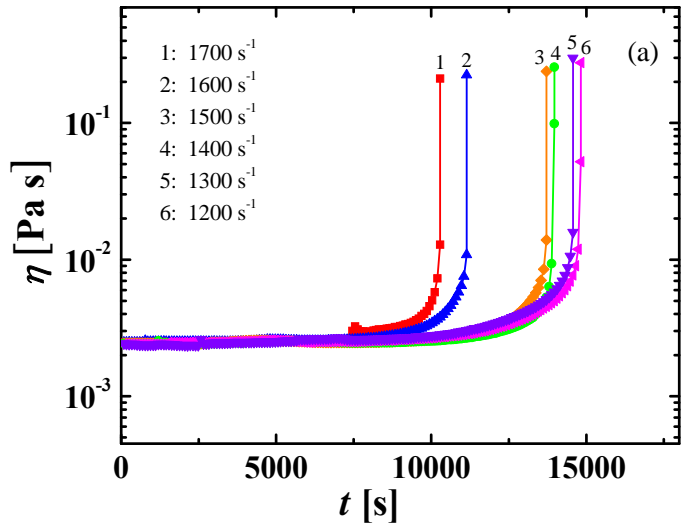

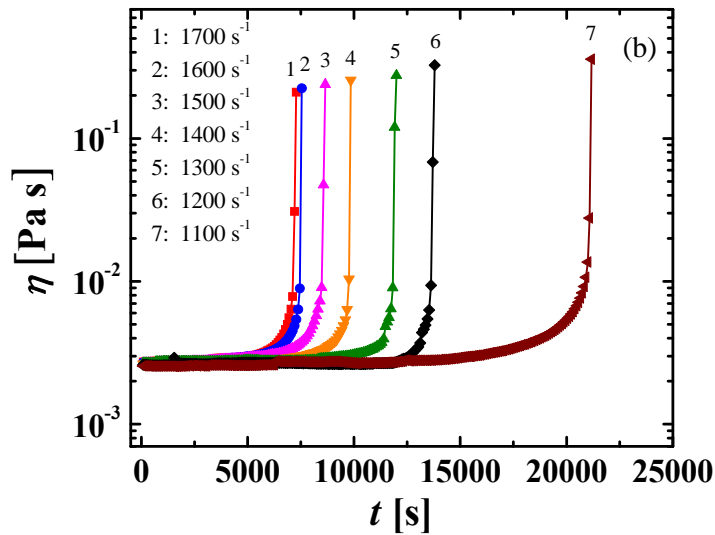

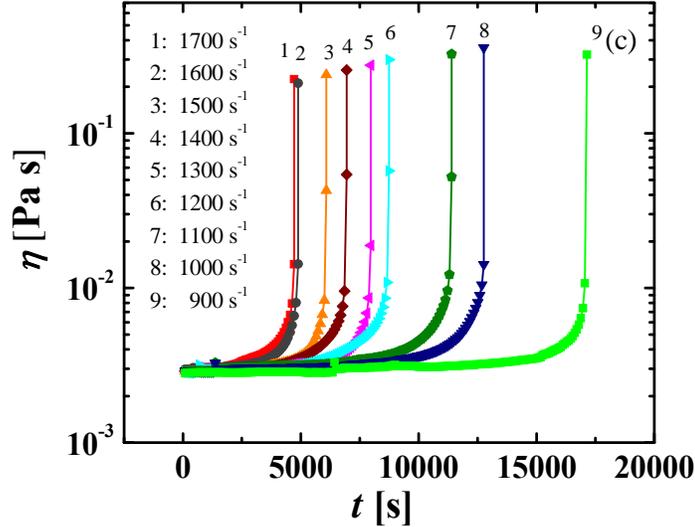

**Figure 3-1: Suspension viscosity as a function of the shearing time under steady shear for charge-stabilized colloids at volume fractions $\phi = 0.19$ (a), $\phi = 0.21$ (b), and $\phi = 0.23$ (c), and at a varying shear rate $\dot{\gamma}$ (see legends). Added electrolyte for all conditions: 17 mM (NaCl). The characteristic time of aggregation ($\tau$) is estimated as shown schematically in (a).**

Consistent with previous observations [7], there is a very sharp, explosive upturn in the suspension viscosity due to the onset of self-accelerated aggregation kinetics as soon as the activation-energy, i.e. the argument of the exponential in Eq. (3.8), vanishes. This happens when on average the formed colloidal clusters reach a shear-activated size, as explained in the previous chapter. A working measure of the characteristic time for aggregation in the experiments ($\tau$) is estimated from the crossing of the asymptotes (according to the protocol introduced by Guery et al. [7]) which is related to the incipient increase of viscosity as a consequence of aggregate formation throughout the system. Rigorously, the theory presented here describes the very initial stage of aggregation where only doublets are formed. However, since the doublets once formed grow further and very quickly to larger clusters, it is very difficult to monitor the conversion of primary particles to doublets for determining the doublet formation rate. The protocol we used to experimentally estimate the characteristic aggregation time is just the most convenient, systematic procedure to evaluate the time scale of a process which is anyways controlled by the doublet formation rate (since aggregation proceeds very rapidly afterwards due to the larger Peclet values of the aggregates with respect to primary particles). Since the viscosity rise is so fast, the overestimation of the true characteristic aggregation time is certainly not dramatic and represents a systematic effect which does not significantly change the agreement and leaves the scaling with $\dot{\gamma}$ unaltered.

We thus obtain a $\tau(\dot{\gamma})$ curve for each $\phi$ investigated. It is seen that, especially at high $\dot{\gamma}$, a modest increase in $\phi$ is able to cause a very significant decrease in the aggregation time. This phenomenon can be attributed to two distinct $\phi$-dependent effects: 1) the effect of collective hydrodynamics which is reflected in a higher effective friction and hence in a higher effective Peclet number; 2) the increase in ionic strength due to increasing the macroion concentration along with $\phi$. In order to compare predictions from our model with these experimental data, one should consider that the characteristic time of a binary encounter between two particles in the effective medium does not correspond yet to the characteristic aggregation time measured in the experiments. The latter is indeed related to the *total* number of collisions per unit time in the system. Under nondilute conditions, colloidal suspensions exhibit a liquid-like structure where each particle is surrounded by a finite number of nearest-neighbours, corresponding to the first peak in the radial distribution function.[1,23] Hence, in the experiment, all binary collisions of each particle with its nearest-neighbours have the same probability to occur. Thus, the *total* frequency of binary collisions relevant for our comparison is given by the collision frequency between two isolated particles in the effective medium times the total number of equally probable collisions, which is directly proportional to the volume fraction. (This is somewhat analogous to the way the total bombardment frequency of gas molecules on a wall is calculated according to the classical kinetic theory of gases [11].) The Arrhenius-like form of Eqs. (7)-(8) is helpful in this respect since it allows us to distinguish the collision *frequency*, $\omega_{1,1} = 8\pi D_{\text{eff}}(\phi)ac_0\sqrt{\alpha Pe_{\text{eff}}(\phi) - \beta U_m''}$, i.e. the pre-exponential factor in Eq. (3.7), from the encounter *efficiency* (the exponential term). Based on these arguments, the total collision frequency in the experimental system (which is relevant to the aggregation time measured), denoted as $\bar{\omega}$, follows upon multiplying the collision frequency for two particles in the effective medium by the total number of (equally probable) collisions in the system. The latter number is given by $zN/2 = 3zV_T\phi/8\pi a^3$, where z is the average number of nearest-neighbours (i.e. of particles in the first coordination shell), and $V_T$ is the total volume of the system. Thus we obtain the following approximate relation for the total binary-collision frequency

$$\bar{\omega} = (3zV_T\phi/8\pi a^3)\omega_{1,1} = (3zV_T\phi D_{\text{eff}}(\phi)c_0/a^2)\sqrt{(\alpha Pe_{\text{eff}}(\phi) - \beta U_m'')} \tag{3.9}$$

According to liquid structure theory, the average number of particles in the first shell is always $z \approx 12$ and does not depend much upon the volume fraction [35]. Of course, in the presence of intense shear flow, the local structure is significantly distorted and anisotropic [10]. However, the higher local density in the two upstream quadrants is counterbalanced by particle depletion in the two downstream quadrants. Hence, despite the significant anisotropic shape of the radial distribution function in shear, due to the balancing of densification and depletion in opposite quadrants, the orientation-averaged number of nearest-neighbours in a disordered suspension is not expected to deviate much from the estimate $z = 12$ that we use in our calculations. Thus, the characteristic time for aggregation to be compared with the experimental data is given by

$$\tau_c \approx \frac{(3zV_T\phi D_{\text{eff}}(\phi)c_0/a^2)^{-1}}{\sqrt{\alpha Pe_{\text{eff}}(\phi) - \beta U_m''}} e^{\beta U_m - 2\alpha Pe_{\text{eff}}(\phi)} \tag{3.10}$$

The effect of macroion (colloid) concentration upon the interaction parameters in the model ($U_m$ and $U_m''$), on the other hand, is more difficult to assess. This is due to the difficulty, in the experimental practice, of accurately determining the electrostatic surface potential ($\psi_0$) of the colloidal particles under such nondilute conditions ($\phi \approx 0.2$) as in our shearing experiments. In view of this, we have left $\psi_0$ (required in the DLVO calculation of $U_m$ and $U_m''$) as the only adjustable parameter. Once $\psi_0$ is fixed, the DLVO-interaction potential curve for the system can be calculated using the Sader-Carnie-Chan formula valid for high surface electrostatic potentials [38]. The van der Waals attractive component of the DLVO interaction is determined using the standard formula for colloidal spheres as is found in the textbooks [3] with the value of Hamaker constant of the polystyrene-acrylate/water system.

The comparison between theoretical estimates from Eq. (3.9) and the experimentally measured values of $\tau_c$, is shown in Fig. 2.

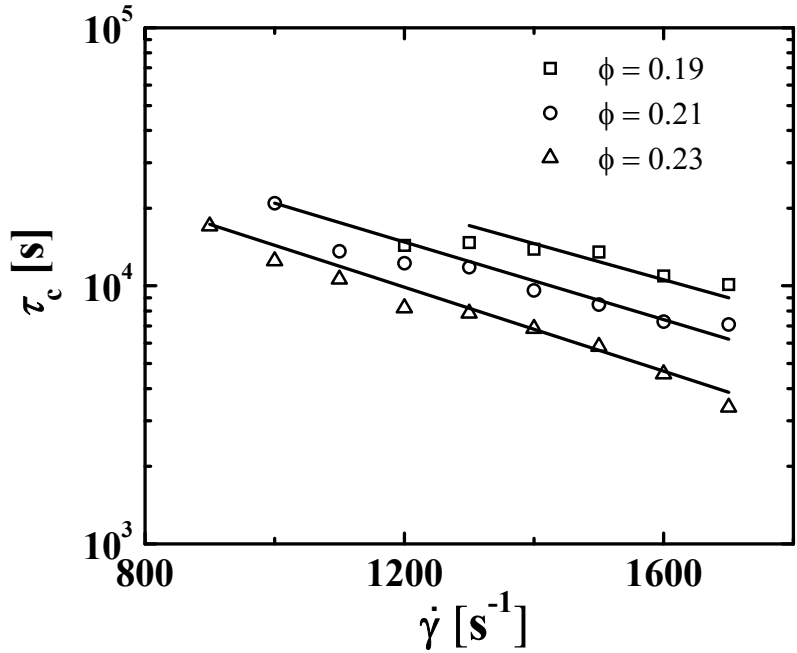

**Figure 3-2: Characteristic aggregation time under nondilute conditions (see legend) as a function of the applied shear-rate. Symbols: data points from the experiments reported in Figure 3-1. Solid lines: theoretical calculations using Eq. (3.10) (see Text).**

For the effective suspension viscosity inside the effective Peclet number we have used the following improved expression for hard-spheres by Mendoza and Santamaría-Holek [36], $\eta(\phi) = \mu[1-\phi/(1-c\phi)]^{-5/2}$ with $c = (1-\phi_c)/\phi_c$, and $\phi_c = 0.7404$ as prescribed for the high-shear branch. An excellent agreement is found using the following values of the surface potential in the calculation of the DLVO interaction (using the Sader-Carnie-Chan formula for the electric double layer repulsion [38]): $\psi_0 = -45.67$ mV ($\phi = 0.19$), $\psi_0 = -45.64$ ($\phi = 0.21$), and $\psi_0 = -45.60$ ($\phi = 0.23$). A slight decrease of $\psi_0$ with increasing $\phi$ is reasonable since an increase in the macroion concentration (by keeping the salt concentration constant) along with $\phi$ brings about an increase of the total ionic strength of the system. These values of $\psi_0$ are still within the confidence interval of the measured colloid $\zeta$-potential under dilute ($\phi = 5 \times 10^{-4}$) conditions, which was found to be $-45.9 \pm 3$ mV. In measuring the $\zeta$-potential, the same background ionic concentration (17 mM of NaCl) as in the shearing experiments was used. Hence, a value of potential under nondilute conditions which is lower than the value under dilute conditions (with the quantity of added electrolyte being the same) is reasonable.

This comparison demonstrates the capability of the model to capture the volume fraction dependence of reaction-limited aggregation kinetics in shear. Furthermore, the comparison indicates that the major effect behind the significant reduction of the aggregation time upon increasing $\phi$ is the increase in the effective friction and thus in the effective Peclet number which controls the encounter efficiency through the exponential term of Eq. (3.10). Indeed, $Pe_{\text{eff}}$ increases nonlinearly with $\phi$ thus causing a strongly nonlinear increase of the encounter efficiency upon increasing $\phi$. The resulting decrease in the absolute value and increase in the slope of the $\tau_c(\dot{\gamma})$ curve predicted by the theory is qualitatively consistent also with previous experimental data by Guery et al. [7]. Finally, the fact that the fitted surface potential is practically constant and changes by less than 0.1mV with $\phi$ suggests that the $\phi$-dependent effect of the ionic strength plays a comparatively minor role.

## 3.5 Overview

In this Chapter we have generalized the two-body rate theory developed in the previous Chapter to the presence of concentration effects with application to the shear-induced aggregation rate of charge-stabilized colloids. Experimental investigations are also reported on nondilute colloidal suspensions in shear flow where the characteristic aggregation time has been measured. The experimental data have been compared with the theoretical predictions. The results of our analysis provide insights into the kinetics of reaction-limited (colloidal) aggregation kinetics under shear in concentrated ($\phi > 0.10$) conditions. In order to account for collective hydrodynamics, we have reformulated the Smoluchowski problem with shear by using an effective medium approach. This introduces an effective friction coefficient which depends upon the colloid volume fraction $\phi$ through the effective suspension viscosity. The Smoluchowski equation is then solved using the boundary-condition proposed in the previous Chapter. By implementing an appropriate expression for the effective suspension viscosity as a function of colloid volume fraction, we calculated the binary encounter rate for concentrate colloidal suspensions. The theoretical calculations are compared with the experimental data of DLVO-interacting colloids (with a potential barrier of approx. 60 $k_BT$) in simple shear in the range $0.19 \leq \phi \leq 0.23$. The theory is in good agreement with the experiments and is able to capture the volume fraction dependence of the characteristic aggregation time. Further, the comparison suggests that the process playing a major role behind the significant reduction of the aggregation time upon increasing $\phi$, is to be identified with the nonlinear increase with $\phi$ of the effective Peclet number which in turn causes a strongly nonlinear increase of the collision efficiency. Hence, many-body hydrodynamic interactions play an active role in enhancing the shear-activated barrier-crossing process which controls the aggregation rate.

In future work, it is hoped that these findings can be applied to situations directly relevant to biological processes where *in vivo* biochemical reactions between biomolecules occur in crowded and mechanically strained environments.

So far we have considered only the initial stage of the aggregation process, i.e. the formation of doublets. In the next chapter we are going to deal with the successive growth kinetics of the colloidal clusters as a function of the shearing time. This will give us insights into how the mesoscopic structures are formed and about their role in determining the macroscopic properties (i.e. the effective viscosity) of colloidal suspensions in shear.

# 4. Kinetics of shear-induced clustering and mesoscopic growth

## *4.1 Introduction*

So far we have investigated the aggregation kinetics under shear without actually characterizing the colloidal clusters formed and their growth process. In this section we will first present a thorough experimental characterization of the cluster formation and growth process. We have found, by means of light scattering, that the clusters growth rapidly reaches a steady-state with time due to the dynamic balance between shear-induced aggregation and shear-induced breakup. The light scattering characterization of the clusters as a function of time along a typical evolution curve (such as those in Figures 3-1) is made possible by the fact that the particles (as well as the clusters) are stabilized by charge-repulsion. Moreover, after the shearing is stopped, the system remains stable with respect to aggregation (and also with respect to breakage, since the aggregates are formed at high shear rate and are thus able to withstand high applied stresses). This enabled us to sample the system and perform off-site light scattering. It would not have been possible with purely attractive colloids, since aggregation would continue also after shearing is switched off. In that case, the only option is the technically more involved in-situ scattering.

Within a relatively short time from the switching on of shearing the cluster population acquires well defined characteristics, in terms of size and morphology, which are preserved during the further time evolution (due to the mentioned dynamic equilibrium between aggregation and breakup). What is going to change from this point on is the number of clusters per unit volume, and, as a consequence, the volume fraction occupied by the clusters. This kinetic process has been characterized in terms of the conversion from primary particles to clusters as a function of time, by combining filtration and light scattering. With this information at hand, from the experiments we could extract the relevant time-evolving parameters of the clusters. We have used these observations to build a phenomenological theory of aggregating suspension viscosity at fairly high shear rates by applying the scaling principle proposed in this thesis. On this basis, we could eventually elucidate the rheopectic (anti-thixotropic) behaviour observed in charge-stabilized colloids of which Figure 3-1 gives a clear exemple.

## 4.2 Materials and methods

### 4.2.1. Shearing conditions, sampling and off-site light scattering

The colloidal system, the small-angle light scattering (SALS) instrument as well as the rheometer setup (used for the shearing and the rheological measurements) have already been introduced in the experimental section of Chapter 3.

For the experiments reported in this section, one typical and reproducible set of operating conditions has been selected. This corresponds to the following parameters:

- volume fraction: $\phi = 0.21$
- ionic strength: 17 mM (NaCl)
- shear rate: $\dot{\gamma} = 1700 \, \text{s}^{-1}$

In order to characterize the system evolution with time, it was necessary to switch off the shearing at different times. Thus it was important to have a reliable reference curve. The latter has been obtained by the method explained in Section 3.2.2.

We proceeded as follows. At regular times after switching on the shearing we sampled the system along the $\eta(t)$ curve after switching off the shearing in the Couette cell of the rheometer. We performed the following procedure for each sampling. We immediately diluted the sample to reach the concentration required for the light scattering analysis. We analyzed the samples by small-angle and static light scattering (SLS) (BI-200SM goniometer, Brookhaven, and an argon-ion laser M95-2, Lexel) in order to characterize the clusters. In fact, performing static light scattering on the very dilute sample (to avoid multiple scattering effects) leads to extracting the average structure factor, $\langle S(q) \rangle$ (where $q = (4\pi n / \lambda_0) \sin(\theta / 2)$ is the wave vector, with $n$ the refractive index of the solvent, $\lambda_0$ the wavelength of the incident beam and $\theta$ the scattering angle), of the clusters from which the system-averaged size and fractal dimension of the clusters can be obtained.

### 4.2.2. Measurements of conversion from primary colloids to clusters

In parallel, we characterized the conversion of primary particles to aggregates/clusters. In order to do that, we sampled from the rheometer and measured the structure factor $\langle S(q) \rangle$, after proper dilution and filtration with a 5 $\mu m$ mesh filter, by light scattering. In this way, we separated all the large clusters. In the filtered samples, we measured the $\langle S(q) \rangle$ of primary particles plus dimers and very small aggregates (oligomers). In order not to overestimate the volume fraction (of the only primary particles present), we applied the following correction. This is based on using the turbidity measurements from SALS using the Lambert-Beer's law that can be expressed as:

$$\tau = N C_{ext} L$$

where $L$ is the path length of the incident light through the samples and $N$ and $C_{ext}$ are respectively the number concentration and the extinction cross section of the particles. $C_{ext}$ is constituted by two terms, the cross section $C_{abs}$ and the scattering cross section $C_{sca}$. The relationship is expressed as:

$$C_{ext} = C_{abs} + C_{sca} \tag{4.1}$$

The $C_{ext}$ values for a primary particle, dimer and trimer are different. For our system they have been computed using the T-matrix method [39] and are equal to $2.18 \times 10^{-13}$, $4.81 \times 10^{13}$ and $7.50 \times 10^{-13}$ m$^2$, respectively. Then, the right-hand side of Eq. (4.1) should be the sum of the contributions from all the three components

$$\begin{aligned}\tau &= [N_1 C_{ext,1} + N_2 C_{ext,2} + N_3 C_{ext,3}]L \\ &= N_{1,T}[x_1 C_{ext,1} + (x_2/2)C_{ext,2} + (x_3/3)C_{ext,3}]L\end{aligned} \tag{4.2}$$

where $N_{1,T}$ is the total number concentration of the primary particles, including those forming the dimers and trimers, $x_1, x_2, x_3$ are the parameters fitted from the $\langle S(q) \rangle$ curve after filtration, which are respectively the volume fraction of primary particles, dimers and trimers. Hence, the turbidity $\tau$ measured from light scattering is divided by the following correction factor

$$C_F = \frac{x_1 C_{ext,1} + (x_2/2)C_{ext,2} + (x_3/3)C_{ext,3}}{C_{ext,1}} \tag{4.3}$$

As we mentioned the turbidity is calculated indirectly by the experimental obscuration value given by the SALS instrument, with the following relation:

$$\tau = -\ln(1 - OBS) \tag{4.4}$$

where $OBS$ is the value of the obscuration from the SALS. By determining the calibration curve of the instrument in terms of $\tau$ as a function of $\phi$, we can get the correct value of volume fraction occupied by the primary colloids present in the system at time $t$, $\phi_p(t)$. From this, it is possible to estimate the conversion from primary particles to clusters at different times as

$$\chi(t) \equiv \frac{\phi - \phi_p(t)}{\phi} = \frac{V_T - V_{pp,T}}{V_T} \approx \frac{N_{pcl,T}(4/3)\pi a^3}{N(4/3)\pi a^3} = \frac{N_{pcl,T}}{N} \qquad (4.5)$$

where $V_T$ denotes the total volume occupied by colloid particles in the system, $V_{pp,T}$ the total volume occupied by primary particles not belonging to clusters, $N_{pcl,T}$ the total number of primary particles which form clusters, and $N_{pp,T}$ the total number of colloid particles in the system.

## 4.3 Kinetics of shear-induced clustering: an off-site light scattering study

### 4.3.1. SALS at different shearing times: characterization of shear-induced clustering

In this section we present the results from the small angle light scattering (SALS) characterization of the aggregating colloidal system. For the conditions selected, the reference time evolution curve of viscosity is shown in Figure 4-1. Great care was put into preparing suspensions under exactly the same conditions of Figure 4-1. The suspensions were sheared always at the same fixed shear rate (1700 s$^{-1}$) and the shearing was then switched off at different elapsed times from the time of switching on the shear (which is taken as $t=0$), along the curve in Figure 4-1.

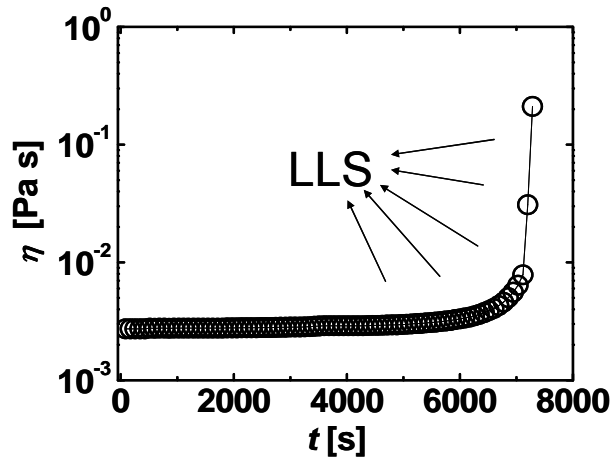

**Figure 4-1: Time-evolution viscosity curve corresponding to the chosen conditions for the growth kinetics study: $\phi = 0.21$, 17 mM (NaCl), $\dot{\gamma} = 1700\,\text{s}^{-1}$. Arrows schematically indicate samplings to laser light scattering (LLS) upon cessation of flow.**

For each of the elapsed shearing times, immediately after switching off the shear, the system was sampled and diluted with deionized (Milli-Q) water to a final volume fraction equal to $2.5 \times 10^{-4}$ for the light scattering. At this volume fraction multiple scattering effects (that otherwise would affect the zero-angle scattering intensity) are absent and the measured scattering curves contain information about the morphology of the clusters present in the system which thus behave as isolated scatterers. The dilution was effected with care in order to precisely reach the same final volume fraction for all the samples. In this way any change in zero-angle scattering intensity is exclusively due to changes in the average size of the clusters (or in their size distribution). The measured scattered intensity $I(q)$ has been divided by the form factor of the primary particles, $P(q)$ (previously measured by means of static light scattering), to obtain the average structure factor of the clusters present in the system, $I(q) = I(0)P(q)\langle S(q)\rangle$ [33,40]. From this relation it is obvious that whenever $\langle S(q)\rangle \approx const$ the system is constituted by primary particles only. The average structure factors corresponding to different shearing times (i.e. to different times in the steady-shear time evolution curve of Figure 4-1) are reported in Figure 4-2.

From these scattering curves, the time evolution of the shear-induced structure-formation process is evident. At the shortest shearing time examined (*t*=412 s) the $\langle S(q)\rangle$ is completely flat indicating the absence of aggregation since the system contains only primary particles (monomers). At successive times, however, the system develops the characteristic features of fractal colloidal aggregation and growth (in dilute systems). The main feature is the appearance of power-law behaviour in a certain range of *q* values $\langle S(q)\rangle \sim q^{-d_f}$ [40]. This is typical of (dilute) suspensions of large fractal aggregates, the power-law exponent $d_f$ being a measure of their fractal dimension. We observe a monotonic increase with time of the zero-angle intensity, $I(0)$. This can be ascribed to the monotonic growth of the average cluster size as well as to the increase with time of the number of clusters. We also note that for shearing times $t \lesssim 7000$ s the scattering curves still exhibit a flat tail in the large *q* range, which then vanishes at higher shearing times. This feature indicates that for $t \lesssim 7000$, besides a significant population of large clusters, the system is still largely dominated by the primary colloids. Therefore a transition from a system dominated by primary colloids to one dominated by clusters occurs at $t \approx 7000$. This value of shearing time agrees well with the value of characteristic aggregation time, $t_c = 7095\,\text{s}$, estimated from the inflection point in the viscosity versus time curve (Figure 4-1) for this system.

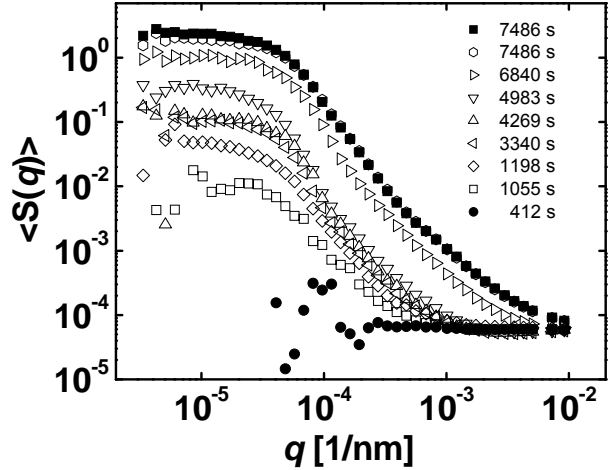

**Figure 4-2: Average structure factor curves from SALS for different shearing times after dilution to volume fraction 2.4x10$^{-4}$.**

From the crossover regime between the low-$q$ plateau and the power-law regime, it is possible to extract the average radius of gyration of the clusters, by applying the well-known Guinier approximation valid at low scattering angles [10]

$$\lim_{q\to 0} I(q) = I(0)\exp\left(-\frac{q^2 R_g^2}{3}\right) \qquad (4.6)$$

where $R_g$ is the average radius of gyration of the clusters and can be readily evaluated from a Guinier plot of $\ln\langle S(q)\rangle$ versus $q^2$ in the regime $qR_g < 1$.

On this basis, from the scattering curves reported in Figure 4-2 we could extract the time evolution of the average radius of gyration of the clusters and of their fractal dimension. The former is reported in Figure 4-3 and the latter in Figure 4-4.

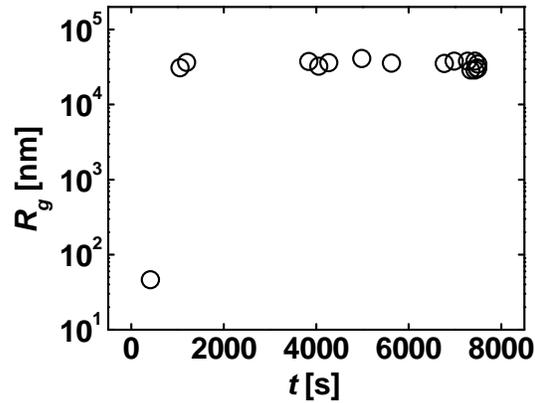

**Figure 4-3: average radius of gyration of the clusters resulting from shear-induced aggregation as a function of the shearing time.**

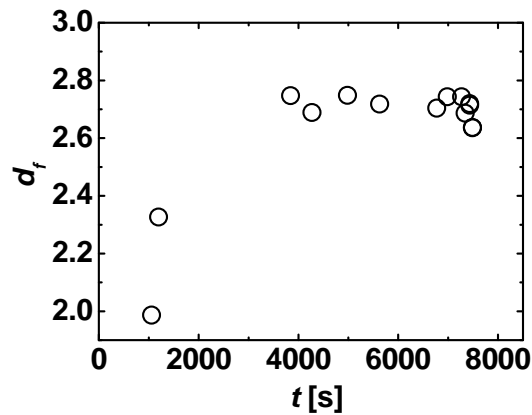

**Figure 4-4: fractal dimension of the clusters resulting from shear-induced aggregation as a function of the shearing time.**

Because of the method used, the average radius of gyration plotted in Figure 4-3 is representative of the cluster population but takes no account of the presence of primary colloids. As a consequence even at $t < \tau_c$ where the system is still numerically dominated by the primary colloids to a large extent, a few but very large clusters (with $R_g \approx 3 \times 10^4 \, \mu m$) coexist with monomers and oligomers (doublets and triplets). This observation will be quantitatively confirmed by the examination of the conversion from monomers to clusters in the next section. In the time evolution in Figure 4-3 it is striking that the cluster size reaches a stationary value at $R_g \approx 3 \times 10^4 \, \mu m$ after a very short transient and that there is practically no deviation from this stationary state value with increasing the shearing time further. This can be explained with the concept of maximally stable size dictated by breakup. Since the clusters are fractal objects, their growth is accompanied by a decrease in the average density inside the cluster. In the clusters the colloid particles are bonded by quite strong bonds since irreversible aggregation occurs in the primary minimum of the DLVO potential (of order $10^1 k_B T$). As a consequence, the clusters are rigid and can be assumed in good approximation to behave as dense, brittle disordered solids. As it is shown in [19], using fracture mechanics arguments, the clusters undergo fracture under a critical stress for which there is no resistance to the unstable propagation of cracks. Since the shear modulus (which gives the resistance to fracture propagation) is a strongly increasing function of the average density inside the cluster, at a certain point in the cluster growth, a value of average volume fraction is reached corresponding to the fracture limit at that applied hydrodynamic stress $\sigma = \eta_0 \dot{\gamma}$, and the cluster breaks up [19]. Thus, for any value of applied stress there will be a maximum stable size of the cluster which varies as a power-law of the applied shear rate: $R_g \sim \dot{\gamma}^p$ [19]. According to this picture, the breakup is a practically instantaneous event (quasi-elastic fracture) and as soon as the critical cluster inner density for breakup has been reached the size should level off to the maximum mechanically stable value. This is indeed what is observed in Figure 4-3.

A similar trend is observed for the average fractal dimension of the clusters, whose time evolution is shown in Figure 4-4. The first (few) clusters that are formed, at $t \approx 1000\,\text{s}$, display $d_f \approx 2$. This value is similar to the case of fractal aggregation with interaction barrier in the absence of flow (reaction limited cluster aggregation). However, at this stage the system is still strongly dominated by the primary particles so that the fractal power-law regime may be affected by the primary particle tail in the structure factor resulting in an *apparent* low $d_f$. Upon increasing the shearing time, the fractal dimension increases and finally reaches a stationary plateau at $d_f = 2.7 - 2.8$. This high value of fractal dimension is the result of flow-induced restructuring processes which lead to fairly compact clusters with a fairly high average inner density inside the cluster (since the latter goes as $\rho \sim (R_g / a)^{d_f - 3}$). Eventually, the leveling off with time of the fractal dimension at $d_f = 2.7 - 2.8$ might be due to the reaching of maximally compacted clusters which are no longer susceptible of further restructuring under the applied shear stress. One should note however that these maximally compacted clusters can still undergo breakup due to crack propagation once they reach a critical size, as just mentioned above [19].

To summarize, we observe a relatively short transient where both the radius of gyration and the fractal dimension of the clusters rapidly grow in time followed in both cases by a stationary state with $R_g \approx 3 \times 10^4 \mu m$ and $d_f = 2.7 - 2.8$, respectively. The transient time is characterized by the following phenomena:

1) cluster growth in size by shear-induced aggregation, implying a decrease of the inner cluster density (due to fractality of the clusters);

2) an increase in the fractal dimension due to shear stress-induced compaction (thus partly compensating the inner density decrease due to cluster growth, according to the fractal scaling $\rho \sim (R_g / a)^{d_f - 3}$)

The transient terminates when a maximally compacted structure is reached, characterized by $d_f = 2.7 - 2.8$. Any further growth (at constant $d_f$) leads to lower values of the average inner density in the cluster which are eventually mechanically unstable with respect to fracture propagation and breakup. This establishes a dynamical balance between aggregation and breakup which keeps the cluster size constant in time around $R_g \approx 3 \times 10^4 \mu m$.

After the stationary state is reached, further aggregation goes into creation of new clusters whose growth is constrained by breakup. Thus, at a rather coarse-grained level, we can say that the number density of large clusters in the system will continue to increase with time due to shear-induced aggregation although their average size remains practically constant.

These considerations have a considerable importance for the understanding and the modelling of the macroscopic properties of colloidal suspensions under shear. We have indeed demonstrated that, after a relatively short transient, the cluster population in the system, generated by shear-induced aggregation, reaches, at $t \sim 1000 \, s^{-1}$, size and morphology which do not change further with time and only the number density of the clusters further evolves (increases). Given the very large size of these clusters (of the order of a few tens of micrometers) with respect to the primary colloid size, it is expected that even when the system is still numerically dominated by primary particles, the presence of a few such large clusters will contribute dramatic effects in terms of hydrodynamic interactions and packing constraints. Accordingly, this should provide, in turn, the dominant contribution to the suspension viscosity. This hypothesis is tested in depth in the following chapter.

### 4.3.2. Experimental determination of shear-induced clustering conversion as a function of the shearing time: results and discussion

As already mentioned, together with the information concerning the morphology of the clusters and its time evolution, we obtained a quantitative measure of the primary particles converted into aggregates (for each sampling from the rheometer at different shearing times). The key result is that for each point in Figure 4-1, we could associate a well defined measure of conversion from primary colloids to clusters, eventually establishing a quantitative relationship between the macroscopic suspension viscosity and the microscopic shear-induced clustering process.

In the range of shearing times $t$ accessible for sampling, the conversion $\chi$ (defined by Eq. (4.5)) has been measured according to the procedure described in section 4.2.2. The time evolution of $\chi$ is plotted in Figure 4-5.

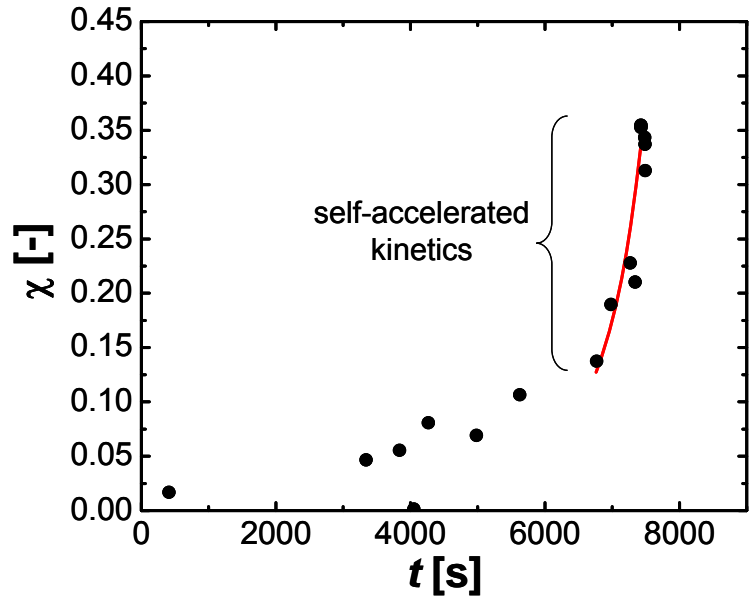

**Figure 4-5: conversion from primary particles to clusters as a function of the shearing time. The red line is a fit to the data in the self-accelerated regime.**

From the plot in Figure 4-5 we can clearly distinguish between two regimes: a first one where conversion increases comparatively weakly as a function of the shearing time and remains below 10-15%, and a second one setting in at $t \approx 7000\,\text{s}$ and $\chi \approx 0.20$, which is characterized by a very sharp increase in the conversion. This value of shearing time well corresponds, once again, to the characteristic aggregation time estimated from the inflexion point in the viscosity versus time curve (Figure 4-1 for the present conditions). Further, the extremely sharp rise of the conversion in this second regime could be well understood with the self-accelerating kinetics concept introduced based on our theory in Chapter 2. In fact, if one considers an average size of the particles in the system, where all size classes including primary colloids are taken into account, there will be an average size (clearly larger than primary colloid size) which grows with the shearing time also after the cluster size growth has reached a stationary value (see the previous section), due to the increase in conversion. When this average size reaches the activated value $a^* = (U_m / 6\pi\alpha\eta_0\dot{\gamma})^{1/3}$ (see Chapter 2), according to our theory a *self-accelerated* aggregation kinetics sets in, leading to a very sharp increase in the conversion.

Moreover, from the analysis of the structure factor from small angle light scattering on the samples after filtration with the $5\,\mu m$ size mesh, it has been possible to obtain more detailed information about the composition of the system. In particular it turns out that for $t < 3000\,\text{s}$ small clusters (oligomers whose size amounts to a few times the primary colloid size) are present in the filtered samples, whereas at higher shearing times only primary particles are present. The analysis has been done by fitting the structure factor of the dilute suspension according to the procedure explained in [41]. Hence, it appears that the population is strongly bimodal, with most colloids belonging to either the primary particle-class and the oligomer-class or to the class of very large clusters (with $R_g \approx 3 \times 10^4 \,\mu m$). This observation can be again explained with our kinetic theory presented in chapter 2. Indeed it turns out from the theory that the activated size in this system amounts to $a^* = (U_m / 6\pi\alpha\eta_0\dot{\gamma})^{1/3} \approx 400\,\text{nm}$ which is compatible with the size of oligomers. Thus, according to the theory, clusters with size $> a^*$ are very unstable and undergo further aggregation practically instantaneously. With further increasing the shearing time, the average system size will grow so that practically all clusters, even the very small ones are highly unstable and therefore disappear from the system leaving just primary particles and large clusters. These considerations can help rationalize the observed absence of a significant amount of clusters with size intermediate between the oligomer size range and the very large clusters.

Finally, we note that the results in Figure 4-5 provide the basis to establish an important link between the shear-induced aggregation, clustering kinetics and the suspension viscosity. Indeed, for each shearing time, by combining Figure 4-1 and Figure 4-5 it is possible to obtain a relationship between the suspension viscosity and the clustering conversion. This will greatly help us to provide a semi-quantitative description of the shear viscosity of aggregating suspensions in the next chapter.

## *4.4 Overview*

We exploited the fact that in charge-stabilized suspensions that are perfectly stable over long times under stagnant conditions, one can straightforwardly sample the system undergoing steady shearing and perform off-site analysis (by putting care into avoiding any shear during handling and measurement) by means of small angle light scattering. For different shearing times, we have sampled the system and diluted it to obtain information about the morphology (size and fractal dimension) of the clusters formed by the shear-induced aggregation. It turns out that both the average size of the clusters and the fractal dimension reach a stationary plateau after a relatively short transient where both are seen to rapidly grow. The appearance of a plateau in the size is explained with the dynamical balance between shear-induced aggregation and shear-induced breakup, while the one in the fractal dimension evolution is likely due to a restructuring process. The latter one leads to maximally compacted clusters whose structure cannot be further modified under the applied shear stress. We have also analyzed the conversion from primary particles to cluster, as measured by filtering the samples and performing light scattering after filtration, and revealed that a very rapid rise sets in after a regime of comparatively weak increase. This feature has been explained by means of the theory developed in Chapter 2 in terms of self-accelerated kinetics making its appearance when the average size of the system population (including *both* primary colloids and clusters) reaches the activated size for shear-induced aggregation. From the conversion analysis we also unveiled that the population is strongly bimodal and is composed mainly by primary colloids (plus, initially, very small oligomers) and very large clusters, with negligible amount of clusters of intermediate size. Also this observation, can be understood by using our theory of shear-induced aggregation: clusters larger than the already mentioned activated size are highly unstable with respect to shear-induced aggregation and undergo further aggregation nearly instantaneously. Therefore, they directly jump into the class of largest clusters that are mechanically stable with respect to breakup.

Finally, we could establish a direct link between the microscopic level, where clustering occurs, and the macroscopic one (investigated through bulk rehology), since to each value of clustering conversion corresponds a value of suspension viscosity. This link will turn out to be essential in the next chapter.

# 5. The shear viscosity of aggregating colloidal suspensions

## *5.1 Introduction*

In this chapter we are concerned with the shear viscosity of aggregating suspensions and also with a first quantitative test of the scaling principle proposed within this thesis (cfr. Chapter 1). Here we briefly recall the main concept. In the presence of aggregation, which leads to mesoscopic structures, the macroscopic properties of colloidal systems in the broad range of intermediate volume fractions between the concentrated limit and the dilute one can be effectively described by the same laws by rescaling them onto the mesoscopic length-scale. This is just equivalent to treating the mesoscopic entities as renormalized or effective particles. A further assumption is that microscopic physics and interactions enter mainly into the aggregation process whereas at the mesoscopic level these are less important and the interactions between mesoscopic particles are the relevant ones. What happens is that the effective volume fraction occupied by porous mesoscopic structures is generally high. At the same time, the direct interactions between primary particles, due to their much shorter range compared to the characteristic mesoscopic length-scale, become less important at the final level of description. These considerations motivate the use of simplified interaction models, such as the hard-sphere model, to describe interactions between mesoscopic bodies.

We will start by applying this coarse-graining scheme to the suspension viscosity of our system. Consistent with this scheme, we consider clusters as effective particles or effective building blocks of the suspension and assume that they behave as hard-spheres. We thus apply hard-sphere theory of suspensions at the cluster (mesoscopic) level. Since we could measure the conversion to clusters experimentally, and thus evaluate the effective volume fraction in the experimental system, this gives us a unique opportunity to test the scaling hypothesis against the experimental measurements in our previously described model colloidal system. The main outcome is that as soon as the clusters occupy a significant volume fraction of the system, their contribution to the dissipated energy (hence to the viscosity) is dominant over the pripary particle contribution in view of the fact that the dissipation is nonlinearly increasing with the cube of the size of the bodies.

The final result is a successful phenomenological description of the shear viscosity of aggregating colloidal systems.

## 5.2 The viscosity of hard-sphere suspensions

Einstein calculated the shear viscosity of a suspension of non-Brownian hard spheres in the dilute limit of volume fractions $\phi \lesssim 0.03$, based on the viscous dissipation that the solvent flow induces around an isolated sphere. He derived the well known formula [42,43]

$$\eta = \eta_0 (1 + [\eta]\phi) \tag{5.1}$$

where, as usual, $\eta_0$ is the solvent viscosity, $\phi$ the volume fraction of the suspension and the intrinsic viscosity is $[\eta] = 2.5$ for spheres. Upon increasing the volume fraction, the drag on one sphere starts to be affected by the motion of a second sphere at close enough distance. This solvent-mediated, mutual disturbance goes under the name of hydrodynamic interaction and the two-body contribution goes as $\phi^2$. Upon further increasing the volume fraction, three-body interactions (proportional to $\phi^3$) will become important as well and so on. Extending Eq. (5.1) to account for three-body interactions proves to be already an intractable task. In the non-dilute regime, besides numerics (e.g. Stokesian dynamics), there are very few analytical techniques to describe the suspension viscosity which is heavily affected by long-range many-body hydrodynamic interactions. The simplest approximate approach is the differential effective medium theory originally due to Arrhenius [44]. This is based on treating the suspension as a *homogeneous* effective continuum with viscosity $\eta(\phi)$. Then, the infinitesimal increment in the medium viscosity can be deduced from Einstein's result for one sphere as

$$d\eta = [\eta]\eta(\phi)d\phi \tag{5.2}$$

leading to the following relation

$$\eta = \eta_0 \exp([\eta]\phi) \tag{5.3}$$

Numerical values for $[\eta]$ have been calculated for spheroids for different values of the ratio between the major semi-axes [45].

Upon reaching very high volume fractions ($\phi \sim 0.5 - 0.6$) the viscosity starts to be affected by crowding phenomena and eventually it is seen to diverge in correspondence to the volume fraction for random close packing of the particles [37]. The volume fraction where the crowding regime begins is however a function of the applied shear rate [46]. In general the crowded regime is significantly shifted to higher volume fractions upon increasing the applied shear rate [46].

## 5.3 The viscosity of aggregating suspensions: coarse-graining and the effective particle model

Our modeling strategy is based on treating the clusters as effective particles whose hydrodynamic interactions are responsible for the observed viscosity of the aggregation suspension. We thus make the following assumptions:

1) the whole system is treated as a bimodal population with one peak due to the primary colloids (thus the background solvent is a suspension of primary colloids) and one due to clusters with an average radius of gyration, $R_g$;

2) the clusters are treated as hard-spheres of radius equal to their average radius of gyration, $R_g$

3) the number of clusters increases with time following the increase in the conversion from particle to clusters, $\chi$.

These assumptions are motivated by the experimental characterization reported in the previous chapter.

Based on this, we can define the volume fraction occupied by the clusters, or effective volume fraction, as

$$\phi_{\text{eff}} \equiv \frac{N_c}{V} \frac{4\pi R_g^3}{3} \tag{5.4}$$

where $N_c$ is the total number of clusters in the system and $V$ is the total volume of the system. Let us define the conversion of primary colloids to clusters as

$$\chi \equiv \frac{\phi - \phi_p}{\phi} = \frac{V_T - V_{pp,T}}{V_T} = \frac{N_{pcl,T} \frac{4}{3}\pi a^3}{N \frac{4}{3}\pi a^3} = \frac{N_{pcl,T}}{N} \tag{5.5}$$

where $V_T$ denotes the total volume occupied by colloid particles in the system, $V_{pp,T}$ the total volume occupied by primary particles not belonging to clusters, $N_{pcl,T}$ the total number of primary particles which form clusters, and $N$ the total number of colloid particles in the system. In the experimental system under study $\phi_p$, $V_{pp,T}$, and $N_{pcl,T}$ are time-dependent quantities, and so is $\chi$, obviously, due to the time evolution of aggregation. Let us also recall the fractal scaling law which relates the number of particles per cluster, $N_{pcl}$, to the radius of gyration of the cluster and to the fractal dimension via [40]

$$N_{pcl} = k_f \left( \frac{R_g}{a} \right)^{d_f} \tag{5.6}$$

where $k_f$ is the prefactor of the fractal scaling law. It thus follows that the effective volume fraction of an aggregating colloidal suspension as a function of the shearing time is given by

$$\phi_{eff}(t) = \chi(t)\phi k_f^{-1} \left( \frac{R_g(t)}{a} \right)^{3-d_f(t)} \tag{5.7}$$

We are now in the position of formulating the scaling behaviour for the suspension viscosity, which reduces to replacing the colloid volume fraction $\phi$ with the effective volume fraction. This leads to

$$\eta = \eta_0 \exp(5\phi_{eff}/2) = \eta_0 \exp\left( [\eta]\chi\phi k_f^{-1} \left( R_g/a \right)^{3-d_f} \right) \tag{5.8}$$

Since in our anti-thixotropic (rheopectic) system the viscosity evolves with time, according to our assumptions, its time dependence will thus in general be given by

$$\eta(t) = \eta_0 \exp([\eta]\phi_{eff}(t)) = \eta_0 \exp\left( [\eta]\chi\phi k_f^{-1} \left( R_g(t)/a \right)^{3-d_f(t)} \right) \tag{5.9}$$

After the stationary state for the cluster morphology evolution is attained, the only time dependent parameter left is the conversion, $\chi(t)$, which thus dictates the time-dependence of viscosity at stationary-state.

In this model the following phenomenological parameters need be estimated experimentally: $\chi$, $R_g$, $d_f$. In general, as we have seen, the latter quantities all depend on time for anti-thixotropic colloidal suspensions [21]. The fractal scaling prefactor is estimated for a given value of fractal dimension by using the empirical correlation $k_f = 4.46 d_f^{-2.08}$ [47]. This correlation has been extracted as the fit to different sets of experimental and simulation data in [47] and is valid also at the high fractal dimensions of interest here. Eq. (5.9) gives a quantitative description of the aggregating suspension viscosity as a function of the mesoscopic structure (cluster) morphology and its time evolution. To make the theory predictive, one should employ the microscopic theory of shear-induced aggregation proposed in Chapter 2 together with an adequate breakup kernel within the evolution (population balance) equation. This is however quite a challenging numerical task and is left for future work.

A final remark regards the solvent viscosity $\eta_0$ which appears in Eqs. (5.8)-(5.9). As we have shown in the previous chapter, the conversion of primary particles to clusters never exceeds $\chi = 0.4$ even for the longest shearing times considered. It follows that the solvent where the clusters are embedded always contains a significant amount of primary particles and its viscosity is thus different from the viscosity of the pure solvent. In order to account for this aspect we set, in a good approximation, $\eta_0 = \eta_{t=0}$, where $\eta_{t=0}$ is the viscosity of the un-sheared suspension. We checked that this approximation does not lead to significant errors.

In the next section we are going to test the predictions from the model, Eq. (5.9), and the validity of the scaling assumptions, in comparison with the experimental data presented in the Chapter 4.

## *5.4 Comparison between experiments and model predictions*

In Figure 5-1 the comparison between the high-shear ($\dot{\gamma} = 1700\,\text{s}^{-1}$) viscosity and the predictions using Eq. (5.9) with $[\eta] = 2.6$ is shown. Note that the experimental shear viscosity has been plotted against the effective volume fraction which has been calculated for each shearing time using Eq. (5.7) and the experimentally determined values of $\chi$, $R_g$, and $d_f$. Hence the time evolution of aggregation causes the effective volume fraction occupied by clusters to evolve with time, and we can assess this with the experimental data at hand. The quantitative agreement with the model is striking and confirms that the simplifying assumptions used are physically motivated. Experimentally it was not possible to reach higher effective volume fractions, which means longer shearing times, since at a certain point, upon stopping the flow, the system appears solid-like (an aspect that will be discussed in depth in the next Chapter) thus making sampling and filtration rather difficult. However, the experimental data agree well with the high-shear viscosity model up to $\phi_{\text{eff}} \sim 0.6$. With monodisperse spheres in the low-shear limit we would observe the viscosity to diverge at packing fractions about 0.64. Here, however, we are dealing with the viscosity at high shear rate where the crowding regime and the divergence of viscosity are known to be shifted to higher volume fractions (about 0.7 with monodisperse spheres) so that also the purely hydrodynamic regime where $\eta = \eta_0 \exp([\eta]\phi_{eff})$ is supposed to apply, extends up to higher volume fractions. Furthermore, we are dealing with a population of spheroids far from being monodisperse, and it is well known that polydisperse spheroids reach random close packing at significantly higher volume fractions than monodisperse spheres [49]. These considerations (high shear rate and polydispersity) can help to explain why the experimental cluster suspension seems to be unaffected by crowding effects (and thus amenable to be described by purely hydrodynamic effects) up to such large volume fractions.

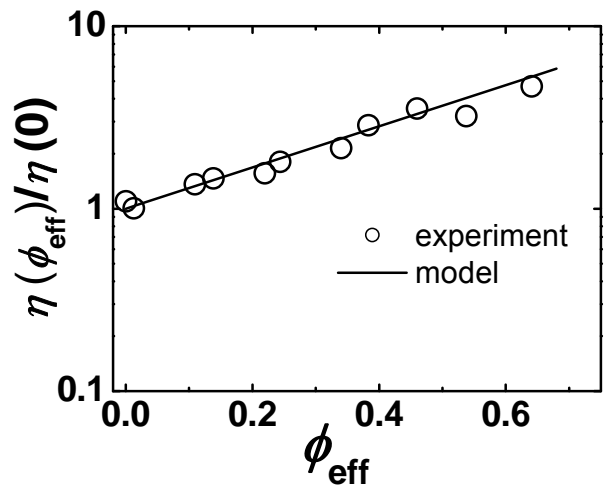

**Figure 5-1: Comparison between the experimentally measured values of high shear viscosity at $\dot{\gamma} = 1700\,\text{s}^{-1}$ for the same conditions of Figure 4-1 and the theoretical relation Eq. (5.9). For each experimental point the effective volume fraction has been calculated using Eq. (5.7).**

In order to better understand what makes the effective hard-sphere treatment work so well for a suspension of clusters, we have conducted optical microscopy (with a Nikon, TMS instrument) to gain more direct insights into the morphology of clusters. An optical micrograph showing typical clusters (in the shearing time regime where both $R_g$ and $d_f$ have reached the stationary plateau) is presented in Figure 5-2.

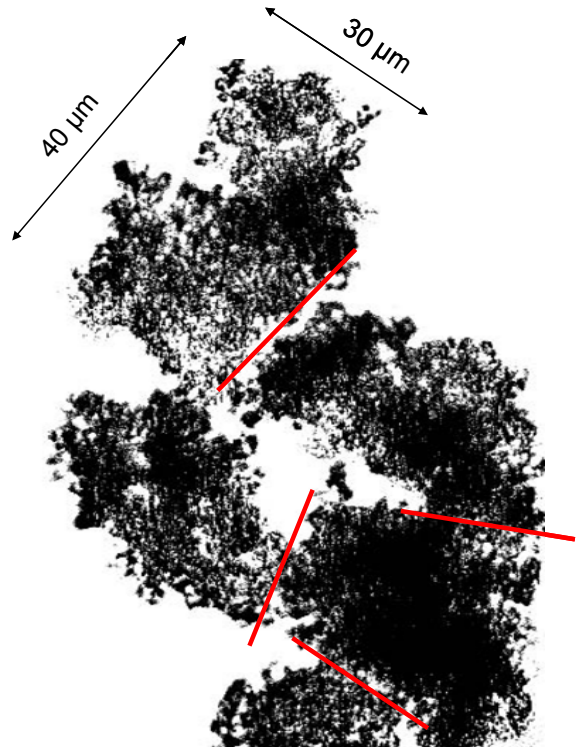

**Figure 5-2: Optical microscopy showing typical clusters present in the system at shearing times >3000 s. Red lines mark the connections between clusters due to weak reversible bonding.**

It is seen that the size of the clusters is in good agreement with the light scattering measurement of the gyration radius ($R_g \approx 35 \mu m$) and the structure is fully compatible with dense fractals characterized by fairly high fractal dimension ($d_f \approx 2.7 - 2.8$ as from light scattering). The fact that the clusters appear interconnected in the micrograph is due, as we could observe, to reversible, perhaps secondary-minimum bonds between the clusters. Hence, thanks to the evidence from microscopy, we can conclude that the clusters are indeed rather compact spheroids with a ratio between the major axes $0.5 \lesssim a/b \lesssim 2$ corresponding to an intrinsic viscosity $2.5 \lesssim [\eta] \lesssim 2.9$ [45]. Varying $[\eta]$ within this range in Eq. (5.9) does not alter the comparison in Figure 5-1 appreciably.

Finally, in Figure 5-3, we report the same results for the high shear viscosity and the corresponding theoretical prediction, together with two measurements of the zero-shear viscosity. Unfortunately, we could not collect more data of the zero-shear viscosity, especially in the lower effective volume fraction regime where viscosity is comparatively lower, because our instrument becomes less sensitive at low values of stress. We can see however, that the zero-shear viscosity experimental points lie much higher than the data obtained at high shear rate. This is expected to be so since at very low shear rate the crowding effects start to dominate the system at sufficiently high volume fractions (in this case: effective volume fractions) and determines a rapid increase of viscosity, eventually making it diverge at the random close packing [48]. To find a confirmation, we have plotted predictions from an improved effective medium differential viscosity theory recently proposed by Mendoza and Santamaria-Holek [36], which is valid at arbitrary volume fractions. Despite the scarcity of experimental data, we can see that they are clearly compatible with the theoretical curve of Mendoza and Santamaria-Holek [48] which reads

$$\eta(\phi_{eff}, \phi_c) = \eta_0 \left( 1 - \phi_{eff} \left( 1 - \left( \frac{1-\phi_c}{\phi_c} \right) \phi_{eff} \right)^{-1} \right)^{-5/2} \quad (5.10)$$

where the only free parameter is the close packing fraction $\phi_c$, chosen here to be $\phi_c = 0.8$. This is a rather high packing fraction for random packings of spheres but similar values have been measured for random packings of polydisperse spheroids [49], to which our aggregates look more similar (see Figure 5-2).

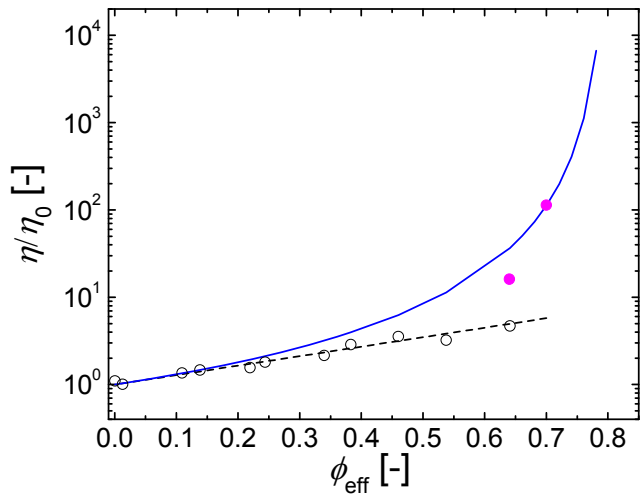

**Figure 5-3: comparison between the measured sheared viscosity and differential effective medium theories rescaled on the cluster level. Filled symbols: zero-shear viscosity ($\dot{\gamma} = 0.0176 \text{ s}^{-1}$). Open symbols: high-shear viscosity ($\dot{\gamma} = 1700 \text{ s}^{-1}$). Solid line: zero-shear differential effective medium theory by Mendoza and Santamaria-Holek [36] accounting for crowding. Dashed line: Eq. (5.9)**.

## 5.5 Understanding and modelling the anti-thixotropy of colloidal suspensions

We have now reached a certain degree of knowledge about the relationship between shear-induced aggregation and clustering and the macroscopic response of the system in terms of its viscosity. We now raise the question if it is possible to use this knowledge to understand and model the anti-thixotropic rheological behaviour of aggregating colloidal suspensions. In particular we are interested in the sharp rise of the viscosity systematically observed with charge-stabilized colloids (see e.g. the rheological curves of Chapter 3).

In principle, Eq. (5.9) provides the missing link between the microscopic transformations occurring in the aggregating suspension, and the macroscopic suspension viscosity determined by viscous dissipation of the clusters in motion under shear.

Since both the clusters average radius of gyration and fractal dimension reach a plateau as a function of time rather rapidly, because the number of clusters during the transient is too small to affect the viscosity, we can rewrite Eq. (5.9) by considering $R_g = R_g^\infty = const$ and $d_f = d_f^\infty = const$, where $R_g^\infty$ and $d_f^\infty$ are the plateau values

$$\eta(t) = \eta_0 \exp([\eta]\phi_{\text{eff}}(t)) = \eta_0 \exp\left([\eta]\chi\phi k_f^{-1}\left(R_g^\infty/a\right)^{3-d_f^\infty}\right) \quad (5.11)$$

Hence the time dependence of the viscosity is uniquely given by and coincides with the time dependence of the conversion from primary colloids to aggregates. In the absence of a predictive theory for the time evolution of the conversion, we can use an empirical fit to the experimental data in Figure 4-5. We note that the viscosity is not significantly affected for $\chi \lesssim 0.1$: we have seen that the clusters are composed by a huge number of primary colloids so that their number is extremely small at low values of conversion, and thus unable to produce macroscopic effects. Let us recall the presence of two distinct regimes in the time evolution of conversion in Figure 4-5, a weakly increasing one (before the characteristic time of aggregation is reached) and the very sharp one revealing self-accelerated kinetics. We found that the self-accelerated regime in Figure 4-5 can be fitted by the following empirical expression

$$\chi(t) = c_1 |t - t_d|^{-c_2} \qquad (5.12)$$

with parameters $c_1 = 1 \times 10^{10}$, $c_2 = 3.2$, and $t_d = 9300$. The last parameter might be interpreted as the value of shearing time at which the self-accelerated kinetics becomes infinitely fast. In Figure 5-4 we have shown the comparison between the experimental curve (at $\phi = 0.21$, $\dot{\gamma} = 1700 \text{ s}^{-1}$, 17mM NaCl) and the curve calculated using Eqs. (5.11)-(5.12).

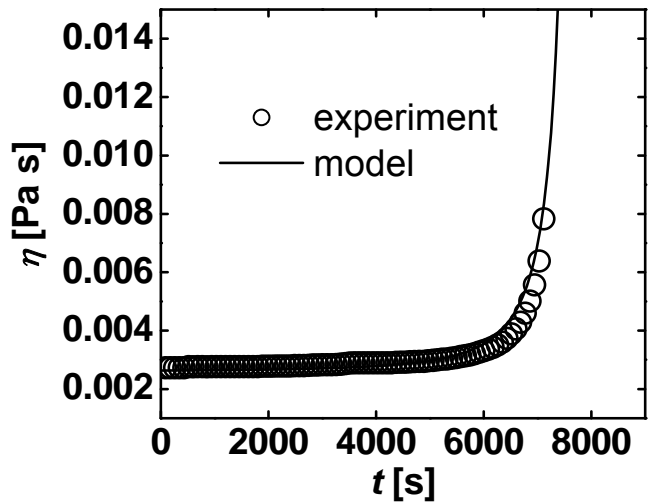

**Figure 5-4: Comparison between the measured viscosity as a function of the time of shearing (circles) at (at $\phi = 0.21$, $\dot{\gamma} = 1700 \text{ s}^{-1}$, 17mM NaCl ) and the phenomenological model predictions (line), Eq. (5.11), calculated using Eq. (5.12).**

From Figure 5-4 we can see that indeed the phenomenological model, which makes use of experimentally determined parameters (but no free parameters), is very successful in accurately reproducing the experimental data.

In particular, this comparison demonstrates that the phenomenological assumptions made are justified. From it we learn that the sharp rise in the viscosity is a direct consequence of the setting in of self-accelerating shear-induced aggregation kinetics which is manifest in the sharp rise of the conversion of primary particles to clusters. In turn, the self-accelerated kinetics can be explained by means of the theory presented in chapter 2 as due to the reaching of an average activated size of clusters which do not see any barrier for aggregation and where the kinetic constant is a strongly nonlinear increasing function (the exponential of the cube) of the cluster size.

## 5.6 Overview

We have seen that aggregating charge-stabilized (DLVO) suspensions display a typically anti-thixotropic rheological behaviour (viscosity increasing with time under steady-shear) due to the shear-induced aggregation (clustering) process. With the results presented in this chapter we have made substantial progress from this generic statement. We have shown how the clusters formed are directly responsible for the behaviour of the viscosity. As was noted in the previous chapter, after a comparatively short transient, the cluster population reaches a well defined morphology, which remains constant with time due to the dynamic balance between shear-induced aggregation and breakup. Hence, after this short transient, there is a unique key parameter left whose time evolution directly controls the time evolution of the macroscopic response: the conversion from primary colloids to clusters due to shear-induced aggregation. In practice, the increase in time of the conversion determines the increase in the number density of clusters (whose average size and fractal dimension were not seen to change any more after the transient). This implies, evidently, a well-defined (conversion-controlled) time evolution of the effective volume fraction occupied by the clusters (treated as hard-spheres of radius equal to their radius of gyration). Hence the viscosity versus time curve can be mapped onto a viscosity versus effective cluster volume fraction curve which spans a very broad range of cluster volume fractions (from close to zero up to over 0.6). This gave us a unique opportunity to test the applicability of hard-sphere theories of suspension viscosity to fractal clusters. The excellent agreement found over all the range of effective volume fractions available demonstrates the validity of the coarse-graining concept: the macroscopic properties are controlled by the hydrodynamic interactions between the weakly-interacting clusters. The latter ones, in this respect, can be effectively modelled as hard spheres (as also suggested by optical microscopy). In particular, the simple differential effective medium approximation (Arrhenius) suffices to provide an excellent description of the high shear viscosity as a function of the effective volume fraction up to values around 0.6. This suggests that the high shear viscosity is unaffected by crowding phenomena in the investigated range.

Due to the technical impossibility to explore higher effective volume fractions here, it is still open the issue of the effective volume fraction at which the high shear viscosity will diverge and whether or not there will be ordering phenomena, as they are known to occur with spheres at high shear (although this seems very unlikely in view of the significant polydispersity and irregular shape details of the clusters) [44,48]. We have also shown some data of zero-shear viscosity at high volume fractions. The data appear compatible with an improved differential effective medium theory recently proposed in the literature, if one sets the random close packing fraction for the clusters at around 0.8. This value, much higher than the well known one (0.64) for monodisperse spheres [2,14], can be explained with the significant polydispersity and irregular shape of the clusters.

Finally, by comparing the phenomenological model based on the Arrhenius approximation rescaled onto the cluster level, and using an empirical fit to the experimental data of conversion as a function of time, it has been possible to quantitatively reproduce the experimental (anti-thixotropic) viscosity versus time curve with good accuracy. This indicates that the sharp increase in the viscosity after the induction time is due to the setting in of self-accelerated shear-induced aggregation kinetics characterized, as discussed in chapter 2, by an activated average size of the clusters.

As we could see, once the effective volume fraction occupied by the clusters becomes significant, the low-shear viscosity begins to differ substantially from the high-shear viscosity values. Also, for large enough values of effective volume fraction, the sample appears solid-like after cessation of flow. Therefore, we are in the presence of a shear-induced liquid-solid transition or shear-induced gelation transition. The features of such transition are investigated in more detail in the next chapter by means of bulk rheology.

# 6. Rheology of aggregating suspensions

## 6.1 Introduction

In the previous chapter we have mainly focused on the *high shear* rheology of aggregating (DLVO) suspensions, where the system, although its viscosity may reach up to several times the viscosity of the stable original suspension (or zero shearing-time viscosity), remains fully fluid under all conditions investigated. This is mainly due to the high shear rate applied, >1000 s$^{-1}$, which keeps the system well fluid and which effectively opposes any tendencies to develop a volume-spanning, stress-bearing structure. However, upon cessation of shear flow ($\dot{\gamma}=0$), and depending, once more, upon the shearing time (i.e. on the position along the viscosity-time curve, Figure 4-1), it is possible to observe a solid-like colloidal gel. Hence, at long enough shearing times (which is equivalent to say: at high enough effective cluster volume fractions), we are in the presence of dynamical arrest upon cessation of flow, with the emergence of well defined elastic, solid-like features. It is worth to anticipate that the key parameter which controls the properties of the gel formed upon the cessation of flow is the effective volume fraction, which evolves with time mainly due to the time evolution of the conversion from primary particles to clusters. We have seen in the previous chapter that the effective volume fraction, a function of the shearing time, is the crucial link between the microscopic level (where shear-induced clustering occurs), and the macroscopic properties.

We conducted shear-sweep rheological tests on the arrested systems after cessation of flow, at different shearing times (thus at different effective volume fractions). These give evidence of the development of a marked shear-thinning behaviour starting from a certain value of shearing time (or effective volume fraction). This behaviour has never been observed at such total colloid volume fractions with such small particles and can only be explained, once more, by recognizing that the macroscopic behaviour is dominated by the clusters (and by the effective volume fraction). Also, there is a window of effective volume fractions where the shear-sweep curves exhibit strong and very steep shear thickening. Our analysis of the rheology curves suggests that such reenetrant shear-thickening shares all features that are typical of dense non-Brownian suspensions. This represents further evidence of the presence of a shear-induced transition from Brownian to non-Brownian rheological behaviour.

## 6.2 Dynamic frequency sweep on sheared suspensions upon flow cessation: gelation transition and the emergence of rigidity

Dynamic frequency sweep curves are reported in Figure 6-1. The tests were immediately performed after stopping the shear flow at different shearing times along the viscosity-time curve, Figure 4-1. Hence, each value of shearing time corresponds to a value of effective volume fraction as established by Eq. (5.7).

The frequency has been given in units of the diffusion time in the dilute limit, $\tau_0 = a^2/D_0$. The bottom curves of elastic and loss moduli correspond to the lowest effective volume fraction considered, $\phi_{eff} \approx 0.17$. At lower values (shorter shearing times) the system appeared completely liquid-like. At $\phi_{eff} \approx 0.17$, despite a rather scattered signal, we observe that $G'$ and $G''$ curves exhibit a similar, power-law like, trend with the frequency and are also close to each other in terms of absolute values. The trend is compatible with a power-law with exponent 0.5 (solid line in Figure 6-1). This kind of response shows interesting similarity to the one observed in physical and chemical gelling systems at percolation where the response is indeed characterized by [50]

$$G'(\omega) \sim G''(\omega) \sim \omega^u \tag{6.1}$$

where the power-law exponent $u$ may be related to the morphology of the network. This suggests that at $\phi_{eff} \approx 0.17$ a system-spanning network may already be present. However, it is most probably a *transient* network where the clusters are interconnected by weak bonds (since as we have seen in the previous chapter the interaction between clusters is weak), thus dynamically forming and breaking. This is consistent with the very low values of storage modulus measured here. This picture is also compatible with our direct and microscopy observations where the clusters, also after dilution, were seen to assemble into mechanically weak super-aggregates that can be disrupted upon agitation. Unfortunately it is not possible to better characterize the nature of the inter-cluster bonding, but all evidence at hand points to a relative weakness of these bonds in comparison with the inertia of the clusters. This is different from the case of the intra-cluster primary-minimum bonds between colloidal particles which impart a high degree of cohesion and mechanical strength to the clusters themselves.

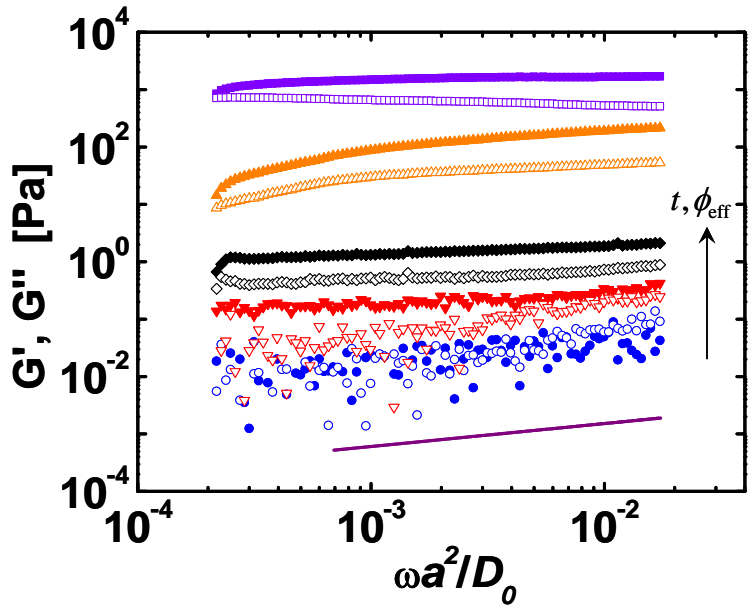

**Figure 6-1: Dynamic frequency sweeps with elastic (storage) modulus $G'$ (filled symbols) and viscous (loss) modulus $G''$ (open symbols) as a function of frequency $\omega$ in units of the diffusion time in the dilute limit, $\tau_0 = a^2/D_0$. The effective volume fraction of the aggregates increases from bottom to top: $\phi_{\text{eff}} \approx 0.17$, $\phi_{\text{eff}} \approx 0.23$, $\phi_{\text{eff}} \approx 0.31$, $\phi_{\text{eff}} \approx 0.56$, and $\phi_{\text{eff}} \approx 0.69$. The solid line shows the typical scaling at percolation $G'(\omega) \sim G''(\omega) \sim \omega^{0.5}$.**

At $\phi_{\text{eff}} \approx 0.17$ and $\phi_{\text{eff}} \approx 0.23$ the system does not appear as rigid yet, which is manifest in the low values of the storage modulus and in the fact that at the higher frequencies the loss modulus is still somewhat larger than the storage modulus. Upon further increasing the effective volume fraction (and the shearing time along our viscosity-time curve, Figure 4-1) there is evidently an increase in the connectivity of the cluster-network which determines an increase in the rigidity and hence in $G'$. At $\phi_{\text{eff}} \approx 0.31$, $G' > G''$ for all frequencies, which is a first indication of the emergence of solid-like response although the two moduli still exhibit very comparable values, especially at the higher frequencies. Finally, at $\phi_{\text{eff}} \approx 0.56$, a fully developed elastic, solid-like behaviour becomes evident in the $G'$ being significantly larger than the $G''$, over all frequencies, including the higher frequency regime. Thus we conclude that the appearance of a fully solid-like response occurs in a range of effective volume fractions roughly identifiable with $0.4 \lesssim \phi_{\text{eff}} \lesssim 0.5$.

Finally at the highest effective volume fraction considered, $\phi_{\text{eff}} \approx 0.69$, the system appears substantially rigid with a storage modulus $\sim 10^3$ Pa which is practically constant with the frequency over the whole range.

It is worth noticing that a fully solid-like response is attained at much lower frequencies, hence at relaxation times much larger than the characteristic relaxation time of the primary colloid particle (in the dilute limit). This observation may support the interpretation that the elastic properties in the arrested aggregated system, just as in the case of the viscosity for the fluid state at high shear, arise at the cluster level rather than at the primary colloid level.

Hence we have seen how dramatically the response of the system can change upon varying the shear rate: the same system (in terms of effective volume fraction of the clusters and shearing time) may appear solid-like at flow cessation ($\dot{\gamma} = 0$) and completely fluid at high shear rate, $\dot{\gamma} \sim 10^3 \text{ s}^{-1}$. In order to draw more insights into this phenomenon it is necessary to assess the effect on the system of varying the applied shear rate. This is done in the next section.

## 6.3 Shear-rate sweep on sheared suspensions after flow cessation: shear thinning and shear thickening

We have conducted shear-rate sweep tests in the shear-rate range $0.0176 \leqslant \dot{\gamma} \leqslant 1569$. The tests were done immediately after stopping the shear flow, at different shearing times (corresponding to different effective cluster volume fractions) along our reference viscosity-time curve in Figure 4-1. The rheological shear-sweep curves are reported in Figure 6-2. The first curve from the bottom corresponds to the lowest effective volume fraction and exhibits a strong shear thinning behaviour terminating with a high-shear plateau at much lower viscosity values than in the zero-shear limit. The observation of shear-thinning over more than two orders of magnitude in viscosity is unprecedented with Brownian particles at colloid fractions $\phi \lesssim 0.3$ [37]. Shear-thinning is due to a shear-induced structural distortion where the particles start to order under the effect of shear, thus loosing the homogeneously disordered structure of the orginal unsheared suspension in favour of a more ordered structure [37]. Hence the tendency of a suspension to develop shear-thinning goes with the cube of the particle size and increases with volume fraction [10,37]. The primary colloid particles used in our experiments are too small to give rise to shear-thinning at volume fraction 0.21. Hence it is evident that shear-thinning is caused directly by the large clusters formed by irreversible shear-induced aggregation which are the relevant coarse-grained building blocks controlling the macroscopic behaviour. In fact, due to the large clusters, the suspension effectively behaves like non-Brownian suspensions for which shear-thinning is observed also at comparatively lower volume fractions (down to values around 0.10) [51]. Indeed, if we consider the effective volume fraction of the bottom curve in Figure 6-2, this is $\phi_{eff} \approx 0.17$, well within the range where shear-thinning is observed with non-Brownian suspensions [51].

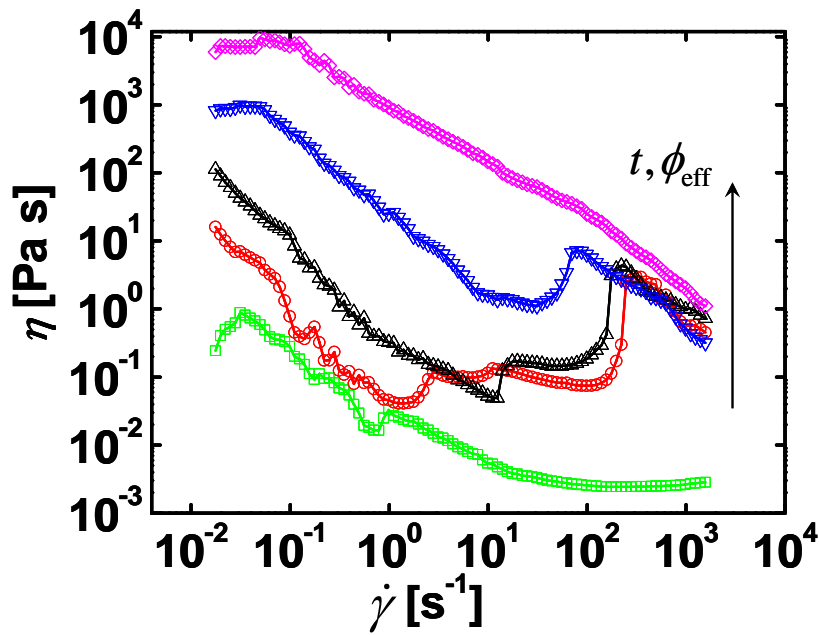

**Figure 6-2: Shear-rate sweeps on the system of Figure 4-1 after cessation of flow at different shearing times. The shearing time and the corresponding effective volume fraction increase from bottom to top. For the first two curves from the bottom: $\phi_{eff} \approx 0.17$, $\phi_{eff} \approx 0.6$, respectively. The effective volume fraction could not be measured for the other curves but keeps on increasing from bottom to top in the range $\phi_{eff} \gtrsim 0.6$.**

This, again, points strongly at the fundamental role that shear-induced aggregation plays in coarsening the system characteristic length-scale, thus making a suspension of Brownian particles effectively behave similar to a non-Brownian suspension.

The next curve from bottom to top in Figure 6-2 refers to the system after flow cessation at a shearing time which corresponds to $\phi_{\text{eff}} \approx 0.6$. This curve appears significantly different from the previous one, in that it has a much higher zero-shear viscosity, which is expected given the much higher effective volume fraction, and it exhibits a strongly marked shear-thickening behaviour at shear-rates $\dot{\gamma} \sim 100\,\text{s}^{-1}$ with a shear-thickening range covering almost two orders of magnitude in viscosity (stress). Further, the onset of shear-thickening appears to be very sharp (discontinuous shear thickening), which suggests, as a microscopic origin of the phenomenon, strong packing-fraction effects rather than the hydrodynamic effects (associated with continuous shear-thickening and where attractions may also play a role). Finally we observe that the experimental data are rather scattered and irregular in the lower shear rate parte of the curve, up to $\dot{\gamma} \sim 10\,\text{s}$. We attribute this to flow inhomogeneity or shear banding, which migt occur below a threshold shear rate value. In the presence of shear banding there might be layers of fluid which are sheared at very high shear rates while others remain un-sheared. A threshold shear rate value for shear banding also of order $\dot{\gamma} \sim 10\,\text{s}$ was recently found experimentally for aggregating colloidal systems by Møller et al. [52].

Upon further increasing the effective volume fraction (as a result of increasing the shearing time at flow cessation) we observe, 1) a reduction in the shear-thickening range, and 2) a shift of the lower boundary to lower stress values, both of which are typical features of shear-thickening in non-Brownian suspensions (see e.g. [51]). Finally, at the highest shearing time (effective volume fraction) considered, there is no signature of shear thickening and the shear sweep curve exhibits shear-thinning over the entire shear-rate range, with no high-shear plateau. The disappearance of shear-thickening upon reaching a very high volume fraction is observed in dense non-Brownian suspensions [53]. In a very recent study it was seen that upon increasing the volume fraction the system develops a yield stress which grows with the volume fraction until it reaches the upper stress boundary for shear-thickening thus effectively hiding the shear-thickening phenomenon which is no longer visible [53]. The generality of this phenomenon was proved by using, besides the packing fraction, also the attraction strength or the external field strength (with particles responsive to magnetic or electric fields) to tune the interplay between yield stress and shear-thickening. It was seen that the shear-thickening may disappear already when the yield stress is lower than the lower stress boundary for shear-thickening. Also in the present case we observe (Figure 6-3), that at the high effective volume fractions there is indeed a substantial yield stress. One should notice that the yield-stress is somewhat masked by an artifact that is due to the non-steady character of the experiment. Below the yield stress the system apparently behaves like a Newtonian fluid. This phenomenon is well known in the literature and has been the object of controversy as to whether true yield stress fluids exist or not [54]. Recently, evidence has been brought by Bonn and co-workers supporting the idea that the low-stress Newtonian behaviour is observed when the measurement time is shorter than the stress-induced aging time of the system [54]. For long enough measurements it was found indeed that the Newtonian viscosity at low stress increases indefinitely with time, thus suggesting that the steady-state of the system is solid and the yield stress does indeed exist. This behaviour is typical of densely packed jammed solids and the fact that we observe it at the higher $\phi_{eff}$ indicates that our system, originally a dilute suspension of colloidal particles, effectively behaves as a dense suspension of large non-Brownian particles.

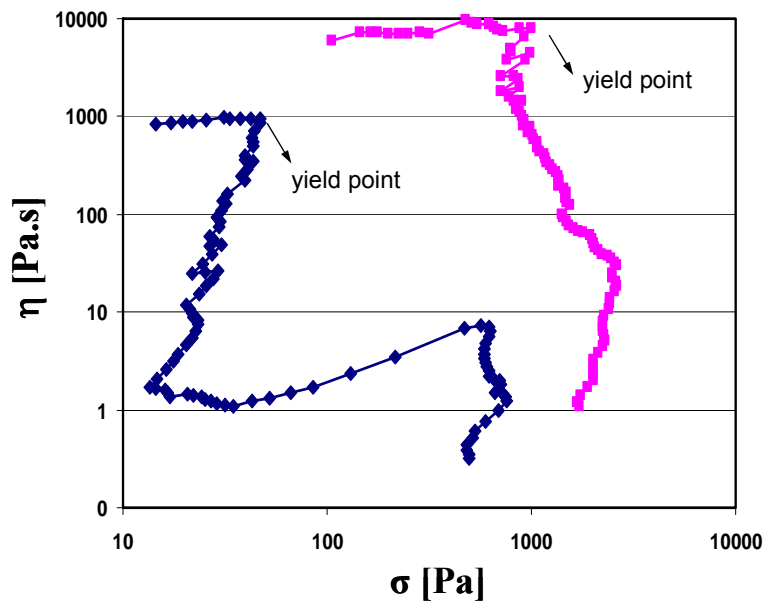

**Figure 6-3: Elimination of shear thickening by increasing packing density. Viscosity versus stress curves for the two highest effective volume fractions considered in Figure 6-2. The lower-lying curve corresponds to lower effective volume fraction (lower shearing time).**

These observations are fully consistent with those of Jaeger and co-workers [53] for the existence of an upper stress threshold at high packing fraction beyond which by the yield stress (induced by the packing) overwhelms the shear-thickening. They also extend the picture to colloidal suspensions aggregating under shear.

We complete our analysis by examining the shear-thickening transition in terms of stress-rate of shear curves (Figure 6-4). It is evident that the high shear-rate part of the shear-thinning regime nicely follows the phenomenological Herschel-Bulkley form commonly used to describe the shear-thinning behaviour [54]

$$\sigma_{HB}(\dot{\gamma}) = \sigma_y + \sigma_1 \dot{\gamma}^{1/2} \qquad (6.1)$$

where $\sigma_y$ is the yield stress, with the characteristic square root dependence upon the shear-rate (continuous line in Figure 6-4). Then the shear-thickening regime sets in with a sharp upturn in the stress and follows the form proposed by Jaeger and coworkers [53]

$$\sigma(\dot{\gamma}) = \sigma_{HB}(\dot{\gamma}) + \sigma_2 \dot{\gamma}^{1/\varepsilon} \qquad (6.2)$$

(dashed line in Figure 6-3). With $\varepsilon \to 0$ the shear-thickening becomes discontinuous.

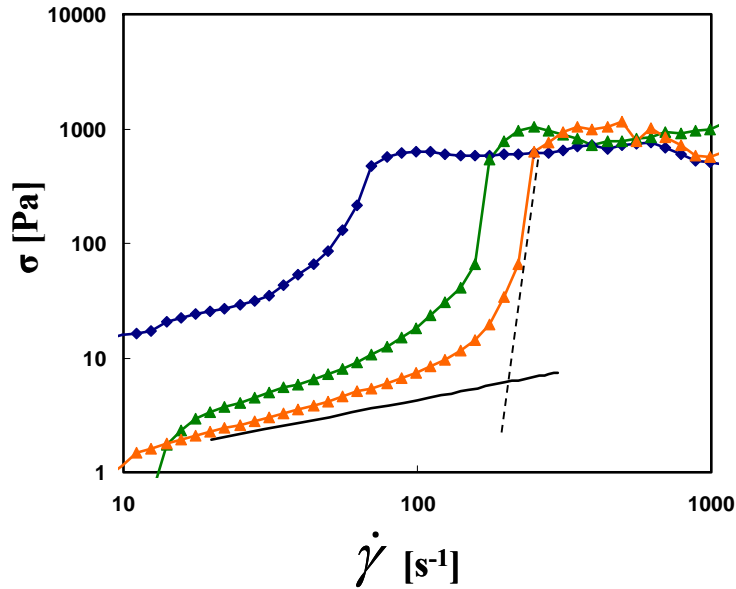

**Figure 6-4: Stress vs shear-rate plot broken up into shear-thinning and shear-thickening components. Solid line: fit to the Herschel-Bulkley law (Eq. 6.1). Dashed line: the term $\sim \dot{\gamma}^{1/\varepsilon}$ as from Eq. (6.2) for the shear-thickening regime.**

The data in Figure 6-4 are compatible with very small values of $\varepsilon$ thus confirming the quasi-discontinuous shear-thickening scenario for our shear-aggregated suspensions. This is in turn a strong indication that the microscopic origin of shear-thickening in this system is more related to packing density effects (obviously on the cluster level) rather than to hydrodynamic effects [53].

## 6.4 Overview

In this Chapter we have investigated the rheological response of colloidal suspensions aggregating under shear, after cessation of shear flow. As we have seen in the previous chapter, stopping the shear flow at different shearing times leaves the suspension in different states of aggregation, in particular with respect to the effective volume fraction occupied by the clusters. Immediately after switching off the shearing we have subsequently performed dynamic frequency and shear-rate sweep tests in order to assess the different rheological responses corresponding to different degrees of aggregation in the system.

From the dynamic frequency sweep test we unveiled the existence of a discontinuous transition from liquid-like to solid-like of the suspension upon increasing the shearing time (hence the effective volume fraction) at which cessation of flow occurs. Rigidity emerges gradually from the system and an initial transient percolation scenario at $\phi_{\text{eff}} \sim 0.2$ is followed by increasingly solid-like features upon increasing the effective volume fraction until a fully solid-like behaviour is observed around $\phi_{\text{eff}} \sim 0.5$ [50].

However, the same system, with respect to the effective cluster volume fraction, appears completely fluid at high shear rate ($\dot{\gamma} \sim 10^3 \, \text{s}^{-1}$) and definitely solid-like upon cessation of flow ($\dot{\gamma} = 0$). This obviously reminds of a strongly non-Newtonian behaviour, which is unprecedented with Brownian particles at the total constant colloid fraction used here [37]. To elucidate this finding, we have carried out shear-rate sweeps immediately after cessation of flow at varying shearing time (effective volume fraction) along the viscosity/time curve in Figure 4-1. At the lowest effective volume fraction studied we found a strong shear-thinning followed by a Newtonian plateau. Upon increasing the volume fraction the shear-thinning regime is found to be followed by a sharp shear-thickening regime. At the highest effective volume fraction studied (well above $\phi_{\text{eff}} = 0.6$) the shear-thickening is no longer visible because the yield stress (which the system develops at sufficiently high packing fractions) becomes comparable to the upper stress boundary for shear-thickening, thus effectively hiding the latter. This mechanism has been very recently reported in systematic studies of shear-thickening and is independent of the microscopic origin of the yield stress [53]. Furthermore, from our analysis it emerges that the shear-thickening in our system is quasi-discontinuous, which strongly suggests that its microscopic origin lies in the packing fraction effects rather than in hydrodynamic or attraction effects [53].

Hence, we gathered plenty of evidence that Brownian suspensions aggregating under shear can develop, depending on the extent of aggregation, rheological responses and behaviours which are very far from those of non-aggregating Brownian systems and very much similar to those of non-Brownian suspensions. The rich phenomenology includes: shear-thinning, shear-thickening and yield stress. All these behaviours could only be explained by considering that the property-controlling entities are the large clusters formed under shear and not the primary colloids. We have demonstrated that the large clusters, at a coarse-grained level, effectively behave like non-Brownian particles thus imparting to the suspension the typical rheological properties of non-Brownian suspensions. This consideration explains the huge diversity of behaviours that are observed in a single colloidal suspension of constant total volume fraction by just varying the degree of aggregation.

# 7. The macroscopic linear response of amorphous solids: affine theory

## 7.1 Introduction

In this chapter we present general results derived by us for the macroscopic elastic constants of amorphous solids, with particular focus on the shear modulus. These results will form the basis for the modeling of the elasticity of arrested colloidal suspensions in the next chapter.

While the structure, elasticity, and lattice dynamics of condensed matter with long-range order (thanks to the intrinsic symmetry of crystalline structures) are fairly well understood [55], the same cannot be said of amorphous solids. Recent advances include the unveiling of connections between disordered solids made of thermal particles (glasses) and granular packings, so that the puzzling properties found in both these classes of materials can be investigated by means of unifying concepts. Well-known examples are the excess of low-frequency modes (the so-called boson peak in the vibrational spectrum) [56], and the inhomogeneity of the elastic response [56-59]: features that have been observed in atomic (and molecular) glasses as well as in granular systems. These phenomena, as recent theoretical studies have proposed, may find their origin in the weak connectivity of amorphous solids [60] as well as in their lack of order [58]. Regarding the former aspect, recently it became clear that coordination plays a fundamental role in determining the mechanical properties of marginally-rigid solids when only central forces are at play. On the other hand, in the case of strongly connected structures or other dense systems where the bonds between building blocks can support bending moments, nonaffinity is often a very small correction to the affine part, thus the affine approximation works relatively well [61,62]. Some technologically important systems appear to belong to this class, e.g. dense networks of semi-flexible polymers, strong attractive colloidal glasses and covalent glasses (e.g., silicon glass) [60,63]. In the present work, we derive explicit expressions for the macroscopic shear modulus of deeply-quenched, arrested states of like particles, using Alexander's Cauchy-Born approach. The validity and application of the results are discussed.

## 7.2 Continuum theory of shear elasticity in random solid states

In Ref. [18], S. Alexander formulated the systematic Cauchy-Born approach for amorphous solids, based on which the Helmholtz free energy at $T=0$ (thus coinciding with the internal energy) can be expanded around a *rigid*, stressed, reference configuration where the set of particle positions is denoted by $\{R\}$. In such low-temperature reference state, as a result of quenching (solidification), particles are *labelled*, in the sense that they occupy well defined and fixed positions on a (disordered) lattice, the set of which represents just one out of $N!$ possible permutations ($N$ being the total number of particles). In other words, permutation symmetry (which is active in the liquid precursor) is broken in the quenching process [18]. Because of this, as opposed to equilibrium fluids, the disorder average for amorphous solids is of non-trivial definition. To avoid this problem, in Alexander's version of Cauchy-Born theory, the expansion (along with the disorder average) is carried out in terms of the *relative deviations* between particles. As shown in [18], this leads to the continuum limit and provides the only systematic application of Cauchy-Born theory to disordered solids. In the following we apply this approach to a generic dynamically arrested (glassy) state composed of spherical particles mutually interacting via two-body central and three-body angular (bond-bending) interactions. This may be a suitable model of well-bonded glassy systems such as atomic (covalent or metallic) glasses or attractive colloidal glasses.

Retaining terms up to second order, and including a three-body angular interaction term, the expansion reads

$$\delta F \equiv F(\{r\}) - F(\{R\}) \simeq \sum_{<ij>} \left.\frac{\partial F}{\partial r_{ij}}\right|_{\{R\}} \delta r_{ij} + \frac{1}{2}\sum_{<ij>} \left.\frac{\partial^2 F}{\partial r_{ij}^2}\right|_{\{R\}} (\delta r_{ij})^2 + \frac{1}{2}\sum_{<ijk>} \left.\frac{\partial^2 F}{\partial \Theta_{ijk}^2}\right|_{\{R\}} (\delta\Theta_{ijk})^2$$

(7.1)

In the first two terms on the r.h.s., the summation runs over all $N_c$ pairs of pair-interacting particles (i.e., over all bonds), and the derivatives are evaluated at the equilibrium distance in the reference rigid state $R_{ij} \equiv |\mathbf{R}_{ij}|$. In the last term (i.e. the angular interaction or bond-bending term) the summation is over pairs of bonds $[ij]$ and $[ik]$, $i \neq j \neq k$, having one common vertex. In our analysis, we will consider the two-body (central) terms and the three-body (bond-bending) terms in Eq. (7.1) separately, starting from the former case.

Expanding in the *relative* distance deviations allows one to define a microscopic displacement field $\mathbf{u}_{ij}$

$$\delta r_{ij} = u_{ij}^{\parallel} + [(u_{ij}^{\perp})^2 / 2R_{ij}] + \mathcal{O}(r_{ij}^3) \tag{7.2}$$

which has a component in the direction of $\mathbf{R}_{ij}$, i.e. $u_{ij}^{\parallel} \equiv (\delta\mathbf{R}_i - \delta\mathbf{R}_j) \cdot \hat{\mathbf{R}}_{ij}$, and an orthogonal component, i.e. $u_{ij}^{\perp} \equiv (\delta\mathbf{R}_i - \delta\mathbf{R}_j)^{\perp}$. In the absence of external forces, substituting Eq. (1.2) into the central-interaction terms in Eq. (7.1), gives

$$\delta F^{(C)} \simeq \sum_{<ij>} \left.\frac{\partial F}{\partial r_{ij}}\right|_{\{R\}} \frac{[(\delta\mathbf{R}_i - \delta\mathbf{R}_j)^{\perp}]^2}{2R_{ij}} + \frac{1}{2}\sum_{<ij>} \left.\frac{\partial^2 F}{\partial r_{ij}^2}\right|_{\{R\}} [(\delta\mathbf{R}_i - \delta\mathbf{R}_j) \cdot \hat{\mathbf{R}}_{ij}]^2 \tag{7.3}$$

The first-derivative terms correspond to the bond-tension or stress terms associated with the *initial* or quenched stresses (which are, generally, dependent upon the aging history). These make an important contribution to the rigidity of weakly-connected (undercoordinated) materials and actually ensure the existence of a rigid reference state around which one can expand [18]. For central interactions, the second-derivative can be written as the bond stiffness $\kappa_{\parallel} \equiv \left.\partial^2 F/\partial r_{ij}^2\right|_{\{R\}}$, also known as Born-Huang term. In the case of a glass, these terms are to be evaluated in the stressed state and in general may differ from the corresponding terms in crystals at true (thermodynamic) equilibrium (which are evaluated exactly at the position of the minimum of the pair potential). Since, however, the interparticle distances in the stressed state are generally not a priori known, a useful and widely used approximation is to evaluate the second derivative of the pair potential at the distance corresponding to the minimum of the potential well, just as for crystals. Though this approximation is usually not justified for amorphous solids, in the case of deep, short-ranged attractive potential wells, the probability of finding the particle is by far the highest near the minimum of the well. As a consequence, the latter is expected to be a good estimate of the average interparticle distance in these systems also in the stressed state, as shown in [64] and in the following chapter. Further, since "initial stresses in glasses must be internal stresses with a zero average" so that "the internal stresses cannot contribute to the overall macroscopic shear rigidity", it follows that "the macroscopic elastic moduli which one measures must be proper Born-Huang shear moduli" [18], as confirmed also within simulation studies [65]. Therefore, the stress terms in the expansion, Eq. (7.3), can be neglected in good approximation [18,64], and one can write

$$\delta F^{(C)} \simeq \frac{1}{2}\kappa_{\parallel}\sum_{<ij>}\overline{[(\delta\mathbf{R}_i - \delta\mathbf{R}_j)\cdot\hat{\mathbf{R}}_{ij}]^2} = \frac{1}{2}\kappa_{\parallel}\sum_{<ij>}\overline{(u_{ij}^{\parallel})^2} \qquad (7.4)$$

where $\overline{...}$ denotes the average over all possible *deviations* (i.e. strain configurations) from the reference state. Introducing a smooth (continuous) displacement field $\mathbf{u}(\mathbf{r})$, to lowest order in the gradient expansion one has

$$\overline{u_{ij}^{\parallel}} \simeq (\mathbf{R}_{ij}\cdot\nabla)\mathbf{u}(\mathbf{r})\cdot\hat{\mathbf{R}}_{ij} = R_{ij}^{-1}R_{ij}^{\alpha}R_{ij}^{\beta}\partial_{\alpha}u_{\beta}, \qquad (7.5)$$

where summation over repeated indices is understood and transposition symmetry is evident. Using this and introducing the affine transformation $\overline{(u_{ij}^{\parallel})^2} \simeq \text{Tr}[(\mathbf{R}_{ij}\otimes\mathbf{R}_{ij})\cdot\mathbf{e}/R_{ij}]^2$ defined by the disorder-averaged linearized symmetric strain tensor $\mathbf{e}\equiv e_{\alpha\beta} = \frac{1}{2}(\partial_{\alpha}u_{\beta} + \partial_{\beta}u_{\alpha})$, with $\overline{u_{ij}^{\parallel}}^2 - \overline{u_{ij}^{\parallel}}^2 \ll \overline{u_{ij}^{\parallel}}^2$, we obtain the continuum limit [18]:

$$\delta F^{(C)} \simeq \frac{1}{2}\sum_{<ij>}\kappa_{\parallel}\overline{(u_{ij}^{\parallel})^2} \simeq \frac{1}{2}\sum_{<ij>}\kappa_{\parallel}\left\{\frac{\text{Tr}[(\mathbf{R}_{ij}\otimes\mathbf{R}_{ij})\cdot\mathbf{e}]}{R_{ij}}\right\}^2 \qquad (7.7)$$

where $\otimes$ denotes the dyadic product. It is easy to find that for an imposed pure shear deformation the above expression reduces to

$$\delta F^{(C)} \simeq \frac{1}{2}\kappa_{\parallel}\sum_{<ij>}4R_{ij}^2\left(\frac{R_{ij}^x}{R_{ij}}\frac{R_{ij}^y}{R_{ij}}\right)^2 e_{xy}^2 \qquad (7.8)$$

For the quenched configuration $\{R\}$, under the assumption that pair (two-body) interactions are much stronger than higher-order multi-body interactions, the summation over pairs of nearest-neighbours can be replaced by the total number of bonds, $N_c$. This implies an average over all possible spatial orientations of the bonds in the reference state $\{R\}$:

$$\delta F^{(C)} \simeq \frac{1}{2}\kappa_{\parallel}N_c\left\langle 4R_{ij}^2(\hat{R}_{ij}^x\hat{R}_{ij}^y)^2 e_{xy}^2\right\rangle_{\Omega} \simeq 2\kappa_{\parallel}N_cR_0^2\left\langle(\hat{R}_{ij}^x\hat{R}_{ij}^y)^2\right\rangle_{\Omega} e_{xy}^2 \qquad (7.9)$$

where $R_{ij}\equiv R_0$ is the average interparticle distance in the reference quenched configuration and $\langle...\rangle_{\Omega}$ denotes the angular average. Introducing the mean coordination $z$, and noting that $N_c/V \equiv \frac{1}{2}zN/V \equiv 3z\phi/\pi R_0^d$, leads to the following form for the free energy density

$$\delta \tilde{F}^{(C)} \simeq 6\pi^{-1}\kappa_{\parallel}z\phi R_0^{2-d}\left\langle (\hat{R}_{ij}^x \hat{R}_{ij}^y)^2 \right\rangle_{\Omega} e_{xy}^2. \tag{7.10}$$

where $d$ is the dimensionality of space. For $d=3$, using spherical coordinates $\hat{\mathbf{R}}_{ij} = (\sin\theta\cos\varphi, \sin\theta\sin\varphi, 1)$, with $R_{ij}^x = \sin\theta\cos\varphi$, $R_{ij}^y = \sin\theta\sin\varphi$ and assuming that the particles have zero degree of spatial correlation, averaging gives

$$\left\langle (\hat{R}_{ij}^x \hat{R}_{ij}^y)^2 \right\rangle_{\Omega} = \frac{1}{4\pi}\int\int d\varphi \sin\theta d\theta (\sin^4\theta \cos^2\varphi \sin^2\varphi) = \frac{1}{15} \tag{7.11}$$

At $T=0$, $\sigma_{\alpha\beta} \equiv \partial\delta F/\partial e_{\alpha\beta}$ and the affine translation-rotation invariant shear modulus for the central-force case can be derived as

$$G^{(C)} \simeq \frac{4}{5\pi}\kappa_{\parallel}z\phi R_0^{-1}. \tag{7.12}$$

Note that this expression has been derived using the definition of linearized strain tensor. In experiments and simulations the engineering strain tensor defined as $\gamma_{xy} \equiv 2e_{xy}$ is often used. This leads to a prefactor $2/5\pi$ in Eq. (7.12) instead of $4/5\pi$ whenever Eq. (7.12) is to be compared with experimental measurements of $G$ where the stress is measured as a function of the engineering strain.

The coordination number $z$ can be estimated from the experimentally determined structure factor, or evaluated, for sufficiently dense glasses, according to the following route. If the glass is dense ($\phi > 0.5$) its structure is homogenous due to mutual impenetrability of the particles and therefore dominated by the hard-sphere component of interaction. As shown by recent experimental studies [66, the result is that dense ($\phi \sim 0.6$) strongly attractive glasses exhibit the same homogeneous structure of purely hard-sphere glasses. Therefore, it is possible to estimate the mean coordination as a function of the packing fraction $\phi$, by calculating the mean coordination of the *hyper-quenched* hard-sphere liquid with the same $\phi$. This is equivalent to integrating the radial distribution function of hard-sphere liquids with a cut-off on the integration determined so as to recover the jamming point of monodisperse hard-spheres (given as $z=6$ at $\phi \simeq 0.64$). This route has been used to interpret experimental data of attractive colloidal glasses in [64] and in the next chapter.

Eq. (7.12) has been obtained under the limiting assumption that, in very attractive systems, the affine approximation leads to a small error. However, in spite of that approximation, in the next chapter it will be shown that Eq. (7.12) gives a rather accurate, quantitative description of the shear modulus of short-ranged attractive (depletion) colloidal glasses such as those studied in [67]. In that case, the affine approximation is justified because the elastic response is dominated by the first linear regime ending with break-up of nearest-neighbour bonds [67].

The more general expansion in Eq. (7.1) involves the three-body bond-bending forces and is somewhat more complex. A suitable model, which satisfies translation-rotation invariance, is the three-body Hamiltonian [68]

$$\delta F^{(B)} = \frac{1}{2}\kappa_\perp \sum_{<ijk>} (\delta\Theta_{ijk})^2 = \frac{1}{2}\kappa_\perp \sum_{<ijk>} \left|(\mathbf{u}_{ij}\times\hat{\mathbf{R}}_{ij} - \mathbf{u}_{ik}\times\hat{\mathbf{R}}_{ik})\cdot(\hat{\mathbf{R}}_{ij}\times\hat{\mathbf{R}}_{ik})/\left\|\hat{\mathbf{R}}_{ij}\times\hat{\mathbf{R}}_{ik}\right\|\right|^2 \quad (7.13)$$

where $\kappa_\perp$ is the local BB stiffness: $\kappa_\perp \equiv \partial^2 F/\partial\Theta_{ijk}^2\big|_{\{R\}}$. Again, following the Cauchy-Born approach of [18] and averaging over all possible strained configurations one can write

$$\left|\hat{\mathbf{R}}_{ij}\times\hat{\mathbf{R}}_{ik}\right|\overline{\delta\Theta_{ijk}} = \left(\overline{\mathbf{u}_{ij}}\times\hat{\mathbf{R}}_{ij} - \overline{\mathbf{u}_{ik}}\times\hat{\mathbf{R}}_{ik}\right)\cdot\left(\hat{\mathbf{R}}_{ij}\times\hat{\mathbf{R}}_{ik}\right), \quad (7.14)$$

which, in tensorial notation, and after expanding in the displacement field, reads

$$\left|\hat{\mathbf{R}}_{ij}\times\hat{\mathbf{R}}_{ik}\right|\overline{\delta\Theta_{ijk}} \simeq (R_{ij}^{-1}R_{ij}^\alpha\partial_\alpha\varepsilon_{\beta\delta\gamma}u_\delta R_{ij}^\gamma - R_{ik}^{-1}R_{ik}^\chi\partial_\chi\varepsilon_{\beta\eta\lambda}u_\eta R_{ik}^\lambda)\varepsilon_{\beta\mu\nu}R_{ij}^\mu R_{ij}^\nu \quad (7.15)$$

As shown in the Appendix A, one has that

$$\left(\overline{\mathbf{u}_{ij}}\times\hat{\mathbf{R}}_{ij}\right)\cdot\left(\hat{\mathbf{R}}_{ij}\times\hat{\mathbf{R}}_{ik}\right) \simeq 2\left[(\mathbf{R}_{ij}^\mathrm{T}\cdot\mathbf{e})\times\hat{\mathbf{R}}_{ij}\right]\cdot\left(\hat{\mathbf{R}}_{ij}\times\hat{\mathbf{R}}_{ik}\right). \quad (7.16)$$

Thus, the disorder-averaged change in the interaction angle can be written as

$$\overline{\delta\Theta_{ijk}} \simeq 2\left[(\mathbf{R}_{ij}^\mathrm{T}\cdot\mathbf{e})\times\hat{\mathbf{R}}_{ij} - (\mathbf{R}_{ik}^\mathrm{T}\cdot\mathbf{e})\times\hat{\mathbf{R}}_{ik}\right]\cdot(\hat{\mathbf{R}}_{ij}\times\hat{\mathbf{R}}_{ik})/\left|(\hat{\mathbf{R}}_{ij}\times\hat{\mathbf{R}}_{ik})\right|$$

(7.17)

which, by making use of Lagrange's identity and rearranging terms, becomes

$$\overline{\delta\Theta_{ijk}} \simeq 2(\sin\Theta_{ijk})^{-1}\left\{\left[(\mathbf{R}_{ij}^\mathrm{T}\cdot\mathbf{e})\cdot\hat{\mathbf{R}}_{ij} + (\mathbf{R}_{ik}^\mathrm{T}\cdot\mathbf{e})\cdot\hat{\mathbf{R}}_{ik}\right]\cos\Theta_{ijk} - \left[(\mathbf{R}_{ij}^\mathrm{T}\cdot\mathbf{e})\cdot\hat{\mathbf{R}}_{ik} + (\mathbf{R}_{ik}^\mathrm{T}\cdot\mathbf{e})\cdot\hat{\mathbf{R}}_{ij}\right]\right\}$$

(7.18)

and, finally,

$$\overline{\delta\Theta_{ijk}} \simeq 2(R_0\sin\Theta_{ijk})^{-1}\left\{\cos\Theta_{ijk}\left[\mathrm{Tr}(\mathbf{R}_{ij}\otimes\mathbf{R}_{ij}) + \mathrm{Tr}(\mathbf{R}_{ik}\otimes\mathbf{R}_{ik})\right]\cdot\mathbf{e} \right.$$
$$\left. - \left[\mathrm{Tr}(\mathbf{R}_{ik}\otimes\mathbf{R}_{ij}) + \mathrm{Tr}(\mathbf{R}_{ij}\otimes\mathbf{R}_{ik})\right]\cdot\mathbf{e}\right\} \quad (7.19)$$

For a pure shear, the above expression reduces to

$$\overline{\delta\Theta_{ijk}} \simeq 4(R_0\sin\Theta_{ijk})^{-1}\left\{\left[(R_{ij}^x R_{ij}^y + R_{ik}^x R_{ik}^y)\cos\Theta_{ijk}\right] - \left[R_{ij}^y R_{ik}^x + R_{ij}^x R_{ik}^y\right]\right\}e_{xy} \quad (7.20)$$

We now take the isotropic average over $\Theta_{ijk}$, thus assuming a flat distribution for $\Theta_{ijk}$: this assumption may be realistic for systems with strong spatial disorder such as e.g. emulsion glasses, colloidal or atomic (metallic and semiconductor) glasses without directional interactions. For molecular network-glasses, however, $\Theta_{ijk}$ will rather be distributed according to the chemistry of the system. With covalent network-glasses, usually the number of angles $\Theta_{ijk}$ is finite and dictated by the valence, thus giving rise to distinct terms in the expansion. Application of this model to specific covalent glasses may be the object of future work. Here we limit our analysis to the case of strong disorder, so that an unbiased average yields

$$\left\langle \overline{\delta\Theta_{ijk}} \right\rangle_\Theta \simeq \left\langle 4(R_0 \sin \Theta_{ijk})^{-1} \left\{ \left[ (R_{ij}^x R_{ij}^y + R_{ik}^x R_{ik}^y) \cos \Theta_{ijk} \right] - \left[ R_{ij}^y R_{ik}^x + R_{ij}^x R_{ik}^y \right] \right\} \right\rangle_\Theta e_{xy}$$
$$\simeq \frac{4}{3} R_0 \sin\varphi \cos\varphi \, (-\sin^2\theta + \cos^2\theta - 4\sin\theta\cos\theta) e_{xy} \quad (7.21)$$

where $R_{ik}^x = \sin(\theta + \Theta)\cos\varphi$ and $R_{ik}^y = \sin(\theta + \Theta)\sin\varphi$ have been used. As before, we assume random orientation of the bonds in the reference state $\{R\}$ so that on average each term in the summation in Eq. (7.13) contributes

$$\left\langle \left\langle \overline{\delta\Theta_{ijk}} \right\rangle_\Theta^2 \right\rangle_\Omega \simeq \frac{16}{9} R_0^2 \left[ \frac{1}{4\pi} \int\int d\varphi \sin\theta d\theta (\sin\varphi\cos\varphi)^2 \right.$$
$$\left. \times (\sin^4\theta + \cos^4\theta + 14\sin^2\theta\cos^2\theta + 8\sin^3\theta\cos\theta - 8\sin\theta\cos^3\theta) \right] e_{xy}$$
$$(7.22)$$

Therefore, using $\overline{\delta\Theta_{ijk} - \overline{\delta\Theta_{jk}}}^2 \ll \overline{\delta\Theta_{ijk}}^2$ for the average over disorder, as well as $\left\langle \overline{\delta\Theta_{ijk}} - \left\langle \overline{\delta\Theta_{ijk}} \right\rangle_\Theta \right\rangle_\Theta^2 \ll \left\langle \overline{\delta\Theta_{ijk}} \right\rangle_\Theta^2$ for the spatial average over the bending angle, linear elasticity leads to

$$G^{(B)} \simeq \frac{124}{135\pi} \kappa_\perp z^{(B)} \phi R_0^{-1} \quad (7.23)$$

where the sum over three-body interactions has been replaced by $\frac{1}{3} zN/V \equiv 2z\phi/\pi R_0^3$.

In Eq. (7.12) and (7.23) the definition of the microscopic bond rigidities ($\kappa_\parallel$ and $\kappa_\perp$, respectively) is clearly different, and the numerical prefactor is also different. In the BB case, the value of the prefactor is especially important, because for (real) network-glasses it also contains information about the chemistry-dependent geometry of the network. Here, the prefactor $(124/135)\pi^{-1}$ has been found for the case of nondirectional bonds and strong disorder but, in the case of real covalent glasses, it depends on the values of the bond-bending angle $\Theta_{ijk}$. For a generic system where both CF and BB interactions are present, as in many real glasses, the shear modulus can be estimated, from the stress-strain relation $\sigma_{\alpha\beta} \equiv \partial\delta F/\partial e_{\alpha\beta}$ where according to Cauchy-Born theory $\delta F \simeq \delta F^{(C)} + \delta F^{(B)}$, as

$$G = G^{(C)} + G^{(B)} \simeq \left(\frac{4}{5\pi}\kappa_{\parallel} z^{(C)} + \frac{124}{135\pi}\kappa_{\perp} z^{(B)}\right)\phi R_0^{-1} \tag{7.24}$$

$G^{(C)}$ and $G^{(B)}$ represent, respectively, the values to which the shear modulus would reduce in the case where respectively purely central interactions and purely bond-bending interactions have to be considered in the free energy.

Eq. (7.24) accounts for the fact that the mean number of bonds per particle which display BB resistance may differ from that of purely CF bonds. Indeed, for real covalent glasses, $z^{(B)}$ is a function of the valence which, in turn, is determined by the specific chemistry of the glass under consideration.

## 7.3 Discussion and applications

Eq. (7.24) has been derived by systematically applying Cauchy-Born theory (with the expansion written in terms of the *relative* deviations) and gives the macroscopic elastic response to shear of amorphous solids with both central-force and bond-bending interactions as a function of average parameters. These are the mean coordination (*z*), the volume fraction ($\phi$), the interparticle interactions (embedded in the Born-Huang term $\kappa$), and the mean separation distance ($R_0$) between nearest-neighbours in the reference (stressed) configuration. The latter, in a solid, is approximately equal to the diameter of the building blocks. We would like to clarify, at this point, the differences of the approach outlined in this work as compared to the various models in the literature, especially within rheological models. In most of these cases, the heterogeneous and heuristic character of the assumptions leads to numerical prefactors inconsistent with each other and generally not quantitatively comparable with the experiments [69]. Moreover, the affine approximation is sometimes applied to weakly bonded materials where nonaffine rearrangements are instead important. Including bond-bending terms in the expansion as done here is crucial thus making possible the application of continuum theory to such materials as strong covalent glasses (Si, Ge) where nonaffine rearrangements are smaller [63].

Furthermore, Eq. (7.24) tells us that the shear modulus of glasses is sensitive to an increase in the average number of covalent bonds (i.e. bonds which can react to shear forces) per atom. Increasing the average number of covalent bonds per atom has two main consequences on the properties of a glass: the stabilization of soft (localized) transverse vibrations and the shift of the Ioffe-Regel crossover to higher frequencies. Hence, both these effects are expected to translate into a decrease of the well-documented excess of modes in the vibrational density of states of glasses, i.e. a decrease in the intensity of the so-called boson peak [70]. Experimentally, a situation where the average number of covalent bonds per atom can be varied (within the range $2 \leq z^{(B)} \leq 4$) is found in chalcogenide alloys Ge$_x$Se$_{1-x}$, where $x$ represents the relative concentration of Ge by varying which the average number of covalent bonds per atom can be varied [71]. It was indeed found experimentally that the boson peak intensity decreases upon increasing $x$, thus $D(\omega_{BP})/\omega_{BP} \propto 1/x^\beta$ where $\beta$ is an exponent of order 1 [72]. According to our result, Eq. (1.24), $G \propto x$. Assuming that this holds even when nonaffinity is important, it follows that $D(\omega_{BP})/\omega_{BP} \propto 1/G^\beta$. The latter relation with $\beta = 1$ has been proposed on the basis of simulations of 2D spin glasses in a recent work [73].

We should note that Eq. (7.24) may help better understand the structure-elasticity properties of (dense) aggregated colloidal systems. In fact, it has been recently shown that polymer latex particles in the micron range display BB rigidity as a consequence of contact adhesion [74,75]. Therefore, for such colloidal systems, the BB stiffness in Eq. (1.23) may be expressed as a function of the surface adhesion parameters according to the experimental findings of [74], where the relation $\kappa_\perp \simeq 6\pi a_c^4 E_0/R_0^3$ was proposed ($a_c$ is the radius of the contact area of adhesion between two particles and $E_0$ is the particle Young's modulus). This is motivated by recognizing that the microscopic elastic constant for shear rigidity, $\kappa_\perp$, is defined here as the energy cost for changing the angle between two bonds with a vertex in common, in the same way as in previous studies on colloidal gels and aggregates, see e.g. [74] and [76]. It is therefore consistent, also from a dimensional point of view, with the single-bond BB rigidity measured in [74]. In particular Eq. (7.24) tells us that the presence of bond-bending forces gives an important contribution to the global shear rigidity. In the case of a colloidal aggregated state with $z^{(B)} = z^{(C)} = 3$, for example, the system is largely overconstrained (since the number of saturated degrees of freedom is $z^{(C)}/2 + z^{(B)}(z^{(B)} -1)/2 = 9$) and nonaffine displacements are small. Thus, our model is expected to be particularly accurate. Situations where $z^{(B)} = z^{(C)} \approx 3$ or larger are commonly encountered in not too diluted aggregated or gelled colloidal dispersions. Thus, combining the results of the present work with those of Ref. [74] for the microscopic description of bond-bending rigidity, may help, in future work, to lay down the groundwork for a comprehensive, quantitative modelling of the elastic properties of aggregated colloidal systems in the semi-dilute to concentrate regime.

## 7.4 Overview

The systematic Cauchy-Born approach to amorphous solids, in the same spirit of Ref. [18], has been applied to evaluate the macroscopic response to shear of deeply-quenched glassy states of spherical particles interacting via a central pair-interaction potential supplemented with an angular (bond-bending) three-body interaction term. Expressions in closed form are derived by making use of the affine approximation. The latter is generally a strong assumption when dealing with disordered systems, but may lead to small errors if the interparticle bonds can support significant bending moments (thus greatly reducing the number of degrees of freedom), as in covalent glasses (e.g. amorphous Si and Ge) [63]. Further, the model has the potentiality to account for the specific chemistry-dependent structure of real glasses. In the case of purely *central* pair interaction potentials, the situation is more complex. There, the affine approximation is of limited application. However, also in the latter case, as will be shown in the next chapter, the formulae derived here can nevertheless yield accurate predictions for colloidal glasses in the limit of strong short-ranged interparticle attraction. In this limit, the observed linear elastic regime is indeed due to stretching of the bonds [67,77], so that, the particles being localized upon strain within the short range of attraction, the assumptions used here yield reasonable predictions.

# 8. The macroscopic elasticity of arrested attractive colloids: coarse-graining of structural inhomogeneity

## *8.1 Introduction*

In this chapter we evaluate the elasticity of arrested short-ranged attractive colloids. We combine an analytically solvable elastic model, based on the theoretical treatment presented in the previous chapter, with a hierarchical arrest scheme into a new approach, which can be used to reveal the presence of and quantify structural heteoregeneity, such as mesoscopic entities which are directly responsible for the macroscopic shear modulus. The results quantitatively predict experimental data in a wide range of volume fractions and indicate in which cases the relevant contribution is due to mesoscopic structures. On this basis we propose that different arrested states of short-ranged attractive colloids can be meaningfully distinguished as homogeneous or heterogeneous colloidal glasses in terms of the length-scale which controls their elastic behavior.

Solutions of short-ranged attractive colloidal particles are the object of intense study due to their technological applications (proteins, paints etc.), as the constituent blocks of nanomaterials, as well as model systems to better understand phase behavior and dynamical arrest of condensed matter [78,79]. A landscape of phases has been observed upon varying the volume fraction $\phi$ or the interaction parameters [80], and, as a matter of fact, extended regions of the phase diagram are still poorly understood. In very dense suspensions ($\phi > 0.5$) the arrested states are spatially homogeneous (i.e. the typical linear size of structural heterogeneity is smaller than the particle diameter $R_0$). Particles are immobilized within the range of attraction, giving rise to bonds that are persistent under strain in the linear regime, and the high density leads to (attractive) glassy states [16,81-85] In the range $0.2 < \phi < 0.5$, the situation is complicated: arrested metastable states can only occur thanks to pronounced structural heterogeneities, typically on length scales larger than $R_0$. Therefore, these arrested states are more related to gelation [86,87], as it is the case at even lower volume fractions, rather than to the caging typical of crowded random media. However, they are also hardly classifiable as classic network gels in view of the different morphology and stress-bearing mechanisms.

For them, different microscopic phenomena should be considered and a new theoretical framework, able to account for the strong spatial heterogeneities and currently still missing, would be desirable. A crucial point, mostly neglected in recent studies, is that arrested states occurring at different volume fractions and attraction strengths do display dramatically diverse mechanical and rheological properties [20,88-91]. That is why their characterization is of true interest in technological applications and advanced material design.

Here we propose a new, more "down to earth", approach to characterize arrested short-ranged attractive colloids, based on their mechanical response. We combine an analytically solvable elastic model with a hierarchical arrest scheme, into an approach that can be applied to attractive systems in a wide range of volume fractions (from random close packing down to $\phi \approx 0.2$). Our model provides a first step in the direction of discriminating the microscopic (primary particle-level) from the mesoscopic (cluster-level) contribution to the macroscopic elasticity. The predictions are given in terms of the shear modulus, a quantity that can be measured in a rheometer, and are directly compared with experimental results. This novel approach can pave the way to a meaningful distinction between different glassy states, on the basis of the length-scale $\widetilde{R}_0$ dominating their elastic response. $\widetilde{R}_0$ varies with volume fraction due to the arising of structural heterogeneities. However, whereas the characterization of structural heterogeneities requires detailed structural information and is therefore often elusive, the variation of $\widetilde{R}_0$ is unambiguously signalled by significant variation of the elastic modulus. Upon connecting the elasticity to the structural features of these materials, our approach provides the missing information in one respect or the other and offers a new insight into the complex physics of arrested attractive colloids.

## *8.2 Model and predictions: the dense regime*

We study a suspension of colloidal particles interacting via a short-range attraction, as the one typically induced by depletion using non adsorbing polymer [78], well above percolation, so that a finite shear modulus is always detectable. In order to distinguish the single-particle contribution from the mesoscopic one, we consider coarse-grained entities, i.e. renormalized particles. Their effective interactions are directly responsible for the macroscopic properties of the system. This scheme can be associated to a double ergodicity-breaking scenario where the arrest occurs in form of a cluster-glass transition [92-96]. A sine qua non is that further coalescence of the clusters, leading to phase separation, is prevented on the time scale of observation [93]. We distinguish between the macroscopic elasticity of the system $G$, the intra-cluster elasticity $G_g$ resulting from mutual interactions between primary particles, and the inter-cluster elasticity $G_c$ resulting from mutual interactions between clusters. Experiments [97,98] suggest that the elasticity of such heterogeneous materials is dominated by the weaker (less rigid) connections.

Let us first consider the case of a homogeneous attractive glass in the low-temperature and high-density ($\phi > 0.5$) region of the phase diagram. Here we focus on the first linear regime reported in rheological measurements [17], which is due to bond-breaking and, for strong attractions, extends nearly up to strains of order 10%.

Hence, we follow the Cauchy-Born approach developed in [18] for amorphous solids and obtain the elasticity from a free energy expansion around a *stressed* (quenched) reference state $\{R\}$ where all particles are *labeled*, as shown in the previous chapter. With only central interactions at play the expansion reads

$$\delta F \equiv F(\{r\}) - F(\{R\}) \simeq \sum_{<ij>} \left.\frac{\partial F}{\partial r_{ij}}\right|_{\{R\}} \delta r_{ij} + \frac{1}{2}\sum_{<ij>} \left.\frac{\partial^2 F}{\partial r_{ij}^2}\right|_{\{R\}} (\delta r_{ij})^2$$

where the sums run over all interacting pairs of particles (bonds) and the derivatives are evaluated at relative distances $\{R_{ij}\}$ in the quenched reference state. Due to the purely internal nature of the stresses, the first term in the r.h.s. does not contribute to the macroscopic elasticity [18,65]. In the case of central forces, considering a pair-interaction with a deep minimum $\varepsilon \gg k_B T$, the bond stiffness is defined by $\kappa \simeq \partial^2 F / \partial r_{ij}^2$. As we have seen in the previous chapter, under the assumption that all pairs of particles are much localized very near the minimum of the potential well, the continuum limit for the microscopic $u_{ij}^{\parallel}$ can be taken by using an affine transformation defined by the average strain tensor $e$ to obtain $\delta F \simeq (1/2)\sum_{<ij>} \kappa_{\parallel} \overline{(u_{ij}^{\parallel})^2} \simeq (1/2)\sum_{<ij>} \kappa_{\parallel} \left\{ R_{ij}^{-1}\text{Tr}[(\boldsymbol{R}_{ij} \otimes \boldsymbol{R}_{ij}) \cdot \mathbf{e}] \right\}^2$. For the quenched reference configuration $\{R\}$, in the case of uncorrelated disorder, the summation over bonds can be replaced by the total number of contacts, which introduces the average coordination number $z(\phi)$ in the final expression of the deformation free energy. We then derive the off-diagonal components of the stress

$$\sigma_{\alpha\beta} \equiv \frac{\partial \delta \mathcal{F}}{\partial e_{\alpha\beta}} = \frac{1}{2}\rho z(\phi)\kappa \frac{\partial}{\partial e_{\alpha\beta}} \left\langle \left(\frac{R_{ij}^{\alpha} R_{ij}^{\beta}}{R_{ij}}\right)^2 e_{\alpha\beta} \right\rangle$$

where <...> denotes the isotropic average. Hence the shear modulus of an arrested phase of volume fraction $\phi = \rho \pi R_0^3/6$, with mean particle diameter $\simeq R_0$, is given by $G_g \simeq (2/5\pi)\kappa_{\parallel} z \phi R_0^{-1}$ (note the difference of a factor two with the expression in the previous chapter since here we consider the engineering strain for comparison with the experiments). The average local structure of dense glassy states depends little on the attraction [66] and does not show significant deviations from the one in the liquid prior to quenching, apart from distortions (buckling) which bring particles into permanent contact. Hence, to estimate the number of short-range contacts which are likely to be involved in the mechanical bonds, we integrate the radial distribution function (rdf) of the dense (hard-sphere) liquid precursor, $g(r)$, up to a value of interparticle gap $l^{\dagger} \approx 0.03$ which allows to recover the critical point of hard-sphere packings, where $z \approx 6$. This result is in derived in Appendix B and is valid under the assumption that the packing of clusters is structurally homogeneous (i.e. the clusters are distributed homogeneously in space). This yields an estimate of the number of permanent (mechanical) contacts created upon quenching (Appendix B). This estimate, and the sensitivity of the results upon $l^{\dagger}$, are discussed, also in comparison with simulations and experiments, in Appendix B. We thus obtain the shear modulus as a function of the rdf:

$$G_g = (48/5\pi)\phi\kappa R_0^{-1} \int_0^{l^{\dagger}} (1+l)^2 g(l;\phi) \mathrm{d}l \tag{8.1}$$

For $g(r)$ near contact ($l < 0.1$) we use liquid theory valid in the dense hard-sphere fluid [99], as explained in the Appendix B, and we calculate $\kappa$ using the Asakura-Oosawa (AO) potential [100,101]. Our predictions have been compared to the experimental data of Ref. [17] for an attractive colloidal glass ($\phi \simeq 0.6$). The shear modulus measured at different frequencies, is plotted as a function of the attraction strength, i.e. the reduced polymer concentration $c_p/c_p^*$ (where $c_p^*$ is the polymer overlap concentration). For strong enough attraction (fully elastic response), the affine approximation is more realistic and our prediction gives a very accurate estimate with an accuracy of a few percent of the measured shear modulus. In fact in Figure 7-1 we compare our predictions to the experimental data of Ref. [17] for an attractive colloidal glass ($\phi \simeq 0.6$). The shear modulus, measured at different frequencies, is plotted as a function of the attraction strength, i.e. the reduced polymer concentration $c_p/c_p^*$: for $0.05 < c_p/c_p^* < 0.2$, the measured shear modulus is significantly dependent upon the frequency, in correspondence of the liquid-like re-entrant region. However, for sufficiently high attraction strength ($c_p/c_p^* > 0.2$) where the frequency dependence is negligible and the affine approximation used to derive Eq. (8.1) in the manuscript becomes more realistic, our prediction for the shear modulus gives an accurate estimate.

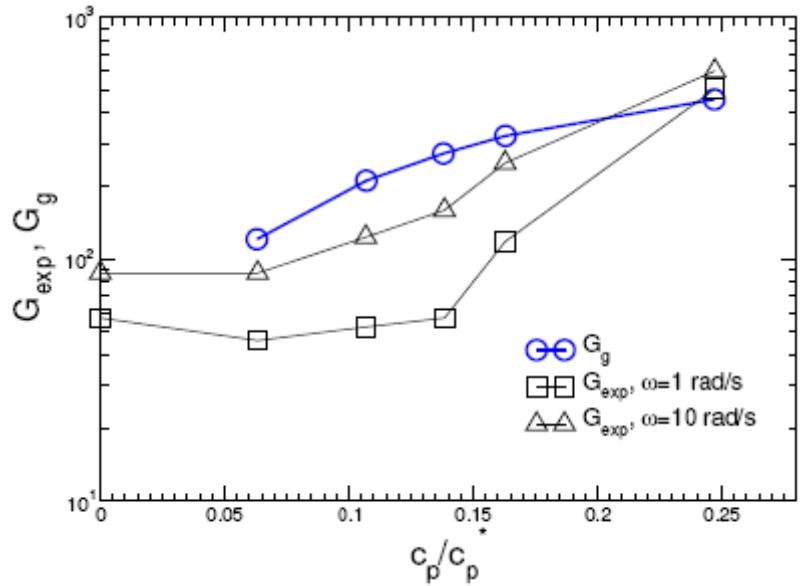

**Figure 8-1: The shear modulus of the attractive glass of [17] as a function of the attraction strength for the experimental data at different frequencies and as predicted by Eq. (8.1), with the same values of physical parameters.**

A more stringent verification of the model requires comparison with simulations of glassy states where the bond-stiffness and the local structure are well-defined. Thus, we performed Molecular Dynamics simulations of a deeply quenched glass obtained from a supercooled Lennard-Jones liquid. We subjected the glass to shear strain and extracted stress-strain curves. We have carried out molecular dynamics simulations on a binary mixture of 2024 particles at 60% volume fraction with size ratio 1:1.2 interacting via a Lennard-Jones potential. The system was first driven into the deepest metastable minimum in the energy landscape at $T$=0.46. Three different glass configurations obtained in this way were then quenched to $T$=0.005 for a period of time of at least $10^6$ MD steps. Simple shear strain experiments were performed at $T$=0.005 using SLLOD equations of motion [102] implemented in LAMMPS [103], making use of Lees-Edwards boundary conditions and a Nose'-Hoover thermostat [102]. Two sets of measurements were done for each configuration: 1) a step-strain measurement were the system is strained (at a fixed shear rate) up to a certain (engineering) strain value $\gamma_{xy}$ and subsequently allowed to relax in time; 2) a continuous-straining measurement where the system is strained continuously in time at a fixed shear-rate up to a maximum strain of 0.1. The shear-rate used was always 0.01. Note again that we use here, as in any other comparison with experiments in the paper, the engineering shear strain $\gamma_{xy}$, which is related to the linearized (theoretical) shear strain by $e_{xy} = \gamma_{xy}/2$. Clearly, as we could verify, the zero-time stress values from the step-strain measurements coincide with the continuous-strain curve. From the relaxation experiments it was seen that the zero-time and the infinite-time values of the stress are practically equal up to strain 3.5-4%. At larger strains the infinite-time values of stress are lower than the instantaneous values due to relaxations. The linear regime in the zero-time stress-strain curve is somewhat larger and extends up to 5% of strain. The shear modulus measured from the linear regime up to 5% and averaged over the three configurations is equal to $19.4 k_B T/\sigma^3 \pm 0.5$. The theoretical stress-strain relation from our model (Eq. 8.1) predicts a value of shear modulus equal to $18.8 k_B T/\sigma^3$, with just a 3% deviation from the measured value, thus in excellent quantitative agreement with the simulation result. The zero-time stress-strain curves of the three configurations are shown together with the theoretical stress-strain curve from our model in Figure 8-2.

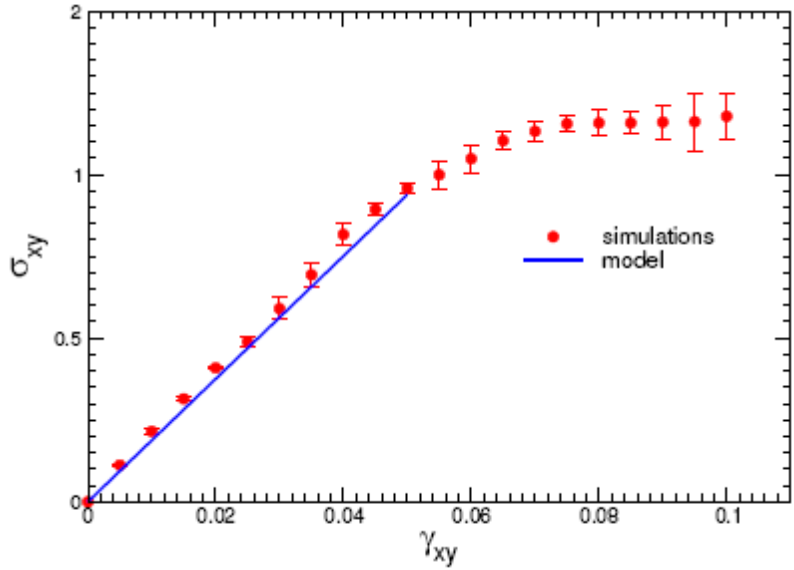

**Figure 8-2: The stress-strain curve for the homogeneous attractive glass as obtained from simulations and model prediction (Eq. 8.1).**

Further, the simulations provide an indication of the mean local structure of the glass. It was found that at the low temperatures investigated the internal energy per atom amounts to about $7k_BT$. This number represents an upper limit for the number of nearest neighbours which are localized within the range of attraction of the potential. This estimate is in agreement with our theoretical estimate of the mean number of mechanical contacts (i.e. the number of particles sitting in the minimum of the well) at 60% volume fraction which is $z = 4.3$. Indeed the number of particles in the minimum of the well has to be lower than the total number of neighbours within the whole attraction range.

These independent comparisons show that the model is quantitatively predictive. They also indicate that the nonaffine rearrangements are actually small in the strong-attraction or deep-quench limit. Such rearrangements become more important upon lowering the attraction, as the system converts gradually into the hard-sphere (repulsive) glass [104], and an overestimation in that regime is expected, due to the break-down of affinity.

## 8.3 Model and predictions: double ergodicity-breaking and the intermediate density regime

For the more dilute regime, we now consider a double-ergodicity breaking scenario, as the one explored in [92-96], with local aggregation of the colloidal particles to form beads (clusters) which, in turn, arrest due to either caging or residual attraction (in the latter case the glass transition would be energy-driven on both levels). The major assumption is that clusters are stabilized from coalescing [93]. They are viewed as compact (spherical or quasi-spherical) renormalized particles of diameter $R_0$, whose effective volume fraction may be identified with the one determined by the spheres enclosing the clusters (i.e. significantly larger than $\phi$). If the cluster linear size is larger than the particle diameter by a factor say less than 10, each contact between clusters is likely to reduce to a single colloid-colloid bond. Upon neglecting: 1) the breakup probability within the cluster, and 2) the effect of long-range repulsion, the effective interaction between clusters obviously reduces to the bare colloid-colloid interaction, in agreement with [93]. Further, the mean coordination will change to $z(\phi_c)$ (where now $\phi_c$ is the cluster volume fraction), but its form can be still determined as in Eq. (8.2) if $\phi_c$ is still in the dense glassy regime dominated by mutual impenetrability. Hence, for the modulus of the cluster-glass we can write:

$$G_c = (2/5)\pi^{-1}\phi_c z(\phi_c)\tilde{\kappa}\tilde{R}_0^{-1} \tag{8.2}$$

with $\tilde{\kappa} \simeq \kappa$ under the assumption of small clusters. Eq. (8.2) gives the elastic modulus of the material, where the macroscopic elasticity is dominated by the mesoscopic level. We test our scheme using the extensive experimental data of Ref. [105] for a system of colloidal silica particles with polystyrene as depletant in organic solvent (decalin), in the range $0.2 \lesssim \phi \lesssim 0.4$, where the ratio of the polymer gyration radius to $R_0$ is the same as in the experiments of [17].

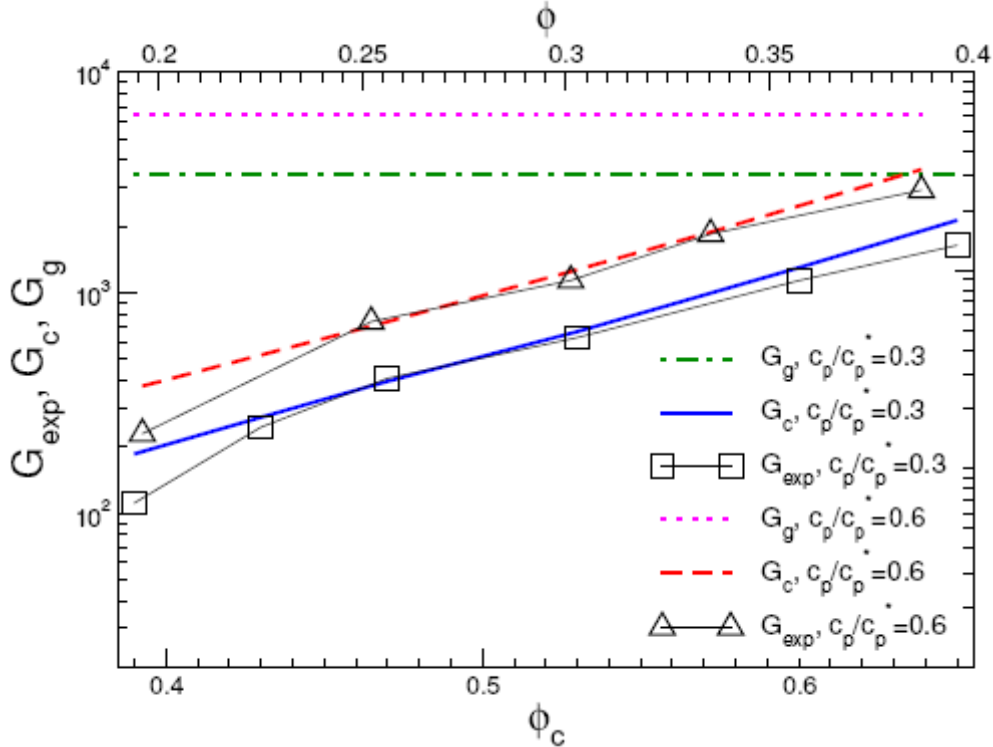

**Figure 8-3: Shear modulus as a function of $\phi_c$ and $\phi$ for the experimental data from [105] and for the model prediction from Eqs. (8.2) and (8.3), with the same physical parameters measured on the experimental system: $R_0/\widetilde{R}_0 = 5.5$, polymer-to-colloid size ratio $\xi = 0.078$.**

In [105], these results are explained in terms of a "dense gel" with compact clusters ($\phi \approx 0.5$ inside the cluster) and $\widetilde{R}_0/R_0 \approx 5$ as determined from the Debye-Bueche plot. In Figure 8-2 we plot the values of the shear modulus measured at 1Hz in [105] for two different attraction strengths, together with our predictions of $G_g$ from Eq. (8.1) and of $G_c$ from Eq. (8.2). According to [105], upon increasing $\phi$ at fixed attraction strength, the cluster volume fraction $\phi_c$ also increases, whereas the volume fraction inside the clusters remains fixed at $\phi \approx 0.5$. Based on this, the data in Figure 8-2 are thus plotted as a function of $\phi$ and $\phi_c$. The excellent agreement between model and experiments in Figure 8-2 confirms the experimental estimate of the mesoscopic length scale and clearly indicates that, differently from the case previously considered, now is $G_c$ that dominates the macroscopic elasticity of the system. The dependence on $\phi_c$ has also been accurately predicted by the model. For lower attractions, the system will gradually cross over towards the repulsive case. We now consider the results recently reported in [106], for a different arrested AO system in the heterogeneous glass regime, i.e. at $\phi = 0.4$. In the experiments, they used light-scattering (LS) to estimate the length-scale of structural heterogeneities and up to $c_p/c_p^* \approx 1$ the results agree with microscopy (MS) data. At higher $c_p/c_p^*$, the LS estimate gives $\widetilde{R}_0/R_0 \approx 1$ or even smaller, contrary to the evidence from MS, which indicates instead a heterogeneity length of 2-3 particle diameters [106]. We have used this experimental input to calculate the prediction of our model for this system and compared it to the experimental value of the shear modulus in Figure 8-3: $G_c$ (solid line) shows a remarkable agreement with $G_{\exp}$ and turns out again to dominate the elastic response (see in comparison $G_g$, dashed line).

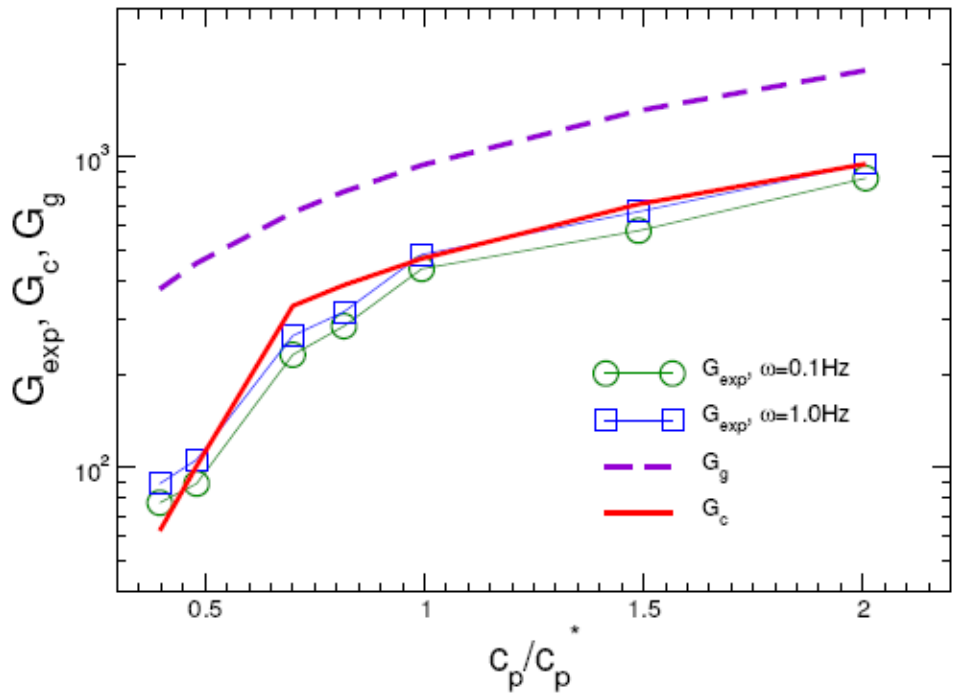

**Figure 8-4: The shear modulus of the attractive glass in [106] as a function of the attraction strength at two different frequencies (symbols) and model prediction from Eq. (8.2) and (8.3) (lines) with polymer-to-colloid size ratio $\xi = 0.08$ according to [106].**

Also here the comparison with experiments has been done at low frequencies and for fully elastic response.

## 8.4 Elastic phase diagram of attractive colloids

The results just discussed lead us to sketch in Figure 8-4 a new qualitative *phase diagram* for the arrested states of short-range attractive colloidal suspensions. As a function of $\phi$ and $k_B T/\varepsilon$, where $\varepsilon$ is the depth of the attractive well, we locate the arrested states in the region where the system displays a non-zero measurable elastic modulus *G* (the continuum line divides fluid from solid states as discussed in [107]). At high enough $\phi$ and large attractions, such state (*homogeneous glass*) is characterized by an elastic modulus dominated by the inter-particle elasticity ($\tilde{R}_0 / R_0 \approx 1$). Upon lowering $\phi$, aggregation produces mesoscopic structural heterogeneities and the mechanical response of the system crosses over towards a regime dominated by the inter-cluster elasticity (*heterogeneous glass*) with $\tilde{R}_0 / R_0 \gg 1$.

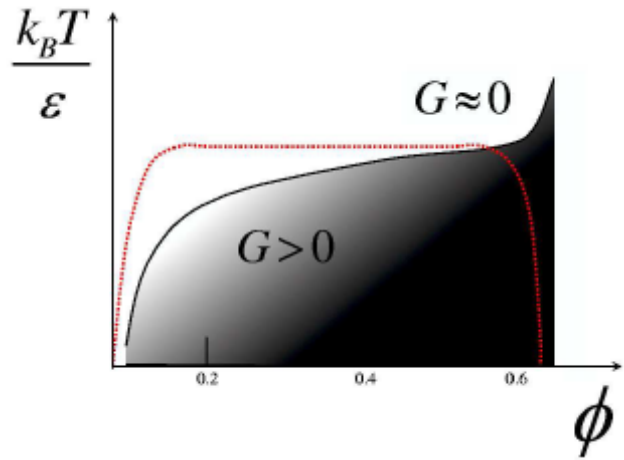

**Figure 8-5: Qualitative phase diagram based on the macroscopic shear modulus *G* and on the typical lenght scale $\tilde{R}_0$ of the mechanically relevant structural heterogeneities. Solid line: demarcation between fluid and solid states. Dotted line: spinodal decomposition line.**

The two regimes will be distinguished by a significant variation of the elastic modulus of the material. Our diagram suggests for the first time a distinction between arrested states of attractive colloidal suspensions in terms of a well defined, directly measurable quantity, the elastic shear modulus, and its length scale dependence. When the structure of the material is known in detail, such distinction can provide information on which part of the structure is relevant to the elastic behavior. Conversely, starting from the mechanical response of the material, one can gain insight into its structural changes. Upon further lowering the volume fraction ($\phi < 0.2$), we expect the elastic response to be dominated by a different length scale, associated to the weakly connected network-like mesoscopic structure and strongly dependent on $\phi$ [86,87]. Due to lower connectivity, nonaffinity is likely to be more pronounced and lead to a sensibly lower shear modulus (Figure 8-3). At very low $\phi$ arrested states may be due to effective directional interactions arising at mesoscopic length scales, leading to open network structures [108-110], where buckling strongly affects the macroscopic elasticity.

## *8.5 Overview*

Our new approach allows for discriminating the microscopic (inter-particle) from the mesoscopic (inter-aggregates) elasticity in arrested short-ranged attractive colloidal suspensions. The predictions quantitatively reproduce the experimental data for different systems and have the ability to signal the existence, as the responsible (building blocks) for the macroscopic elasticity, of mesoscopic entities (larger than the primary particle or microscopic length scale) associated with structural heterogeneity. Hence we gain a new insight into the complex physics of arrested *states* in attractive colloids, especially in the range $0.2 < \phi < 0.5$, hitherto poorly understood. In fact, a characterization based on structural heterogeneities requires detailed structural information and it is often elusive. Instead, we propose that homogeneous and heterogeneous arrested states can be distinguished in terms of the length scale $\widetilde{R}_0$, which controls the macroscopic elastic modulus. Our results show that changes in $\widetilde{R}_0$ produce well detectable variations of the elastic modulus in agreement with experiments.

## 9. Concluding remarks

In this work both theoretical and experimental consideration has been given to the macroscopic properties of colloidal amorphous (i.e. structurally disordered) states of matter in relation to the underlying structure and interactions. A major source of complexity in these systems is the interplay between different phenomena at different length-scales. Hence, in order to tackle this complexity, we have systematically made use of a physically-motivated coarse-graining principle. Coarse-graining essentially consists of integrating out a specific set of degrees of freedom, usually those related to the primary building-blocks of the system. The starting point of our analysis is the hypothesis that, in structurally inhomogeneous systems, where mesoscopic structures emerge naturally, it is the relevant static mesoscopic length-scale the one which controls the macroscopic properties. This is contrary to what happens in structurally homogeneous systems where it is instead the microscopic length-scale which usually determines the macroscopic properties. Hence, the mesoscopic structures can be treated as the effective building blocks of the system, by integrating out the degrees of freedom of the primary (microscopic) particles. The latter ones however still play a role in that they determine the features of the mesoscopic structures. We have systematically applied this coarse-graining principle and the hierarchic approach to the problem, of central importance in soft matter, of the response of colloidal systems to an external shear field.

We have first considered *fluid* colloidal suspensions of interacting particles. In the presence of shearing fields and an attractive component of colloid interaction, the interplay between direct interactions and shear results in the irreversible aggregation with formation of mesoscopic clusters. This aggregation process, due to the presence of the external field, is peculiar and differs substantially from aggregation processes in the absence of field (such as e.g. diffusion limited or reaction limited colloidal aggregation [3]). We have thus proposed a theory based on the Smoluchowski equation with shear that disentangles the microscopic interplay between shear, Brownian motion and interactions. The result is the analytical prediction of the colloidal stability of DLVO-interacting colloids in linear flows. The theory compares well with numerical results in terms of the binary aggregation rate as well as with experiments in terms of the characteristic time of the aggregation process. Moreover, our experiments, where DLVO-interacting colloids are sheared in a rheometer, reveal that the suspension viscosity increases explosively after a lag time. The latter can be predicted (even quantitatively with good accuracy) using our theory which also explains the explosive rise of viscosity observed in the experiments as due to a self-accelerated aggregation kinetic regime setting in when the size of the clusters reaches an average *activated* value.

If the microscopic level is thus essential in the structure-formation process leading to mesoscopic inhomogeneities (clusters), its role is not so directly evident in the determination of the macroscopic properties of the system, in this case the suspension viscosity. We have shown, through a combination of experiments and modeling, that the suspension viscosity can be described entirely in terms of hydrodynamic dissipation induced by the clusters effectively behaving as coarse-grained ellipsoids. We have proposed and experimentally validated a theory of viscosity in aggregating suspensions where the clusters resulting from the aggregation process entirely dominate the macroscopic viscosity due to their many-body hydrodynamic interactions. The macroscopic viscosity depends on the geometry and number density of the clusters while the microscopic (primary colloidal particle) level enters only indirectly into play through the shear-induced cluster formation process. Due the orders-of-magnitude length-scale separation between primary particles and mesoscopic clusters, as soon as the latter ones occupy a significant volume fraction of the system, they control the viscosity since the dissipation induced by their motion is very much larger compared to that of the primary particles. This observation confirms the applicability of the coarse-graining hypothesis in the description of the macroscopic viscosity of DLVO-interacting suspensions in shear.

A further test of the applicability of the coarse-graining principle was given by considering other features of the rheological response as a function of the aggregation-degree of the system. Dynamic frequency sweep tests have revealed that, after cessation of flow, the clusters are able to connect in a rigid space-spanning structure which can bear stresses, thus displaying a solid-like behaviour. The evolution of the viscoelastic response, in particular the emergence of rigidity as a function of the effective volume fraction of the clusters, is reminiscent of gelation processes in the absence of field [50,106], so that we may refer to this phenomenon as "shear-induced gelation". Furthermore, shear-sweep tests reveal that for effective cluster volume fractions $\phi_{\text{eff}} > 0.10$, where the system exhibits solid-like response, the rheological response changes from Brownian-like, with modest shear-thinning followed by a Newtonian plateau at high shear, to a fully non-Brownian-like response exhibiting all the hallmarks of dense non-Brownian suspensions, including a spectacular re-entrant shear-thickening and yield-stress behaviour. Such a shear-induced transition from a Brownian-like to a non-Brownian-like rheological response represents a further confirmation that the mesoscopic structures effectively dominate the macroscopic response above a critical effective volume fraction.

In the case of dynamically arrested colloidal suspensions (colloidal glasses) we would exepect a similar scenario since also in that case structural inhomogeneities are important (especially upon changing the volume fraction from random close packing downwards). In this case, the macroscopic response one is interested in (also in view of the many technological applications of these systems) is the shear modulus. Hence, as the first step in our analysis, theoretical consideration has been given first to the elastic response of (dense) structurally *homogeneous* glasses. We thus came up with an affine theory relating the macroscopic response to the microscopic size, density and interactions of the building blocks. The accuracy of the predictions has been demonstrated by means of simulations in the case of strong attractive bonds (as can be formed through deep quenches). Application of the coarse-graining hypothesis allows one to derive expressions for the shear modulus of structurally *inhomogeneous* systems at *lower* volume fractions. Also in this case, the length-scale of structural inhomogeneities is assumed to be the characteristic length-scale of the building blocks, while the degrees of freedom of the primary particles are integrated out. The predictions of the coarse-grained model are tested in comparison with different sets of experimental data from the literature and good quantitative agreement has been found in all cases.

The emerging picture from our study is that the complexity arising from the presence of structural inhomogeneities in disordered states of matter, and thus of different phenomena at different length-scales, may be effectively disentangled in those cases where the structural inhomogeneity appears associated with rigid mesoscopic structures herewith referred to as clusters. We have demonstrated that in those cases the macroscopic properties can be described by treating the clusters as the effective building blocks of the system by integrating out the degrees of freedom of the primary particles (e.g. the primary colloidal particles in colloidal states). The microscopic physics remains however essential for our understanding of the formation and of the features of the mesoscopic level. This leads to a hierarchic approach that we hope can be applied in the future to achieve a unified description of the properties of amorphous states [129], possibly including many-body atomic, molecular and fermionic systems [130].

# Appendix A

# Derivation of Eq. (7.16)

One can decompose the gradient expansion of the smooth displacement field $\mathbf{u}(\mathbf{r})$ into an explicitly symmetric part (i.e. the disorder-averaged symmetric strain tensor) and an antisymmetric one as:

$$\overline{\mathbf{u}_{ij}} \equiv \overline{\mathbf{u}_i - \mathbf{u}_j} \simeq (\mathbf{R}_{ij} \cdot \nabla)\mathbf{u} = \left[ \mathbf{R}_{ij}^T \cdot \mathbf{e} + \frac{1}{2}(\nabla \times \mathbf{u}) \times \mathbf{R}_{ij} \right] \tag{A.1}$$

Using the well-known identities:

$$(\nabla \times \mathbf{u}) \times \mathbf{R}_{ij} = -\mathbf{R}_{ij} \times (\nabla \times \mathbf{u}) = -\nabla(\mathbf{R}_{ij} \cdot \mathbf{u}) + (\mathbf{R}_{ij} \cdot \nabla)\mathbf{u} \tag{A.2}$$

Eq. (A.1) can be rewritten as:

$$\overline{\mathbf{u}_{ij}} \simeq (\mathbf{R}_{ij} \cdot \nabla)\mathbf{u} = \left[ \mathbf{R}_{ij}^T \cdot \mathbf{e} - \frac{1}{2}\nabla(\mathbf{R}_{ij} \cdot \mathbf{u}) + \frac{1}{2}(\mathbf{R}_{ij} \cdot \nabla)\mathbf{u} \right] \tag{A.3}$$

Rearranging terms:

$$\overline{\mathbf{u}_{ij}} \simeq (\mathbf{R}_{ij} \cdot \nabla)\mathbf{u} = 2\left[ \mathbf{R}_{ij}^T \cdot \mathbf{e} - \frac{1}{2}\nabla(\mathbf{R}_{ij} \cdot \mathbf{u}) \right] \tag{A.4}$$

Hence, we can rewrite Eq. (16) as:

$$\left(\overline{\mathbf{u}_{ij}} \times \hat{\mathbf{R}}_{ij}\right) \cdot \left(\hat{\mathbf{R}}_{ij} \times \hat{\mathbf{R}}_{ik}\right) \simeq 2\left\{[(\mathbf{R}_{ij}^T \cdot \mathbf{e}) \times \hat{\mathbf{R}}_{ij}] - \frac{1}{2}\left[\nabla\left(\mathbf{R}_{ij} \cdot \mathbf{u}\right) \times \hat{\mathbf{R}}_{ij}\right]\right\} \cdot \left(\hat{\mathbf{R}}_{ij} \times \hat{\mathbf{R}}_{ik}\right) \tag{A.5}$$

For the antisymmetric parts, we can use the identity $\nabla(\mathbf{R}_{ij} \cdot \mathbf{u}) \times \hat{\mathbf{R}}_{ij} = \nabla \times (\mathbf{R}_{ij} \cdot \mathbf{u})\hat{\mathbf{R}}_{ij} - (\mathbf{R}_{ij} \cdot \mathbf{u})\nabla \times \hat{\mathbf{R}}_{ij}$, where the second term on the RHS is clearly zero. Therefore, the term $[\nabla \times (\mathbf{R}_{ij} \cdot \mathbf{u})\hat{\mathbf{R}}_{ij}] \cdot (\hat{\mathbf{R}}_{ij} \times \hat{\mathbf{R}}_{ik})$, making use of Lagrange's identity, is seen to be zero

$$[\nabla \times (\mathbf{R}_{ij} \cdot \mathbf{u})\hat{\mathbf{R}}_{ij}] \cdot (\hat{\mathbf{R}}_{ij} \times \hat{\mathbf{R}}_{ik}) = (\nabla \cdot \hat{\mathbf{R}}_{ij})[(\mathbf{R}_{ij} \cdot \mathbf{u})\hat{\mathbf{R}}_{ij} \cdot \hat{\mathbf{R}}_{ik}] - (\nabla \cdot \hat{\mathbf{R}}_{ik})[(\mathbf{R}_{ij} \cdot \mathbf{u})\hat{\mathbf{R}}_{ij} \cdot \hat{\mathbf{R}}_{ij}] = 0$$
(A.6)

because $\hat{\mathbf{R}}_{ij}$ and $\hat{\mathbf{R}}_{ik}$ are constant vectors. Hence, Eq. (7.16) is verified.

# Appendix B

# Packing fraction-dependent mean coordination and structural inhomogeneity in amorphous solids

In this Appendix we formulate the model for the mean coordination number as a function of packing fraction used in this thesis for estimating the elastic response of amorphous solids. The central result derived in this Appendix, Eqs. (B.8)-(B.10), has been used in Chapter 8 and in the modeling of the breakup scaling of colloidal aggregates in Ref. [19].

Dynamical arrest in amorphous condensed and soft condensed matter represents an open problem where the arising of rigidity is closely associated with subtle structural changes [14,120,121]. Therefore, it is essential to define measures in order to properly describe the structure as well as to follow its transformations. The simplest yet most important topological parameter which is used to describe the structure of amorphous condensed matter is the mean coordination number [14]. Its definition does not gather general consensus and highly depends upon the system under consideration. From the experimental viewpoint, structural information can be obtained by either radiation scattering or by optical (e.g. microscopy) techniques (which can be applied for example to colloidal and granular systems). While scattering provides structural information in Fourier space, microscopy and direct imaging give quantitative, real space information about the local coordination. A practical way to define a mean coordination number is to identify it with the mean number of contacts, which in practice means counting the number of particles physically touching a tagged particle [122]. Some uncertainty may still arise as to the degree of "touching" between particles, so that some arbitrariness is unavoidable [122]. From the theoretical standpoint, most theories of condensed matter structure produce as a result the familiar radial distribution function (rdf), $g(r)$ [2,14]. Integrating $g(r)$ over a value of shell width around the tagged particle gives the most likely number of particles to be found within that radial distance. Also in this case, however, different choices are possible for the integration boundary and this introduces again a substantial degree of arbitrariness in the characterization of the local structure. Besides these problems, there is the issue of characterizing the structure of amorphous systems at a more coarse-grained level where the appearance of structural heterogeneity usually plays an important role [120]. This has done, for example, in terms of the statistics and structure of the voids, including remoteness and the void size distribution, which have been usefully employed in simulation studies of fractal colloidal aggregates and gels as well as colloidal glasses [123,124]. In this work we want to elucidate whether it is possible to give a description of structural heterogeneity in terms of the mean coordination number, which is a more directly accessible quantity. Moreover, we also use the mean coordination number to assess the effect on structure of different microscopic interaction potentials (attractive rather than merely repulsive).

To this aim, we focus our attention on amorphous model (colloidal [125] and Lennard-Jones) systems. We start by proposing an analytical formula to estimate the mean number of contacts in purely hard-sphere connected (jammed) systems with *homogeneous* spatial distribution of the particles. Then we consider the arising of structural heterogeneity as due to the addition of an attractive component of interaction, in simulated Lennard-Jones glasses. In both cases we measure the mean contacts number $z$, after defining it properly, at different volume fractions $\phi$ and compare it with the theoretical predictions for the homogeneous system. On the basis of such comparisons we propose a criterion to indirectly assess the presence of structural heterogeneity in terms of the deviation of the measured $z(\phi)$ from the theoretical prediction for homogeneous systems presented here.

A deep and rapid enough quench of a supercooled liquid is able to cause the freezing-in of the liquid structure almost instantaneously, so that the resulting (solid) glassy state presents a spatial organization which cannot be distinguished, in practice, from that of the liquid snapshot at the quenching time. In the theoretical framework proposed by Alexander [18], this would correspond to the assumption that the set of interparticle distances in the *rigid* reference glass state, $\{\mathbf{R}\}$, is the same or approximately the same as the set of interparticle distances in the liquid snapshot, $\{\mathbf{r}\}$, at the quenching time $t_q$

$$\{\mathbf{R}\}_{glass} \approx \{\mathbf{r}(t_q)\}_{liquid} \tag{B.1}$$

In Alexander's approach, it is clear that the situation is more complicated than this simplistic picture because the particle positions in the liquid snapshot correspond to an unstable structure which is out of mechanical equilibrium [18]. Even for an ideal, infinitely fast quench, where diffusive motion is suppressed, a solid-like relaxation process accompanies the quench and is responsible for the creation, through a hierarchy of restructuring and buckling phenomena, of the mechanically stable structure which characterizes the final amorphous solid [18]. It is obviously impossible to theoretically describe such complex structural reorganization phenomena. What we attempt to do in the following is to find an analytical route which, starting from the liquid structure, leads to the number of contacts in the final arrested state.

The simplest case that one can think of is that of a liquid of hard spheres (where the interaction reduces to mutual impenetrability at contact). Upon quenching (at zero applied pressure) there is a unique *T*=0 state which is mechanically (marginally) stable and corresponds to the jamming point of hard spheres at volume fraction $\phi \approx 0.64$, denoted as point *J*, where (marginal) rigidity is expressed through the isostatic condition $z = 2d = 6$, where *z* is the mean coordination, and *d*=3 is the dimensionality of space. It was shown already by Maxwell that a structure possesses rigidity when the total number of degrees of freedom is at least equal to the total number of constraints $N_c$. That is, for a system of *N* particles, $N_c \geq 3N - 6$ [117]. As shown in recent literature, point *J* displays properties of a critical point where the Maxwell criterion is only marginally satisfied (isostaticity), and for large systems one thus has $z = 2N_c / N = 6$ [113]. At volume fractions lower than point *J*, the quenched metastable states are usually very fragile at $T > 0$ (and mechanically unstable at $T = 0$), unless an attractive interaction is present, since the elasticity of hard-sphere glasses then is due exclusively to particle caging. If, on the other hand, an attractive interaction is there, substantially rigid states can be obtained even at $\phi \ll 0.64$ [106]. In the latter case, however, the extent of attractive interaction also affects the spatial organization so that attraction favours the appearance of structural heterogeneity (typically in the form of clusters) and a higher probability of finding particles near contact (which manifests itself in higher nearest-neighbour peaks in the radial distribution function) [106].

Here we start by deriving an accurate analytical approximation for the short-range part of the radial distribution function of dense *hard-sphere* liquids which represent the most homogeneously disordered system known. Integrating the liquid rdf over a certain value of shell width may certainly give a flavour of how the probability of finding neighbours at that distance from the tagged particle changes as a function of density, but does not correspond to any well-defined quantity from a physical point of view. Instead we submit the liquid snapshot structure to a so called "hyperquenching" protocol [60]. This may be visualized as the process of expanding the particle size in the liquid snapshot until the particles come into contact with each other and each particle gains on average a certain finite number of contacts. An essential constraint to be satisfied is that the system must be an isostatic random close packing at point J, the jamming point of hard-spheres [56]. Mathematically speaking, such hyperquenching protocol can be realized by integrating the rdf of the hard-sphere liquid with an integration boundary which allows to recover point J at $\phi = 0.64$. The result of the integration with the boundary determined in this way is an estimate of the mean contact number, z, as a function of the volume fraction, for the arrested homogeneous hard-sphere system which recovers the isostatic packing at $\phi = 0.64$, i.e. point J.

The structure of a (disordered) fluid of N particles with diameter $\sigma$ is entirely described via the n-particle distribution function, defined by

$$g_N^{(n)}(\mathbf{r}^n) = \frac{\rho_N^{(n)}(\mathbf{r}_1,...,\mathbf{r}_n)}{\prod_{i=1}^{N} \rho_N^{(1)}(\mathbf{r}_i)} \tag{B.2}$$

For a uniform system $\rho_N^{(1)}(\mathbf{r}) = N/V = \rho$, i.e. the number density of the fluid. If the system is homogeneous and isotropic, we have $g_N^{(2)}(\mathbf{r}_1,\mathbf{r}_2) = g(r)$, i.e. the radial distribution function (rdf). The delta function representation of the rdf gives clear evidence of its geometrical meaning in terms of (random) correlations between particles positioned around a reference one [35]

$$\rho g(r) = \frac{\rho^2}{N} \int g_N^{(2)}(\mathbf{r},\mathbf{r}')d\mathbf{r}' = \left\langle \frac{1}{N} \sum_{i=1}^{N} \sum_{j=1}^{N} \delta(\mathbf{r} - \mathbf{r}_j + \mathbf{r}_i) \right\rangle \tag{B.3}$$

The direct correlation function, c(r), is defined by the Ornstein-Zernike convolution relation $h(r) = c(r) + \rho \int h(r')c(|\mathbf{r} - \mathbf{r}'|)d\mathbf{r}'$ which can be solved within the Percus-Yevick closure, $c(r) \approx [1 - e^{\beta U(r)}]g(r)$, where U is the interaction potential, the solution being piecewise analytic [35]. Since we are interested in the near-contact region, we can approximate the PY solution in the range $1 \leqslant r/\sigma \lesssim 1.1$ using the following formula [99],

$$g'(r/\sigma';\phi') = (1+\phi'/2)/(1-\phi')^2 - (9/2)\phi'(1+\phi')/(1-\phi')^3 (r/\sigma' - 1) \tag{B.4}$$

Correcting for the contact-value at high density and for the phase shift according to the Verlet-Weis prescription, leads to [35]

$$g(r/\sigma;\phi) = g(r/\sigma';\phi') + \delta g_1(r/\sigma) \tag{B.5}$$

where $\phi' \approx \phi - \phi^2/16$ and $\sigma' = \sigma(\phi'/\phi)^{1/3}$. The short-range term can be expressed as:

$$\delta g_1(r/\sigma) = \delta g_1(x) = (A/x)\exp\left[-\alpha(x-1)\right]\cos\left[\alpha(x-1)\right] \quad \text{(B.6)}$$

where $x = r/\sigma$, and $A$ and $\alpha$ are coefficients which are only functions of $\phi$ and contain the contact value of the rdf, $g(1;\phi)$. The advantage of this formulation is that the contact behaviour of the PY solution can be modulated by setting $g(1;\phi)$ equal to the value predicted by equations of state valid in the high-density regime, such as the Carnahan-Starling, as well as by numerical simulations. Eqs. (B.4)-(B.6), as can be easily verified, are in excellent agreement with the most recent exact rdf from computer simulations in the range $1 \leqslant x \lesssim 1.1$ [127]. However, it can be noted from the above equations that the rdf is given exclusively as a function of geometrical parameters, independent of temperature (as a consequence of there being no energy parameter in the hard-sphere interaction potential).

The integral of $\rho g(r)$ over a volume element $d^3r$ is just the number of atoms in that volume element and volume integration yields the total number of atoms minus the one at the origin, $\int \rho g(\mathbf{r})d^3r = N-1$. The volume integration of $\rho g(r)$ for isotropic systems

$$z = \rho \int_\sigma^r 4\pi r^2 g(r) dr = 24\phi \int_1^x x^2 g(x) dx \quad \text{(B.7)}$$

gives the number of particles $z$ in a shell of thickness $r - \sigma$ (or $x - 1$) around a given particle.

Seen from a different perspective, the integral in Eq. (B.7) represents a mapping between the original liquid state, described by $g(r)$ and a state with a mean number of *contacts* given by $z$. The latter state is obtained by expanding the size of the original particles in the liquid snapshot by an amount such that each particle end up being, in the final state, in physical contact with $z$ neighbours. We refer to this process as "hyperquenching" of the precursor liquid and we can schematically visualize it in two steps as follows:

(i) quenching the dense (hard-sphere) liquid instantaneously such that thermal motion is suppressed and the liquid snapshot structure is frozen-in;

(ii) increase the particle diameter from $\sigma$ to $\hat{\sigma} = \sigma + \varepsilon\sigma$, where $\hat{\sigma}$ is the sphere diameter in the jammed configuration while $\varepsilon$, the hyperquenching parameter, is a (small) number.

What one obtains from step (i) is clearly a configuration corresponding to the frozen-in liquid snapshot. Then, in step (ii), the reference particle comes into *contact* with a number of nearest neighbours which depends upon the value of $\varepsilon$. Note that since collisions among particles may occur in the liquid, at step (i) some particles may already be at contact. This requires displacing the contacted particles by an amount $\varepsilon\sigma$ at step (ii). Such an effect can be ignored if the $\varepsilon$ value is small in comparison with $\sigma$.

We thus proceed by first constructing the formula which yields the nearest-neighbours number in the liquid as a function of the dimensionless gap (or shell width) and volume fraction, and subsequently submit it to the hyperquenching procedure. The integral in Eq. (B.7) can be rewritten in the interparticle gap $l = (r-\sigma)/\sigma = x-1$, as $z(\varepsilon;\phi) \equiv 24\phi\int_0^\varepsilon (1+l)^2 g(l)dl$. Then, applying Eqs. (B.4)-(B.6), one can integrate analytically and we arrive at the following expression

$$z(\varepsilon;\phi) \equiv 24\phi\int_0^\varepsilon (1+l)^2 g(l)dl = 2\phi\left\{\varepsilon\left[4\frac{1+\phi'/2}{(1-\phi')^2}(3+3\varepsilon+\varepsilon^2) - \frac{9}{2}\phi'\frac{1+\phi'}{(1-\phi')^3}\varepsilon(6+8\varepsilon+3\varepsilon^2)\right]\right. \quad \text{(B.8)}$$
$$\left. +6(A/\alpha)\exp(-\alpha\varepsilon)\left[\alpha\exp(-\alpha\varepsilon)+\sin\alpha\varepsilon+\alpha(1+\varepsilon)(\sin\alpha\varepsilon-\cos\alpha\varepsilon)\right]\right\}$$

which gives the number of nearest-neighbours in the liquid within the dimensionless gap $\varepsilon$. Since we are interested in the high-density regime, we can evaluate the Verlet-Weis parameters, *A* and $\alpha$, using the contact-value of the rdf, $g(1;\phi')$, from the Hall equation of state for the fluid branch of the high-density hard-sphere system (which employs up to 7 virial coefficients and is thus valid in the high-density regime) [128], and we obtain

$$A = (3/4)\phi'^2(1-0.7117\phi'-0.114\phi'^2)/(1-\phi')^4 - [1-(\phi'/2)]/(1-\phi')^3$$
$$+ 0.4074(\phi'-0.9378)(1.3948-0.7365\phi'+\phi'^2)$$
$$\times (1.7988+1.9685\phi'+\phi'^2)/[(-0.7405+\phi')(1.2947-2.1357\phi'+\phi'^2)],$$

$$\alpha = 18(1-0.7117\phi'-0.114\phi'^2)/(1-\phi')[1-(\phi'/2)]$$
$$\times \{[1-(\phi'/2)]/(1-\phi')^3\}/\{0.4074(\phi'-0.9378)(1.3948-0.7365\phi'+\phi'^2) \quad \text{(B.9)}$$
$$\times (1.7988+1.9685\phi'+\phi'^2)/[(-0.7405+\phi')(1.2947-2.1357\phi'+\phi'^2)]\}$$

At this point, we can fix the hyperquench by assigning to $\varepsilon$ a value which recovers the critical point *J* of hard-spheres. Thus, from the latter condition we find that $z=6$, i.e. the isostatic solid, is obtained at $\phi = 0.64$ if one sets

$$\varepsilon = 0.03325 \quad \text{(B.10)}$$

This fixes the hyperquenching protocol in that now Eqs. (B.8)-(B.10) produce a unique

$z(\phi)$ curve which passes through point *J* (i.e. through $z=6$ at $\phi=0.64$). For illustration purposes we have plotted the dependence of *z* upon $\varepsilon$ at $\phi=0.64$ in Figure B-1.

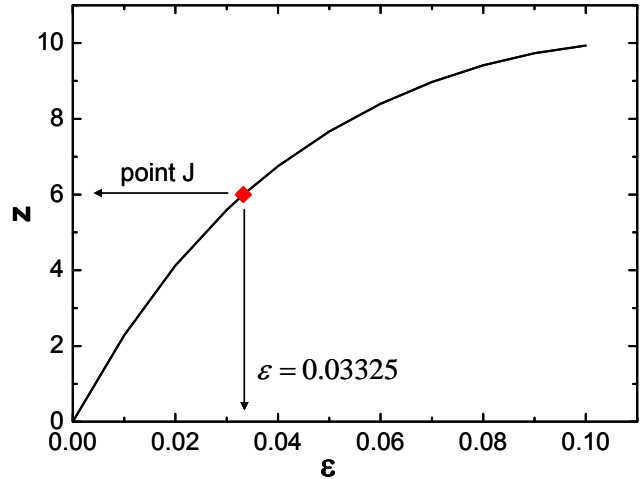

**Figure B-1: The contact number in the hyperquenched system as a function of the shell width $\varepsilon$ calculated from Eqs. (B.8)-(B.9) for $\phi=0.64$. The diamond identifies the value of the parameter $\varepsilon$ which recovers the isostatic jamming point of hard-spheres (*z*=6), that is $\varepsilon=0.03325$.**

In the presence of attraction, we may expect two distinct scenarios depending upon the volume fraction. In the first one, density is high enough that restructuring driven by attraction is strongly hindered (frustration), so that the structure is spatially homogeneous and a mean coordination number for the contacts close to the value predicted by Eqs. (B.8)-(B.10) should be expected at the end of the quench. In this case, $z<6$, the system is rigid due to initial stresses and harmonically restoring forces [64]. In the second scenario, density is not high enough (i.e. the system is not frustrated enough) to forbid larger scale restructuring and aggregation of the particles into locally denser regions (clusters), accompanied by a reduction in the internal energy of the system. Therefore when structural heterogeneity becomes important, the mean coordination is expected to be higher than the model prediction.

In order to test these speculations, also from a quantitative point of view, as well as the model itself, we have carried out simulations on model glassy systems with attraction. Molecular Dynamics (MD) simulations have been carried out on a binary mixture of 2024 atoms at with size ratio 1:1.2 interacting via two different cut-off Lennard-Jones-type potentials with exponents 20-10 and 60-30. The system was first driven into the deepest metastable minimum in the energy landscape at $T \approx 0.46$. At least three different glass configurations for each potential obtained in this way were then quenched to *T*=0.005 for a period of time of at least $10^6$ MD steps. Further, the simulations provide an indication of the mean local structure of the glass. A readily accessible quantity in this respect is provided by the average internal energy per particle, $<E_{int}>$.

For systems with an attractive component of interaction, there is a close relationship between mean coordination and the macroscopic elastic response, as shown also in Chapter 8 and in [64]. Attraction provides indeed restoring forces which can oppose an externally applied deformation, thus conferring rigidity to the system. It seems natural then to identify the mean coordination number with the number of mechanical contacts which are involved in the restoring forces. It is likely that the neighbours that give a major contribution to stress-bearing are those localized near the minimum of the potential well. In fact, in the Cauchy-Born approach for amorphous solids by Alexander [18], if one neglects non-affine relaxations (which are usually a small contribution for strongly-bonded glasses) the dominant term in the free energy expansion is the one involving the second derivative of the interaction potential evaluated at the minimum of the potential well. The neighbours which are positioned in the low-energy tail of the potential, instead, are expected to play a negligible role in the stress-bearing mechanism. Therefore we calculate the mean number of contacts in glasses with attraction

$$z \equiv \frac{1}{N}\sum_{l=1}^{N}\sum_{m=1}^{nn} w_m^{(l)} = \frac{1}{N}\sum_{l=1}^{N}\sum_{m=1}^{nn} U_m^{(l)}/U_0 \qquad (B.11)$$

i.e. for each particle *l* the number of mechanical contacts is determined as the weighted average number of its neighbours where for each *m*th neighbour of *l* the weight $w_m^{(l)}$ is given by its interaction potential energy level, $U_m^{(l)}$, normalized by the depth of the potential well, $U_0$. *nn* is the number of neighbours around the *l*th particle. In the case of truncated Lennard-Jones potentials, as they are often used in simulations, this is just the number of particles within the cut-off distance. For a schematic illustration of the terms in Eq. (B.11) see Figure B-2. The expression Eq. (B.11) corresponds to the internal energy per particle, $<E_{int}>$, as calculated from our MD simulations.

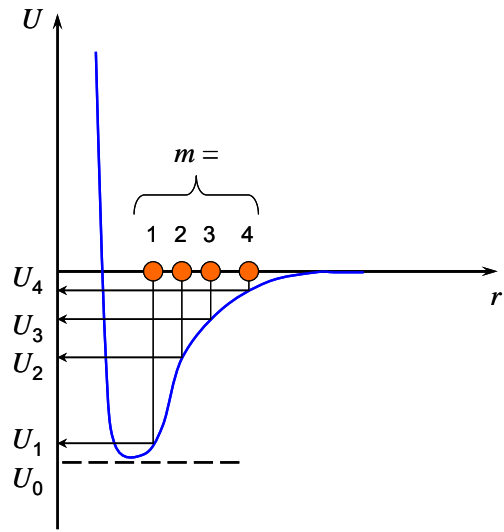

**Figure B-2: Schematic representation of an interaction potential with attraction (such as e.g. for Lennard-Jones systems). The arrows indicate the energy levels corresponding to the positions on the radial distance axis occupied by the neighbours. See Text and Eq. (B.11).**

In Figure B-3 distributions for the fraction of particles, $n(z)$, within a distance equal to the minimum of the potential-well having coordination number $z$, are shown for the two Lennard-Jones potentials investigated. We observe that in both cases the distribution tends to significantly broaden toward higher values of $z$ upon lowering the volume fraction below $\phi = 0.6$. Especially in the case of the 20-10 potential the distribution clearly develops a tail at high values of $z$ which tends to grow upon lowering the volume fraction. This indicates that, at lower volume fractions, the geometric frustration being less severe, particles can more easily rearrange during the quenching process to aggregate locally with a significantly higher value of $z$ (nearly by a factor two). It is clear that for this reason structural heterogeneity must increase upon lowering the volume fraction. Therefore, one can describe the arising of structural heterogeneity upon decreasing $\phi$ in terms of the formation of locally very dense regions (clusters) characterized by a coordination number significantly higher.

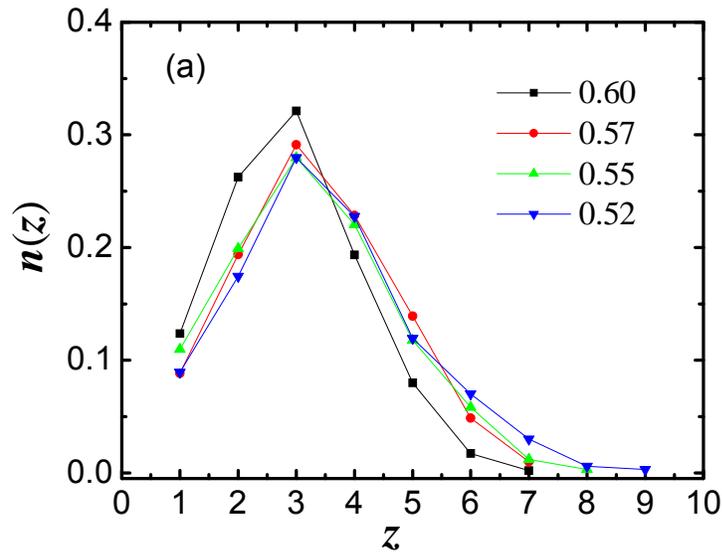

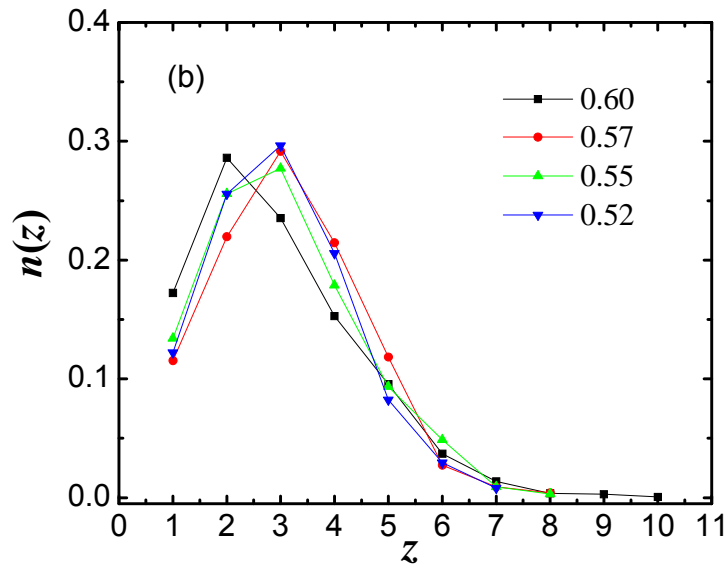

**Figure B-3: Fraction of particles having a number of neighbours *z* within the minimum of the potential well at different volume fractions (see legend). (a) Lennard-Jones 20-10 potential. (b) Lennard-Jones 60-30 potential.**

In Figure B-4 we report the theoretical $z(\phi)$ curve and the average values for the Lennard-Jones glasses evaluated according to Eq. (B.11). We observe that the *z* values for the Lennard-Jones systems are rather close to the theoretical predictions around $\phi = 0.6$ (where indeed we could not see signatures of structural inhomogeneity in Figure B.3) but then decrease much less rapidly upon decreasing $\phi$ and become increasingly higher with respect to the theoretical prediction.

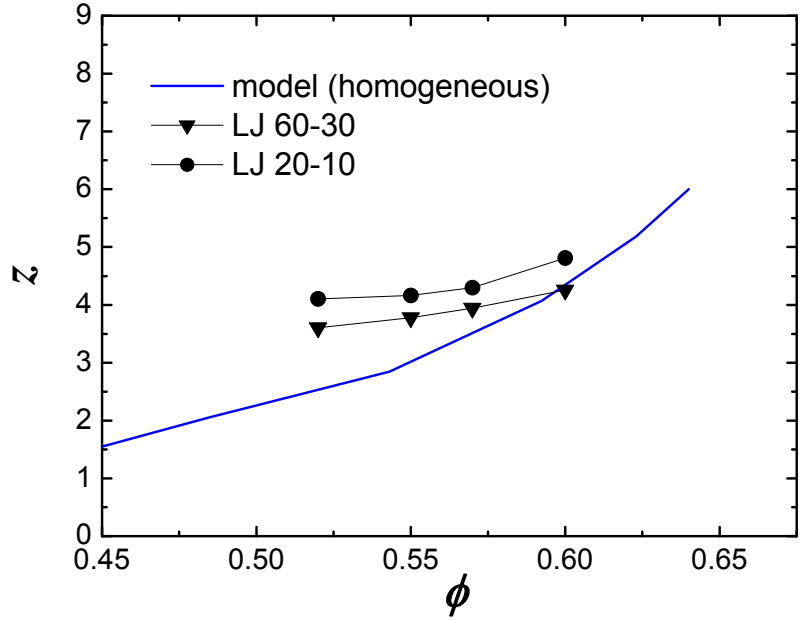

**Figure B-4: Comparison between the contact number as a function of volume fraction curve calculated from Eqs. (8)-(10) (solid line), and the contact numbers evaluated from simulations of Lennard-Jones glasses according to Eq. (11) (symbols).**

Based on what suggested by the analysis of Figure B.3, this can be explained if one thinks that, upon decreasing $\phi$, the locally denser regions still retain a very high value of contact number and make an important contribution to the measured sample-averaged value even at the lowest $\phi$. A much weaker dependence of the average $z$ upon $\phi$, as compared to the perfectly spatially homogeneous case, and the consequent higher values of $z$, could be then interpreted as hallmarks of significant structural inhomogeneity. The deviation increases upon decreasing $\phi$ as the structural inhomogeneity becomes more important. This suggests that the $z(\phi)$ curve of a specific glassy system can be used to assess the extent of structural heterogeneity in the system by comparing it with the theoretical curve, Eqs. (B.8)-(B.10), corresponding to the highest degree of structural *homogeneity*. The structural *inhomogeneity* is expected be more pronounced the more the $z(\phi)$ curve will depart from the theoretical one.

We derived an analytical formula (using liquid theory and the properties of the jamming point of hard-spheres) to estimate the mean number of contacts in hard-sphere systems as a function of the volume fraction. The predictions apply whenever the distribution of the particles in space is *homogeneously* disordered. Furthermore, we have investigated the contact number versus volume fraction in model amorphous solids where a significant attractive component of interaction is present (Lennard-Jones glass). In these cases we estimate the average contact number from the internal energy per particle. In the measured distributions of the contact number, we found clear signatures of the development of structural inhomogeneity upon lowering the volume fraction corresponding to a significant deviation from the homogeneous curve for $\phi < 0.6$. At higher $\phi$, geometric frustration due to packing constraints overwhelms the tendency to develop inhomogeneity induced by attraction and the theoretical estimate may have a universal validity (with spherical particles) independent of the microscopic interaction potential.

In conclusion, we have proposed an effective method to assess structural inhomogeneity in disordered media with spherical particles, in terms of the comparison of the mean contacts number measured as a function of $\phi$, and a theoretical curve obtained for homogeneous systems. Based on the cases examined, we suggested that values of *z* larger than the theoretical ones indicate significant structural inhomogeneity, which usually grows upon further lowering the volume fraction.

# References

[1] P.G. de Gennes, *Scaling Concepts in Polymer Physics* (Ithaca, Cornell University Press, 1979).

[2] J.M. Ziman, *Models of Disorder* (Cambridge, Cambridge University Press, 1979)

[3] M. Kleman and O.D. Lavrentovich, *Soft Matter Physics* p.552 (New York, Springer, 2003).

[4] K.M.N. Oates *et al.*, *Journal of the Royal Society-Interface* **3**, 167-174 (2006).

[5] S. Rammensee, U. Slotta, T. Scheibel, and A.R. Bausch, *Proc. Natl. Acad. Sci. U.S.A.* **105**, 18 (2008).

[6] H.J. Jin, D.L. Kaplan, *Nature* **424**, 1057-1061 (2003).

[7] J. Guery, E. Bertrand, C. Rouzeau, P. Levitz, D.A. Weitz, and J. Bibette, *Phys. Rev. Lett.* **96**, 198301 (2006).

[8] W. Götze and L. Sjogren, *Rep. Prog. Phys.* **55**, 241 (1992).

[9] C.B. Holmes, M.E. Cates, M. Fuchs, and P. Sollich, *J. Rheol.* **49**, 237 (2005).

[10] J.K.G. Dhont, *An Introduction to Dynamics of Colloids* (Amsterdam, Elsevier, 1996).

[11] E.M. Lifshitz and I.P. Pitaevskii, *Physical Kinetics* (New York, Pergamon Press, 1981).

[12] J. Bergenholtz, J. F. Brady and M. Vicic, *J. Fluid Mech.* **456**, 239 (2002).

[13] G.K. Batchelor and J.T. Green, *J. Fluid Mech.* **56**, 401 (1972).

[14] K. Binder and W. Kob, *Glassy Materials and Disordered Solids: An Introduction to their Statistical Mechanics* (World Scientific, Singapore, 2005).

[15] C. Brito and M. Wyart, *Europhys. Lett.* **76**, 149 (2006).

[16] K.N. Pham *et al.*, *Science* **296**, 5565 (2002).

[17] K.N. Pham, G. Petekidis, D. Vlassopoulos, S.U. Egelhaaf, W.C.K. Poon and P. Pusey, *J. of Rheol.* **52**, 649 (2008).

[18] S. Alexander, *Physics Reports* **296**, 65 (1998).

[19] A. Zaccone et al., *Phys. Rev. E* **79**, 061401 (2009).

[20] P. Coussot, *Rheometry of Pastes, Suspensions, and Granular materials: Applications in Industry and Environment* (Wiley, New York, 2005).

[21] P. Coussot, Q.D. Nguyen, H.T. Huynh, and D. Bonn , *J. Rheol.* **46**, 573 (2002).

[22] V.G. Levich, *Physicochemical Hydrodynamics* (Prentice-Hall, Englewood Cliffs, NJ, 1962).

[23] J.K.G. Dhont, *J. Fluid Mech.* **204**, 421 (1989).

[24] B.J. Ackerson and N.A. Clark, *Physica A* **118**, 221-249 (1983).

[25] S. Melis, M. Verduyn, G. Storti, M. Morbidelli, and J. Bałdyga, *AIChE J.* **45**, 7 (1999).

[26] R.A. Lionberger, *J. Rheol.* **42**, 843 (1998).

[27] G.I. Barenblatt, *Scaling, Self-similarity, and Intermediate Asymptotics*, pp. 39-43 (Cambridge University Press, Cambridge, 1996).

[28] H. Masliyah and S. Bhattacharjee, *Electrokinetic and Transport Phenomena*, pp. 446-447 (Wiley, Hoboken NJ, 2006).

[29] H. Risken, *The Fokker-Planck Equation* (Springer, Berlin, 1996).

[30] B.V. Derjaguin and L.D. Landau, *Acta Physicochimica USSR* **14**, 633 (141); E.J.W. Verwey and J.Th.G. Overbeek, *Theory of the Stability of Lyophobic Colloids* (Elsevier, New York, 1948); M. Vanni and G. Baldi, *Adv. Colloid and Interface Sci.* **97**, 151 (2002).

[31] S. Chandrasekhar, *Rev. Mod. Phys.* **15**, 1 (1943).

[32] H.A. Kramers, *Physica* **7**, 284-304 (1940).

[33] A. Zaccone, H. Wu, M. Lattuada, M. Morbidelli, *J. Phys. Chem. B* **112**, 1976 (2008).

[34] M. von Smoluchowski, Z. Phys. Chem. **92**, 129 (1917).

[35] J.P. Hansen and I.R. McDonald, *Theory of Simple Liquids* (Academic Press, London, 1986).

[36] C.I. Mendoza and I. Santamaria-Holek, *J. Chem. Phys.* **130**, 044904 (2009).

[37] R.G. Larson, *The Structure and Rheology of Complex Fluids* (Oxford University Press, New York, 1999).

[38] J.E. Sader, S.L. Carnie, D.Y.C. Chan, *Journal of Colloid and Interface Science*, **171**, 46 (1995).

[39] M.I. Mishchenko, L.D. Travis, *J. Quant. Spectrosc. Radiat. Transf.* **60**, 309 (1998).

[40] M. Lattuada, P. Sandkuhler, H. Wu, J. Sefcik, M. Morbidelli, *Adv. Colloid Interface Sci.* **103**, 33 (2003).

[41] H. Wu, A. Tsoutsoura, M. Lattuada, A. Zaccone, M. Morbidelli, *Langmuir* **26**, 2761 (2010).

[42] A. Einstein, *Ann. Phys.* **19**, 289 (1906).

[43] A. Einstein, *Ann. Phys.* **34**, 591 (1911).

[44] S. Arrhenius, Biochemical Journal **11**, 112 (1917).

[45] C.G. de Kruif, E.M.F. van Iersel, A. Vrij, W.B. Russel, *J. Chem. Phys.* **83**, 4717 (1985).

[46] L.D. Landau and E.M. Lifshitz, *Fluid Mechanics* (Oxford, Pergamon Press, 1987).

[47] L. Ehrl, M. Soos, M. Lattuada, *J. Phys. Chem. B* **113**, 10587 (2009).

[48] Z. Cheng, J. Zhu, P.M. Chaikin, S, Phan, W.B. Russel, *Phys. Rev.* E **65**, 041405 (2002).

[49] P.M. Chaikin, A. Donev, W.N. Man, F.H. Stillinger, S. Torquato, *Industrial & Engineering Chemistry Research* **45**, 6960 (2006).

[50] H.H. Winter and M. Mours, *Adv. Polym. Sci.* **134**, 167 (1997); E. Del Gado, A. Fierro, L. de Arcangelis, and A. Coniglio, *Phys. Rev. E* **69**, 051103 (2004).

[51] E. Bertrand, J. Bibette, V Schmitt, *Phys. Rev. E* **66**, 060401 (2002).

[52] P.C.F. Møller, S. Rodts, M.A.J. Michels, and D. Bonn, *Phys. Rev. E* **77**, 041507 (2008).

[53] E. Brown, N.A. Forman, C.S. Orellana, H. Zhang, B.W. Maynor, D.E. Betts, J.M. DeSimone, H.M. Jaeger, *Nature Mat.* **9**, 220 (2009).

[54] H. A. Barnes, *J. Non-Newt. Fluid Mech.* 81,133 (1999); P.C.F. Møller, A. Fall, and D. Bonn, *EPL* **87**, 38004 (2009).

[55] M. Born and H. Huang *Dynamical Theory of Crystal Lattices* (Oxford, Oxford University Press, 1954).

[56] C.S. O'Hern, L.E. Silbert, A.J. Liu, and S.R. Nagel, *Phys. Rev. E* **68**, 011306 (2003).

[57] B.A. DiDonna, and T.C. Lubensky, *Phys. Rev. E* **72**, 066619 (2005).

[58] A. Lemaitre, and C.E. Maloney, *J. Stat. Phys.* **123**, 415 (2006)

[59] C.E. Maloney, *Phys. Rev. Lett.* **97**, 035503 (2006).

[60] M. Wyart, *Annales de Physique* **30,** 1 (2005).

[61] M. Das, F.C. MacKintosh, and A.J. Levine, *Phys. Rev. Lett.* 2007 **99**, 038101 (2007).

[62] A. Zaccone, M. Lattuada, H. Wu, M. Morbidelli, *J. Chem. Phys.* **127**, 174512 (2007).

[63] M. Wyart *et al.*, *Phys. Rev. Lett.* **101**, 215501 (2008).

[64] A. Zaccone, H. Wu, and E. Del Gado, *Phys. Rev. Lett.* **103**, 208301 (2009).

[65] A. Tanguy *et al.*, *Phys. Rev. B* **66**, 17 (2002).

[66] L.J. Kaufman and D.A. Weitz, *J. Chem. Phys.* **125** 074716 (2006).

[67] K.N. Pham *et al.*, Europhys. Lett. **75,** 624 (2006).

[68] M. Sahimi, *Physics Reports* **306,** 214 (1998).

[69] R. Buscall *et al.*, *Journal of the Chemical Society-Faraday Transactions* I **78,** 2889 (1982).

[70] G. Ruocco, *Nature Materials* **7,** 842 (2008).

[71] J.C. Phillips, *J. of Non-Cryst. Sol.* **43,** 37 (1981).

[72] W.A. Kamitakahara *et al.*, *Bull. Am. Phys. Soc.* **32,** 812 (1987).

[73] H. Shintani and H. Tanaka, *Nature Materials* **7** 870 (2008).

[74] J.P. Pantina and E.M. Furst, *Phys. Rev. Lett.* **94** 138301 (2005).

[75] V. Becker and H. Briesen, *Phys. Rev. E* **71,** 061404 (2009).

[76] Y. Kantor, and I. Webman, *Phys. Rev. Lett.* **52,** 1891 (1998).

[77] P.N Pusey *J. Phys.: Condens. Matter* **20**, 494202 (2008).

[78] D. Frenkel, *Science* **296**, 65 (2002).

[79] V.J. Anderson and H.N.W. Lekkerkerker, *Nature* **416**, 811 (2002).

[80] W.C.K. Poon, A.D. Pirie, M.D. Haw and P.N. Pusey, *Physica A* **235**, 110 (1997).

[81] J. Bergenholtz, W.C.K. Poon and M. Fuchs, *Langmuir* **19**, 4493 (2003).

[82] E. Zaccarelli *et al.*, *Phys. Rev. E* **63**, 031501 (2001).

[83] K.S. Schweizer, *J. Chem. Phys.* **123**, 244501 (2005).

[84] A. M. Puertas *et al.*, *J. Chem. Phys.* **127**, 144906 (2007).

[85] Y.-L. Chen and K.S. Schweizer, *J. Chem. Phys.* **120**, 7212 (2004).

[86] A. Coniglio *et al.*, *J. of Phys.: Condens. Matter* **16**, S4831 (2004).

[87] A. Coniglio *et al.*, *J. of Phys.: Condens. Matter* **18**, S2383 (2006).

[88] H.M. Wyss *et al.*, *J. Am. Ceram. Soc.* **88**, 2337 (2005).

[89] F. Cardinaux *et al.*, *Phys. Rev. Lett.* **99**, 118301 (2007).

[90] M.H. Lee, and E.M. Furst, *Phys. Rev. E* **77**, 041408 (2008).

[91] C.O. Osuji C. Kim, D.A. Weitz, *Phys. Rev. E* **77**, 060402(R) (2008).

[92] P.N. Segre *et al.*, *Phys. Rev. Lett.* **86**, 6042 (2001).

[93] K. Kroy, M.E. Cates and W.C.K. Poon, *Phys. Rev. Lett.* **92**, 148302 (2004).

[94] M.E. Cates *et al.*, *J. of Phys.: Condens. Matter* **16**, S4861 (2004).

[95] E. Del Gado *et al. Europhys. Lett.* **63**, 1 (2003).

[96] E. Del Gado *et al. Phys. Rev. E* **69** 051103 (2004).

[97] H. Wu and M. Morbidelli, *Langmuir* **17**, 1030 (2001).

[98] W.H. Shih *et al.*, *Phys. Rev. A* **42**, 4772 (1990).

[99] D. Henderson and E.W. Grundke, *J. Chem. Phys.* **63**, 602 (1975).

[100] S. Asakura and F. Oosawa, *J. Chem. Phys.* **22**, 1255 (1954).

[101] H.N.W. Lekkerkerker *et al.*, *Europhys. Lett.* **20**, 559 (1992).

[102] M.P. Allen and D.J. Tildesley, *Computer Simulations of Liquids* (Clarendon Press, Oxford, 1987).

[103] S.J. Plimpton, *J. Comp. Phys.* **117**, 1 (1995).

[104] J.J. Crassous *et al.*, *J. Chem. Phys.* **128**, 204902 (2008).

[105] S. Ramakrishnan, Y.-L. Chen, K. S. Schweizer and C. F. Zukoski, *Phys. Rev. E* **70**, 040401(R) (2004).

[106] M. Laurati, G. Petekidis, N. Koumakis, F. Cardinaux, A.B. Schofield, J.M. Brader, M. Fuchs, S.U. Egelhaaf, *J. Chem. Phys.* **130**, 134907 (2009).

[107] V. Trappe *et al.*, *Nature* **411**, 772 (2001).

[108] A.D. Dinsmore *et al.*, *Phys. Rev. Lett.* **96**, 185502 (2006).

[109] E. Del Gado and W. Kob, *Phys. Rev. Lett.* **98**, 028303 (2007).

[110] E. Del Gado and W. Kob, *J. Non-Newt. Fluid Mech.* **149**, 28 (2008).

[111] C.E. Maloney and A. Lamaitre, *Phys. Rev. E* **74**, 016118 (2006).

[112] D.C. Wallace, *Thermodynamics of Crystals* (New York, Wiley, 1972).

[113] M. Wyart, S.R. Nagel, T.A. Witten, *Europhys. Lett.* **72**, 486 (2005).

[114] M. van Hecke, *J. of Phys.: Condens. Matter* **22**, 033101 (2009).

[115] E. Del Gado, P. Ilg, M. Kröger, H.C. Öttinger, *Phys. Rev. Lett.* **101**, 095501 (2008)

[116] W.G. Ellenbroek, E. Somfai, M. van Hecke, W. van Saarloos *Phys. Rev. Lett.* **97** 258001 (2006).

[117] J.C. Maxwell, *Philosophical Magazine* **27**, 294 (1864).

[118] W.G. Ellenbroek, Z. Zeravcic, W. van Saarloos, M. van Hecke, *Europhys. Lett.* **87**, 34004 (2009).

[119] H.A. Makse *et al. Phys. Rev. Lett.* **83**, 5070 (1999).

[120] P. G. de Gennes, *Comptes Rendus Physique* **3**, 1263 (2002).

[121] L. Cipelletti and L. Ramos, *J. of Phys.: Condens. Matter* **17**, R253 (2005).

[122] T. Aste, M. Saadatfar and T.J. Senden, *Phys. Rev. E* **71**, 061302 (2005).

[123] M.D. Haw, *Soft Matter* **2**, 950 (2006).

[124] R.M.L. Evans and M.D. Haw, *Europhys. Lett.* **60**, 404 (2002).

[125] H. Tanaka, S. Jabbari-Farouji, J. Meunier, and D. Bonn, *Phys. Rev. E*, **71** 021402 (2005).

[126] D. Henderson and E.W. Grundke, *J. Chem. Phys.* **63**, 601 (1975).

[127] A. Trokhymchuk *et al.*, *J. Chem. Phys.* **123**, 024501 (2005).

[128] K.R. Hall, *J. Chem. Phys.* **57**, 2252 (1972).

[129] J. Frenkel, *The Kinetic Theory of Liquids* (Oxford, Clarendon Press, 1946).

[130] B.I. Shklovskii and A.L. Efros, *Electronic Properties of Doped Semiconductors* (Berlin, Springer, 1984).

# CURRICULUM VITAE
## of Alessio Zaccone,
## born 07/09/1981 in Alessandria, Italy

# Personal Information

Nationality/status: Italian, not married

Place/date of birth: Alessandria (Italy), 7$^{th}$ of September 1981

Office address: Department of Chemistry and Applied Biosciences,
ETH Zurich
W. Pauli Str. 1, Hoenggerberg, HCI
8093 Zurich, Switzerland

# Professional Experience and Education

**01/2006-2010:** Research and Teaching Assistant and PhD Student, Department of Chemistry, ETH Zurich, Zurich, Switzerland.

**10/2005**: MSc graduation (Laurea) with full marks (grade: 110 out of 110) in Chemical Engineering from Politecnico di Torino, Turin, Italy.

**03/2005 – 09/2005**: MSc Thesis in Hydrodynamics at the Technical University of Berlin, Germany.

**09/1996 – 07/2000:** Graduation with full marks from Liceo Classico "G. Plana", Alessandria, Italy.

# Refereed Journals Publications

1) A. Zaccone, D. Gentili, H. Wu, and M. Morbidelli,
"Shear-induced reaction-limited aggregation kinetics of Brownian particles at arbitrary concentrations" The Journal of Chemical Physics 132, 134903 (2010).

2) A. Zaccone and E. Del Gado,
"On mean coordination and structural heterogeneity in model amorphous solids", The Journal of Chemical Physics 132, 024906 1-6 (2010).

3) H. Wu, A. Tsoutsoura, M. Lattuada, A. Zaccone, and M. Morbidelli
"Effect of temperature on high shear-induced gelation of charge-stabilized colloids without adding electrolytes", Langmuir 26, 2761–2768 (2010).

4) A. Zaccone, H. Wu, and E. Del Gado
"Elasticity of short-ranged attractive colloids: homogeneous and heterogeneous glasses", Physical Review Letters 103, 208301 1-4 (2009).

5) A. Zaccone, D. Gentili, H. Wu, and M. Morbidelli
"Theory of activated-rate processes under shear with application to the shear-induced aggregation of colloids", Physical Review E 80, 051404 1-8 (2009).

6) A. Zaccone
"The shear modulus of metastable amorphous solids with strong central and bond bending interactions", Journal of Physics: Condensed Matter 28, 285103 1-6 (2009).

7) A. Zaccone, M. Soos, M. Lattuada, H. Wu, M. U. Babler, and M. Morbidelli
"Breakup of dense colloidal aggregates under hydrodynamic stresses", Physical Review E 79, 061401 1-6 (2009).

8) H. Wu, A. Zaccone, M. Lattuada, and M. Morbidelli
"Liquid-like to solid-like transition of repulsive polymer colloids through high shear in a microchannel", Langmuir 25, 4715-4723 (2009).

9) A. Zaccone, H. Wu, M. Lattuada, and M. Morbidelli
"Charged molecular films on Brownian particles: structure, interactions and relation to stability", Journal of Physical Chemistry B 112, 6793-6802 (2008).

10) A. Zaccone, H. Wu, M. Lattuada, and M. Morbidelli
"Correlation between colloidal stability and surfactant adsorption/association phenomena studied by light scattering", Journal of Physical Chemistry B 112, 1976-1986 (2008).

11) A. Zaccone, H. Wu, A. Portaluri, M. Lattuada, and M. Morbidelli
"Mechanically stirred single-stage column for continuous gelation of colloidal systems", AIChE Journal 54, 3106-3115 (2008).

12) A. Zaccone, H. Wu, M. Lattuada, and M. Morbidelli
"Theoretical elastic moduli for disordered packings of interconnected spheres", The Journal of Chemical Physics 127, 174512 1-9 (2007).

13) A. Zaccone, A. Gäbler, S. Maaß, D. Marchisio, and M. Kraume
"Drop breakage in liquid-liquid stirred dispersions-modelling of single drop breakage", Chemical Engineering Science 62, 6297-6307 (2007).

14) S. Maaß, A. Gäbler, A. Zaccone, A. R. Paschedag, and M. Kraume
"Experimental investigations and modelling of breakage phenomena in stirred liquid/liquid systems", Chemical Engineering Research and Design 85 (A5) 703-709 (2007).

# Conference Proceedings

1) S. Maaß, A. Gäbler, M. Wegener, **A. Zaccone**, A. Paschedag, and M. Kraume
"Drop Breakage and Daughter Drop Distribution in Stirred Liquid/Liquid Systems and their Modelling within the Population Balance Equation"**. Proceedings of *12th European Conference on Mixing, Bologna,* 27.-30. June 2006**

2) H. Wu, **A. Zaccone,** M. Lattuada, and M. Morbidelli
"Rheological properties and structure of gels generated from stable polymer colloids through high shear in a microchannel". *15th International Congress on Rheology/80th Annual Meeting of the Society-of-Rheology Monterey, CA,* June 2008

# Invited Lectures

1) "Colloidal science at the interface of physical chemistry, soft matter and engineering", University of Cambridge, Department of Chemical Engineering, Cambridge, United Kingdom, 26/10/2006

2) "Elasticity of colloidal aggregates under turbulent flows" National University of Singapore, Department of Chemical Engineering, Singapore, 25/10/2007

3) "Adsorption of surfactant on colloidal particles studied by laser light scattering" Chinese University of Hong Kong, Department of Chemistry, 27/10/2007

4) "Elasticity of dense colloidal aggregates and its application to problems in the rheology of dispersions" ETH Zurich, Switzerland, Polymer Physics Seminar, 16/04/2008

5) "On a coarse-graining concept in colloidal physics with application to colloidal fluids, gels, and glasses in shearing fields" ETH Zurich, Switzerland, Computational Physics Seminar, 30/09/2009

6) "Shear-induced jamming and rheology of dilute Brownian suspensions" Lorentz Institute for Theoretical Physics, University of Leiden, The Netherlands, 23/10/2009

7) "Activated-rate processes in shear and application to shear-induced phenomena in soft matter" Helmholtz Institute for Materials and Energy (formerly Hahn-Meitner Institute), Berlin, Germany, 16/11/2009

# Oral Conference Contributions

1) A. Zaccone, H. Wu, M. Lattuada, and M. Morbidelli "Adsorption of ionic surfactants on polymer colloids studied by light scattering and its effect on colloidal stability" November 2007, AIChE Annual Meeting, Salt Lake City, USA

2) A. Zaccone, H. Wu, M. Lattuada, and M. Morbidelli "Study of short-range repulsive hydration forces between surfactant-coated colloidal particles" September 2007, European Colloid and Interfaces Society Conference, Geneva, Switzerland

3) A. Zaccone, H. Wu, M. Lattuada, and M. Morbidelli "Effect of hydration repulsion on the shear-induced aggregation of colloids" September 2008, European Colloid and Interfaces Society Conference, Krakow, Poland

4) A. Zaccone, H. Wu, and E. Del Gado "Modeling the elastic response of amorphous solids with attraction" November 2008, AIChE Annual Meeting, Philadelphia, USA

5) A. Zaccone, H. Wu, and E. Del Gado "Elasticity of short-ranged attractive colloids: homogeneous and heterogeneous glasses" June 2009, American Chemical Society-Colloids&Interfaces Division Meeting, New York, USA.